\documentclass[11pt,a4paper]{article}
\usepackage{jheppub}
\usepackage{comment} 
\renewcommand\beforetochook{\pagestyle{myplain}\pagenumbering{roman}\begingroup\small}
\renewcommand\afterTocRuleSpace{\par\endgroup\clearpage}
\usepackage[T1]{fontenc}
\usepackage{lmodern,microtype,amsthm,mathtools,mathrsfs}
\usepackage{booktabs,array,tabularx,longtable,graphicx,xcolor,tikz}
\usetikzlibrary{arrows.meta,decorations.markings,positioning,calc}
\usepackage{enumitem,caption,xurl,adjustbox,float}
\hypersetup{colorlinks=true,linkcolor=blue!45!black,citecolor=blue!45!black,urlcolor=blue!45!black,pdftitle={From boundary data to bound states IV: Waveforms}}
\definecolor{routea}{RGB}{32,88,150}
\definecolor{routeb}{RGB}{192,110,25}
\definecolor{bound}{RGB}{25,120,88}
\definecolor{rem}{RGB}{170,42,42}
\tikzset{arr/.style={postaction={decorate},decoration={markings,mark=at position .56 with {\arrow{Stealth}}}},parrow/.style={->,dashed,thick,>=Stealth}}
\setlist{itemsep=3pt,topsep=4pt}
\numberwithin{equation}{section}
\newcommand{\ii}{i}
\newcommand{\dd}{\mathrm d}
\newcommand{\uret}{u_{\rm ret}}
\newcommand{\E}{\mathcal E}
\newcommand{\Ord}{\mathcal O}
\newcommand{\Korb}{\mathcal K}
\newcommand{\Cnorm}{\sqrt{\pi/5}}
\preprint{DESY-26-173\, \\\phantom{~}}
\title{From boundary data to bound states IV: Waveforms}
\author[a]{\large Gregor K\"alin}
\author[b]{\large and Rafael A. Porto}
\affiliation[a]{Paul Scherrer Institut, CH-5232 Villigen PSI, Switzerland}
\affiliation[b]{Deutsches Elektronen-Synchrotron DESY, Platanenallee 6, 15738 Zeuthen, Germany.}
\emailAdd{gregor.kaelin@psi.ch}\emailAdd{rafael.porto@desy.de}

\abstract{We extend the boundary-to-bound (B2B) dictionary to gravitational waveforms, deriving frequency-domain relations between mass- and current-type modes for hyperbolic and elliptic motion. We identify the continuation and contour conditions required to recover bound harmonics from scattering data. In the adiabatic approximation, the construction applies to source multipoles evaluated on local-in-time equations of motion to all orders in the Post-Newtonian (PN) expansion. Propagation through the mass monopole, including tails and their iterations, introduces multiplicative factors that do not modify the gravitational-wave frequency and therefore naturally preserve the B2B relations. We provide two paradigmatic examples of the B2B correspondence, at the Newtonian and 1PN level, respectively. Finally, we investigate the reconstruction of bound harmonics from truncated post-Minkowskian (PM) scattering waveforms. We highlight the limitations of a direct implementation of the B2B map to waveforms computed to a finite order in $G$ and explore a PN-informed bootstrap approach. Within a specified class of functions, we reconstruct the instantaneous bound mass-quadrupole harmonic at 1PN from scattering data through $G^3$. We also show how the bootstrap approach, combined with the B2B map in the conservative sector, allows us to incorporate a series of corrections to all orders in the velocity.}
\begin{document}
\maketitle

\section{Introduction}
\label{sec:conventions}
Gravitational scattering offers a relativistic description of the two-body problem in which the interaction can be expanded in Newton's constant without expanding in velocity.  Building on the worldline effective field theory (WEFT) approach \cite{Goldberger:2004jt,Porto:2005ac,Rothstein:2014sra,Porto:2016pyg,Goldberger:2022ebt}, worldline frameworks \cite{Kalin:2020mvi,Kalin:2022hph,Mogull:2020sak,Jakobsen:2022psy} and scattering-amplitude methodologies \cite{Cheung:2018wkq,Kosower:2018adc,Bern:2019nnu,Herrmann:2021tct} have turned the post-Minkowskian (PM) expansion into a successful arena to study gravitational dynamics in the relativistic regime. The relativistic scattering impulse is known through fourth PM (4PM) order for arbitrary mass ratios \cite{Dlapa:2022lmu,Jakobsen:2023hig}, and also restricted to conservative-only terms \cite{Bern:2019nnu,Kalin:2020fhe,Bern:2021dqo,Dlapa:2021vgp,Jakobsen:2023ndj}. At fifth order, the expansion in the mass ratio has yielded complete scattering dynamics at first self-force (SF) order \cite{Driesse:2024xad}, including radiative effects \cite{Driesse:2024feo}. Partial results are also known through 2SF order \cite{Driesse:2026qiz,Bern:2025wyd} (see also \cite{Porto:2024cwd,Porto:2026fsd}). Furthermore, calculations in worldline approaches and amplitude methods give access to the scattering waveform~\cite{Mougiakakos:2021ckm,Jakobsen:2021smu,Brandhuber:2023hhy,Herderschee:2023fxh,Georgoudis:2023lgf,Brunello:2025eso,Georgoudis:2023eke,Georgoudis:2024pdz,DeAngelis:2026kew}, with detailed comparisons to the multipolar post-Minkowskian (MPM) formalism \cite{Bini:2023fiz,Bini:2024rsy}. 

The question for inspirals, however, remains:  {\it How can we harness scattering information to describe gravitational-wave (GW) radiation from bound orbits?}\vskip 4pt 
For conservative dynamics, a possible route is to use scattering information to reconstruct the form of an effective metric or two-body Hamiltonian \cite{Neill:2013wsa,Damour:2016gwp}. The latter can then be used to describe bound motion. Although powerful, this approach still relies on gauge-dependent intermediate steps. Inspired by the remarkable connection between the distance of closest approach and the turning points in hyperbolic and elliptic motion \cite{Kalin:2019rwq},  the boundary-to-bound (B2B) dictionary between gauge-invariant observable quantities was constructed: first in the conservative regime by relating scattering and bound radial actions \cite{Kalin:2019rwq,Kalin:2019inp}, and later on in the radiative regime by connecting integrated GW losses in an adiabatic approximation (for `source' quantities) \cite{Cho:2021arx}---both via an analytic continuation in energy and angular momentum. The separation of local dynamics from nonlocal hereditary effects is essential to this program. For instance, in the conservative sector, tail interactions are nonlocal in time starting at 4PN order \cite{Damour:2014jta,Galley:2015kus}, which must be subtracted prior to implementing the B2B analytic continuation. The separation between the local and nonlocal tail contributions in the PN expansion was carried over in \cite{Bini:2020nsb,Bini:2020hmy} to 6PN order, whereas the introduction of memory and second-order radiation-reaction effects was studied in detail in \cite{Porto:2024cwd,Almeida:2024lbv,Porto:2026fsd} to 5PN order, and extended to 6PN in \cite{Brunello:2026mkv}. Recent PM calculations within the WEFT framework isolate the local dynamics at 4PM and at 5PM/first-self-force (1SF) order, with complementary higher-self-force results to 10SF for nonlocal tail effects \cite{Dlapa:2024cje,Bini:2024tft,Dlapa:2025biy,Dlapa:2026oyq}. \vskip 4pt

Since the instantaneous GW fluxes can be also reconstructed from one-point functions, the existing B2B map for the radiated (sourced) energy \cite{Cho:2021arx} suggests that an extension to the waveforms themselves may be possible. Following this idea and building on the analytic continuation between
hyperbolic and elliptic motion established in~\cite{DD}, the authors
of~\cite{Adamo:2024oxy} explored the continuation of scattering waveforms in the
time domain. Their analysis, however, was limited to tree level and through 1PN
order. Moreover, as emphasized in \cite{Adamo:2024oxy}, their continued expression in the PM regime does
not recover periodic bound motion, which instead requires a resummation scheme.\vskip 4pt  

In this paper we focus instead
on a map connecting scattering Fourier transforms to the discrete
harmonics of bound motion. We formulate a B2B waveform relation
for arbitrary mass- and current-type multipoles of a nonspinning
planar binary that is valid to all PN orders. Following the notation of
\cite{Kalin:2019rwq,Kalin:2019inp,Cho:2021arx},
the waveform B2B correspondence takes the schematic form
(for the mass-type harmonics)
\vspace{-0.1cm}
\begin{equation}\label{eq1}
\vspace{-0.1cm}
 U^{b}_{\ell m,k}({\cal E}^-,J)
 =\frac{1}{T_r}
 \left[
 S^{U}_{\ell m}({\cal E}^-,J,\omega)
 -(-1)^{\ell+k}
 S^{U}_{\ell,-m}({\cal E}^-,-J,\omega)
 \right]_{\omega=\omega_{km}}.
\end{equation}
Here $U^{b}_{\ell m,k}$ is the bound Fourier coefficient,
$S^{U}_{\ell m}$ the scattering Fourier transform,
${\cal E}^-<0$ the binding energy, $J$ the angular
momentum, $T_r$ the radial period, and $\omega_{km}$ the bound frequencies.  An analogous map holds for the current harmonics. The B2B continuation is performed in two steps:
 first it follows the energy route, and secondly the root-exchanging $J\to-J$ path, but for the associated 
{\it reflected} mode, $m \to -m,$ and including the $(-1)^\ell$ $J$-parity factor. The notation specifies the endpoint of the
 continued branch, not necessarily a direct substitution. These continuations are performed at
fixed frequency, before setting $\omega \to \omega_{km}$.  The origin of the extra phase can be understood from the (one-passage) orbital condition: ${\rm exp}\left[-\tfrac{i}{2} (\omega_{km} T_r - m\Delta\phi)\right]= e^{-i \pi k}$, where $\Delta \phi/2\pi=\Korb$ and $2\pi(\Korb-1)$ is the periastron advance over a full radial period.\vskip 4pt 

In this paper, we formalize the above steps and establish the validity of the B2B correspondence in \eqref{eq1} within a general class of local PN
Hamiltonians and source multipole moments---thus excluding nonlocal conservative tail effects. By contrast, propagation tails and repeated interactions with the (stationary) mass monopole act as frequency-dependent `form factors' for each source multipole---yielding so-called `radiative' multipoles\cite{Blanchet:2013haa,Ross:2012fc,Porto:2016pyg}. These can therefore be readily incorporated into the B2B relation in \eqref{eq1}. As a direct application of the B2B map we study Newtonian and 1PN order, where the analytic continuation can be performed explicitly. 

These realizations then led us to the  practical question: {\it How much of the bound waveform can be recovered from scattering data to a finite PM order?}\vskip 4pt  Unlike previous perturbative B2B realizations \cite{Kalin:2019rwq,Kalin:2019inp,Cho:2021arx}, the scales in the waveform problem immediately illustrate how a truncated expression cannot be directly extrapolated through the B2B dictionary. To resolve this issue, we explore here a {\it bootstrap} approach through a frequency-domain completion that is constrained by PN information. We introduce a
frequency-dependent combination including a phase term that reduces to the B2B relation at the bound harmonics. At 1PN, a finite class of frequency
functions together with the known precession contribution allows scattering data to fix the remaining coefficients.\vskip 4pt The same strategy suggests how to incorporate into the waveforms some of the full velocity dependence inherited from PM data: after aligning the radiative frames, the tree and one-loop results constrain a frequency-domain completion whose Newtonian and 1PN limits are already known. This naturally leads us to a bootstrap implementation in which the analytic structure of the special (Bessel) functions entering at a given PN order are then used as a {\it background} in which the conservative B2B dictionary \cite{Kalin:2019rwq,Kalin:2019inp} can supply all-orders-in-velocity information at a given order in $G$.\vskip 4pt  Building on the 1PN waveforms, we identify the clock contribution in the tree and loop and follow the angular terms supplied by conservative dynamics. In the probe limit, the clock generates logarithmic terms that agree with scattering data through $G^5$, upon PN expanding \cite{Geralico:2026kbm,Geralico:2026efi}. The square of the conservative precession factor also fixes an angular contribution that agrees with the $G^5$ result through 2PN. These comparisons isolate contributions whose frequency dependence is guided by the PN calculation, while the conservative clock and precession retain their relativistic energy dependence. Reconstructing the full waveform, however,  requires additional information. We will discuss further details of the bootstrap approach elsewhere.\vskip 4pt

This paper is organized as follows. In \S\ref{sec:emission} we fix our conventions.
In \S\ref{sec:theorem} we develop the waveform B2B relation, separating the analytic continuation of the local source, contour matching and propagation tails. We apply in
\S\ref{sec:applications} the B2B construction to Newtonian
and 1PN waveforms. In \S\ref{sec:PM-bound} we explore what finite-order PM data determine
and how to implement a PN-informed bootstrap approach to reconstruct bound waveforms. We conclude in \S\ref{sec:conclusions} with a summary and an outlook. The appendices contain detailed contour arguments and
calculations underlying the correspondence. Throughout this paper we follow the notation of previous B2B studies  \cite{Kalin:2019rwq,Kalin:2019inp,Cho:2021arx}, see table below. Except for $\Delta\phi$, which we use here for the total change in the angle, and the separate symbol $P_\infty$, which prevents confusing momentum with the boost factor, $p_\infty =\sqrt{\gamma^2-1}$, often used in the PM waveform literature \cite{Bini:2024rsy}. 
  \begin{table}[h!]
\centering\small
\begin{tabularx}{\linewidth}{@{}lX@{}}\toprule
Symbol & Definition and role ($c=1$)\\\midrule
$M,\mu,\nu$ & Total mass $M=m_1+m_2$, $\mu=m_1m_2/M$, $\nu=\mu/M$.\\
$E,\E,\Gamma,\epsilon$ & Total energy $E=E_1+E_2$, $\E=(E-M)/\mu$, $\Gamma=E/M=1+\nu\E$, and $\epsilon=-2\E>0$ at the bound endpoint.\\
$\gamma,p_\infty$ & $\gamma=(E^2-m_1^2-m_2^2)/(2m_1m_2)
=1+\E+\nu\E^2/2$; $p_\infty=\sqrt{\gamma^2-1}$ is dimensionless.\\
$P_\infty,b$ & Physical centre-of-mass momentum $\mu p_\infty/\Gamma$;
impact parameter $b=J/P_\infty$.\\
$J,j$ &  Orbital angular momentum $J$; $j=J/(GM\mu)$.\\
$s$ & Bound continuation parameter: $p_\infty\to\ii s$,
$s^2=1-\gamma^2>0$.\\
$r_-,r_+$ &  Periastron and apoapsis roots; $r_{\min}$ is the scattering root.\\
$R_{\rm obs}, r$ & Observer's far-zone radius, orbital separation.\\
$t_{\rm obs},\uret,t$ & Observer coordinate time, radiative retarded time, and orbital source time.\\
$T_r,\Omega_r,\Omega_\phi$ & Bound radial period in orbital time, $2\pi/T_r$, and $\Delta\phi/T_r$.\\
$\Korb,k,p$ & $\Korb=\Omega_\phi/\Omega_r$; integer radial label $k$;
the Newtonian harmonic label $p=k+m$.\\
$\Korb_s$ & Coefficient multiplying the angular anomaly in hyperbolic motion; its continuation gives $\Korb$.\\
$\omega_{km}$ & $\omega_{km}=k\Omega_r+m\Omega_\phi$; for $22$,
$\omega_p=[p+2(\Korb-1)]\Omega_r$.\\
$X,e$ & $X=j\sqrt{\epsilon}$, $e^2=1-X^2$: Newtonian reference variables after energy continuation.\\
$x$ & $x=\omega J/(\mu\epsilon)$: dimensionless frequency held fixed in the $1/X$ expansion.\\
$I_L\,,\ J_L$ & STF mass and current source moments; $L=i_1\cdots i_\ell$.\\
$U_L,V_L;\ U_{\ell m},V_{\ell m}$ & Radiative STF moments and their spin-weighted angular coefficients.\\
$\mathcal M$ & Stationary ADM mass monopole: $\mathcal M=E=\Gamma M$.\\
$S^U_{\ell m},S^V_{\ell m}$ & Scattering transforms of the mass and current radiative modes.\\
$U^b_{\ell m,k},V^b_{\ell m,k}$ & Corresponding bound coefficients, normalized by $1/T_r$.\\
\bottomrule\end{tabularx}
\end{table}

 \newpage
\section{Gravitational wave emission}\label{sec:emission}
\subsection{Radiative modes}
The source is described by symmetric-trace-free (STF) mass moments
$I_L(t)$ and current moments
$J_L(t)$, with $L=i_1\cdots i_\ell$ and
$\ell\geq2$ for radiation. At Newtonian order, for point masses labelled by
$A$ with centre-of-mass positions $\mathbf x_A$ and velocities $\mathbf v_A$,
\begin{equation}
 I_L^{\rm N}=\sum_A m_A x_A^{\langle i_1}\cdots x_A^{i_\ell\rangle},
 \qquad
 J_L^{\rm N}=\sum_A m_A x_A^{\langle i_1}\cdots x_A^{i_{\ell-1}}
       (\mathbf x_A\times\mathbf v_A)^{i_\ell\rangle}.
 \label{eq:source-multipoles-general}
\end{equation}
Angle brackets mean STF projection over the enclosed indices. In the WEFT approach, the multipole moments are matched coefficients coupled to the electric and magnetic components of the Weyl tensor. They acquire  
PN corrections from the matter and near-zone gravitational fields \cite{Blanchet:2013haa,Goldberger:2009qd,Ross:2012fc,Porto:2016pyg,Amalberti:2024jaa}.
Restoring $c$, to illustrate the power counting, we have (for nonspinning sources)
\begin{equation}
 I_L=I_L^{\rm N}+c^{-2}I_L^{\rm 1PN}+c^{-4}I_L^{\rm 2PN}+\cdots,
 \qquad
 J_L=J_L^{\rm N}+c^{-2}J_L^{\rm 1PN}+c^{-4}J_L^{\rm 2PN}+\cdots.
 \label{eq:source-PN-counting}
\end{equation}
Time derivatives must use equations of motion at the same level of accuracy. For local sources, logarithms as well as the associated poles in dimensional regularization must be retained to obtain consistent results~\cite{Goldberger:2009qd,Amalberti:2024jaa}. The radiation field is described by the radiative moments $U_L,V_L$.
For a single emitting multipole, separate direct emission from its
hereditary propagation through the stationary mass monopole:
\begin{align}
 U_L(\uret)&=U_L^{\rm src}(\uret)+U_L^{\rm hered}(\uret),&
 U_L^{\rm src}(\uret)&=I_L^{(\ell)}(t)\big|_{t=\uret},\\
 V_L(\uret)&=V_L^{\rm src}(\uret)+V_L^{\rm hered}(\uret),&
 V_L^{\rm src}(\uret)&=J_L^{(\ell)}(t)\big|_{t=\uret}.
 \label{eq:radiative-source-structure}
\end{align}
A parenthesized superscript denotes orbital-time derivatives. We restrict ourselves to hereditary effects in the linear emission (tail) sector only. Couplings of several time-dependent
multipoles can generate additional hereditary terms
\cite{Blanchet:2013haa,Porto:2024cwd,Porto:2026fsd}, which are outside of the scope of our work here.\vskip 4pt 

To fix the angular normalization, let $\mathbf n$ be the observer direction
and $\bar{\mathbf m}=(\mathbf e_\vartheta-\ii\mathbf e_\varphi)/\sqrt2$
the polarization vector in the standard spherical basis. We write
$n_{L-2}=n^{i_1}\cdots n^{i_{\ell-2}}$ and take $\epsilon_{123}=+1$. We follow the conventions in~\cite{Bini:2024rsy}:
\begin{align}
 \sum_{m=-\ell}^{\ell}U_{\ell m}\,{}_{-2}Y_{\ell m}(\mathbf n)
 &=\frac{1}{\ell!}\bar m^i\bar m^j n_{L-2}U_{ijL-2},\\
 \sum_{m=-\ell}^{\ell}V_{\ell m}\,{}_{-2}Y_{\ell m}(\mathbf n)
 &=-\frac{2\ell}{(\ell+1)\ell!}
   \bar m^i\bar m^j n^c n_{L-2}\epsilon_{cdi}V_{jdL-2}.
 \label{eq:STF-mode-projection}
\end{align}
 In particular,
the current projection already includes its complex polarization phase, such that no additional $-\ii$ multiplies~$V_{\ell m}$. All moments in this equation are evaluated at $\uret$. The strain is given by:
\begin{align}
 h_+-\ii h_\times&=\sum_{\ell=2}^{\infty}\sum_{m=-\ell}^{\ell}
       h_{\ell m}\,{}_{-2}Y_{\ell m},\\
 h_{\ell m}(t_{\rm obs},R_{\rm obs})
 &=4G[U_{\ell m}+V_{\ell m}]/R_{\rm obs}
   +\Ord(R_{\rm obs}^{-2}).
 \label{eq:strain-general}
\end{align} 

As an example, take
$I_{ij}^{\rm N}=\mu x^{\langle i}x^{j\rangle}$ and define the scalar
$Q(t)=I_{xx}(t)-I_{yy}(t)-2\ii I_{xy}(t)$. Then
$Q^{\rm N}=\mu(x-\ii y)^2$, and
\begin{equation}
\label{eq:Q0}
 U_{22}^{\rm N}(\uret)=\frac{\Cnorm}{2}
 \left.\frac{\dd^2 Q^{\rm N}(t)}{\dd t^2}\right|_{t=\uret}.
\end{equation}
The same expression with the 1PN source and equations of motion gives
the complete instantaneous $22$ mode through 1PN order.

\subsection{Spectra versus harmonics}
\label{sec:observables}

Our Fourier convention and bound normalization is given by:
\begin{align}
 S^U_{\ell m}(E,J;\omega)&=\int_{-\infty}^{\infty}
 e^{\ii\omega\uret}U_{\ell m}(\uret;E,J)\dd\uret,\label{eq:Sdef}\\
 U^b_{\ell m,k}(E,J)&=\frac1{T_r}\int_0^{T_r}
 e^{\ii\omega_{km}\uret}U_{\ell m}(\uret;E,J)\dd\uret,
 \label{eq:bounddef}
 %S^V_{\ell m}(E,J;\omega)&=\int_{-\infty}^{\infty}
% e^{\ii\omega\uret}V_{\ell m}(\uret;E,J)\dd\uret,\\
% V^b_{\ell m,k}(E,J)&=\frac1{T_r}\int_0^{T_r}
% e^{\ii\omega_{km}\uret}V_{\ell m}(\uret;E,J)\dd\uret.
 %\label{eq:current-def}
\end{align}
and similarly for $V_{\ell m}$ modes. The scattering transform integrates an entire encounter, while the bound coefficient averages over one radial period $T_r$. We choose $t=0$ and $\phi=0$ at periastron. The period is measured in orbital coordinate time, normalized at infinity. After a full radial period the radial variables return to their initial values and the angle advances by $\Delta\phi$. A local mode proportional
to $e^{-im\phi}$ therefore obeys
\begin{equation} U_{\ell m}(t+T_r)=e^{-im\Delta\phi}U_{\ell m}(t),\end{equation} and similarly for
$V_{\ell m}$. Hence,
\begin{equation}
 \omega_{km}T_r-m\Delta\phi=2\pi k \, \longrightarrow \, \omega_{km}=[p+m(\Korb-1)]\Omega_r,\quad p=k+m\in\mathbb Z \,.
 \label{eq:periodphase}
\end{equation}
The periodicity condition plays a key role in fixing the relative phase of the analytic continuation of scattering contributions.\vskip 4pt 

Changing to observer time at fixed $R_{\rm obs}$ gives $\dd\uret=\dd t_{\rm obs}$ and a common Fourier phase. For instance, for the leading  term,
\begin{equation}
 \int\dd t_{\rm obs}\,e^{\ii\omega t_{\rm obs}}h_{22}(t_{\rm obs},R_{\rm obs})
 =\frac{4G}{R_{\rm obs}}e^{\ii\omega R_{\rm obs}}S^U_{22}(\omega)\,.
\end{equation}
We use the Fourier transform in retarded time throughout, so this observer phase is absent from both the scattering  and bound coefficients. We align the source clock with radiative time,
$t=\uret$, at the evaluation point. This identifies their origins and
rates. (A hereditary term samples earlier source times. We return to this point in \S\ref{sec:tails}.) Therefore, for an instantaneous mode, we can perform a direct change of integration variable,
\begin{equation}
 \int\dd\uret\,e^{\ii\omega\uret}U^{\rm inst}_{\ell m}(\uret)
 =\int\dd t\,e^{\ii\omega t}U^{\rm inst}_{\ell m}(t), 
\end{equation}
and similarly for $V_{\ell m}$, such that the integral is evaluated on the source orbit. In our derivations below we suppress the ``inst''  (or ``src'') label. Hereditary propagation is treated separately in \S\ref{sec:tails}. In addition, the notation $U^b_{22,p}$ or $V^b_{21,p}$ below uses $p$ rather than $k$. Hence, the same $p$ implies $k=p-2$ and $k=p-1$ for each channel, respectively.

\subsection{Monopole tails}
\label{sec:emission-frequency}
In our B2B map we separate emission
from propagation. In the WEFT approach for PN sources the near-zone dynamics is matched
onto time-dependent moments coupled to radiation fields
\cite{Goldberger:2004jt,Goldberger:2009qd,Porto:2016pyg}. Potential modes have
$k^0\sim v/r$, $|\boldsymbol k|\sim1/r$, whereas radiation has
$k^0\sim|\boldsymbol k|\sim v/r$. Integrating out the former determines
the local dynamics and multipole coefficients, while propagation through the
exterior field determines the response~\cite{Galley:2015kus,Porto:2024cwd,Almeida:2024lbv,Porto:2026fsd}. 

The  B2B dictionary acts first on the local source emission. Propagation through the background is a separate problem. For the
 source monopole,
$\mathcal M \equiv \Gamma M$, GW emission can be organized in terms of new, often denoted as `radiative' multipole moments \cite{Blanchet:2013haa},
\begin{align}
 \widetilde U^{\rm {\cal M}}_{\ell m}
 &=\mathcal D^U_\ell\widetilde U^{\rm src}_{\ell m},&
 \mathcal D^U_\ell&=1+\mathcal T^U_{\ell,1}
                      +\mathcal T^U_{\ell,2}+\cdots,\\
 \widetilde V^{\rm {\cal M}}_{\ell m}
 &=\mathcal D^V_\ell\widetilde V^{\rm src}_{\ell m},&
 \mathcal D^V_\ell&=1+\mathcal T^V_{\ell,1}
                      +\mathcal T^V_{\ell,2}+\cdots\,,
 \label{eq:general-tail-organization}
\end{align}
where tildes denote Fourier transforms.  The term $\mathcal T_{\ell,1}$ is the standard tail effect \cite{Galley:2015kus,Blanchet:1992br},
$\mathcal T_{\ell,2}$ the tail-of-tail, and higher terms contain the subsequent monopole insertions \cite{Goldberger:2009qd,Marchand:2016vox,Blanchet:2023sbv,Henry:2021cek}. The expansion parameter in these expressions is given by $G {\cal M} \omega$ \cite{Goldberger:2009qd}. At fixed harmonic and
eccentricity, for a bound PN orbit with characteristic speed $v$, this parameter is of order $v^3$. Hence, these sectors first enter at relative 1.5PN, 3PN, and so on and so forth~\cite{Blanchet:1992br,Goldberger:2009qd,Porto:2012as,Marchand:2016vox}.\vskip 4pt Within each multipole and parity channel, the
 exterior gravitational field gives the same response for scattering and
bound sources at the same value of the energy and frequency. (Nonlinear hereditary couplings of several time-dependent moments,
including memory, are not covered by this multiplicative statement
\cite{Porto:2024cwd,Almeida:2024lbv,Porto:2026fsd}.) Therefore, propagation corrections can be readily incorporated into the B2B dictionary. See \S\ref{sec:tails} for more details.

\section{The waveform B2B correspondence}
\label{sec:theorem}

The basic idea is essentially the same as for the previous realizations of the B2B map: One continued
encounter starts at periastron, the other at apoapsis, and their common
exterior passages cancel. The waveform, however, adds a further twist to the problem.
Its source and Fourier phase must agree on those passages. In order to set up the B2B connection, we first
establish the continuation of the local source, then determine the
relative phase, and finally ask whether the path difference is one bound
period. Keeping these steps separate will identify where a remainder may arise. 
The analytic continuation is performed at fixed (and independent) values of $\omega$, such that the bound frequency
is imposed only after we arrive at the endpoint. The specific order plays an important role, notably because the spectral
lines themselves depend on the orbit parameters. 

\subsection{Radial time}
\label{sec:local-source}\label{app:local}

For either local source channel, rotational symmetry separates the
azimuthal dependence of the waveform (suppressing the local-source label)
\begin{equation}
 U_{\ell m}=A^U_{\ell m}(r,p_r,J)e^{-\ii m\phi},
 \label{eq:Adef}
\end{equation}
and similarly for $V_{\ell m}$ modes. We suppress the channel
label $U$ or $V$ in $A_{\ell m}$ and $S_{\ell m}$ when the same
argument applies to both. The Fourier transform then becomes
\begin{equation}
 S_{\ell m}(E,J;\omega)=\int_{-\infty}^{\infty}\dd t\,
 A_{\ell m}(r(t),p_r(t),J)e^{\ii\omega t-\ii m\phi(t)}.
 \label{eq:local-scattering-time}
\end{equation}
Continuing this transform in energy and angular momentum at fixed
$\omega$ requires control of both the local mode and its integration
path. The local dynamics determines the first, while writing the transform
in radius makes the turning points and the two passages explicit. In principle, hyperbolic encounters and elliptic orbits have different global
histories, yet their local equations of motion and (unevaluated) source multipole moments are
the same \cite{Cho:2021arx}. Changing the energy changes the solution selected by the initial
data. This observation supplies the first ingredient of the B2B map, which relies on a common integrand prior to performing an analytic continuation on the energy and angular momentum.

\paragraph{A common integrand.} In order to show that the integrands for both the unbound and bound motion are indeed two sides of the same coin we proceed as follows. At a fixed PN order, we assume the existence of a local-in-time Hamiltonian
$\mathcal H(r,p_r,J)$, even in $p_r$ and $J$, and source moments
$I_L,J_L$, analytic in a complex neighbourhood of
the relevant phase-space points. Rational factors are allowed
away from their poles. Radial logarithms occur in the source
multipoles~\cite{Amalberti:2024jaa} and in the potential-region contribution
to the local orbital dynamics~\cite{Porto:2017dgs,Foffa:2019yfl,Dlapa:2021npj}.
The latter accompany infrared poles, while the former accompany ultraviolet ones, whose matching involves conservative
tail~\cite{Galley:2015kus} and tail-of-tail \cite{Goldberger:2009qd,Amalberti:2024jaa} effects at leading order.\footnote{For classical black holes in four-dimensional general
relativity, the static tidal Love numbers and their renormalization group flow vanish~\cite{Porto:2016zng}, and therefore no additional logarithms are present.} In our general analysis we do not include the nonlocal-in-time part of the Hamiltonian.\vskip 4pt 

We denote partial derivatives by subscripts, for example $\mathcal H_r=\partial\mathcal H/\partial r$.
The Hamiltonian equations then read
\begin{equation}
 \frac{\dd}{\dd t}(r,p_r,\phi)
 =\left(\mathcal H_{p_r},-\mathcal H_r,\mathcal H_J\right).
 \label{eq:analytic-flow}
\end{equation}
The analytic properties of the right hand side make a regular local solution analytic
in its initial data and orbital parameters. Continuing these data along a path
that avoids singularities continues the orbital solution. The source
moments are functions of the same phase-space point. For any such
function, $f(r,p_r,\phi,J)$, we have \begin{equation}
 \frac{\dd f}{\dd t}=\left(\mathcal H_{p_r}\partial_r
 -\mathcal H_r\partial_{p_r}+\mathcal H_J\partial_\phi\right) f\,.
 \label{eq:flow-derivative}
\end{equation}
 Finite numbers of derivatives, needed to evaluate radiative multipole moments, combined with
angular projections preserves local analyticity. Both mass and current
amplitudes in \eqref{eq:Adef} consequently can be analytically continued across the orbital motion. This is the local ingredient behind the time-domain continuation of~\cite{Adamo:2024oxy}. It does not yet establish the continuation of
the integral in \eqref{eq:local-scattering-time}.\vskip 4pt 
\begin{figure}[t!]
\centering
\begin{tikzpicture}[>=Stealth,font=\small]
\begin{scope}
\node at(0,3.4){Parameter path at fixed energy};
\draw[darkgray,->](-2.6,0)--(2.8,0)node[right]{$\Re X$};
\draw[darkgray,->](0,-.4)--(0,2.6)node[above]{$\Im X$};
\draw[routeb,dashed,very thick,arr](2.1,0)arc(0:180:2.1);
\foreach \x/\lab in {1/{1},-1/{-1}}{\fill[rem](\x,0)circle(2pt)node[below=4pt]{$\lab$};}
\fill(2.1,0)circle(2pt)node[below=4pt]{$X_0$};
\fill(-2.1,0)circle(2pt)node[below=4pt]{$-X_0$};
\node at(0,-.95){Newtonian reference: $X_0>1$};
\end{scope}
\begin{scope}[xshift=7cm]
\node at(0,3){Exchanging roots};
\node at(0,2.25){$\mathcal H(r_\star,0,J(\theta))=E$};
\node (a) at(-1.55,1.15){$r_-$};
\node (b) at(1.55,1.15){$r_+$};
\draw[parrow,routeb](a)--node[above=6pt]{$e\longrightarrow-e$}(b);
\node at(0,.2){$\displaystyle r_\mp=\frac{GM(1\mp e)}{\epsilon}$};
\node[align=center]at(0,-.95){};
\end{scope}
\end{tikzpicture}
\caption{On the left: The root-exchanging path described in \eqref{eq:J-route}. We use the dimensionless variable $X = j \sqrt{\epsilon}$, such that $e^2=1-X^2$. On the right: The resulting analytic continuation from $e$ to  $-e$ exchanges the two roots. An additional analytic continuation to the bound regime, with $e<1$, is required to connect to the physical domain.}
\label{fig:Jpath}
\end{figure}
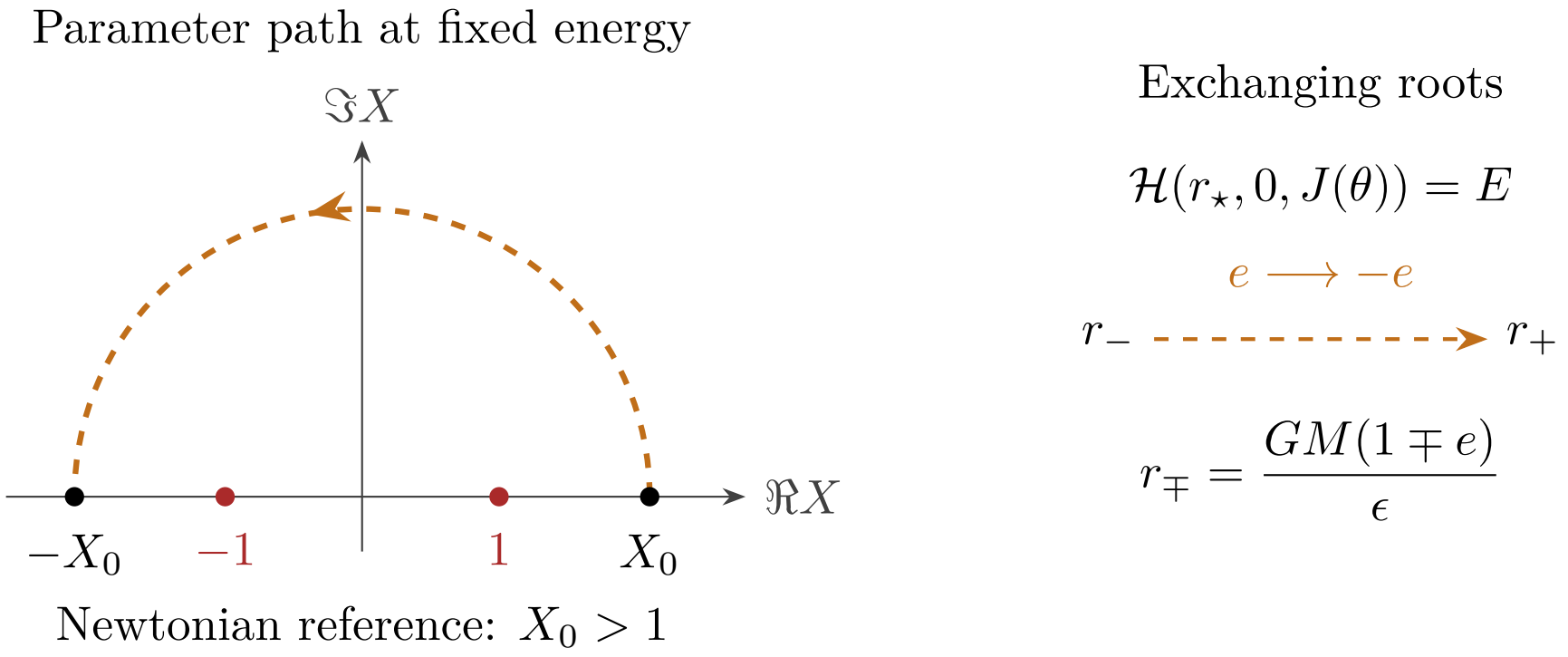

\paragraph{Turning point(s).}  Turning points are roots of
\begin{equation}
 \mathcal H(r_\star,0,J)=E.
 \label{eq:turning-root-definition}
\end{equation}
If $\mathcal H_r(r_\star,0,J)\ne0$, this root is locally analytic in
$E,J$, with $\partial_Er_\star=1/\mathcal H_r$. It is therefore followed along
the parameter path. We choose $t=\phi=0$ at the scattering turning point $r_{\min}$,
where $p_r=0$. For example, the Newtonian scattering root is
\begin{equation}
 r_{\min}=\frac{GMj^2}{1+e_s},
 \qquad e_s=\sqrt{1+2\E j^2}.
 \label{eq:local-peri}
\end{equation}
The energy continuation takes $e_s$ to
$e=\sqrt{1-\epsilon j^2}$, with $\epsilon=-2\E>0$. Once restricted to the physical branch, $0<e<1$, this would give the physical periastron. The subsequent root-exchanging $J$ continuation carries $e\to-e$. Therefore, it takes periastron to apoapsis\cite{Kalin:2019rwq,Kalin:2019inp}:
\begin{equation}
 r_-=\frac{GMj^2}{1+e}
     =\frac{GM}{\epsilon}(1-e),
 \qquad
 r_+=\frac{GMj^2}{1-e}
     =\frac{GM}{\epsilon}(1+e).
 \label{eq:local-bound-roots}
\end{equation} 
Let us stress the importance of specifying the analytic continuation.
Simply replacing $J$ by $-J$ leaves $J^2$ unchanged, whereas following
the square-root branch along the prescribed path exchanges the two
roots. The path, rather than its endpoints alone, therefore determines
the continuation.\footnote{Whenever the observable admits an analytic expansion in
integer powers of $1/J$ that can be continued term by term at fixed
energy, the continuation acts as $J^{-n}\to(-1)^nJ^{-n}$.
The B2B dictionary can then be implemented directly on the expanded
quantities~\cite{Kalin:2019rwq,Kalin:2019inp}.}  As shown in Fig.~\ref{fig:Jpath} for the Newtonian parameters, we follow the path
\begin{equation}
 e=\sqrt{1-X^2},\quad X=j\sqrt{\epsilon},
 \quad X(\theta)=X_0e^{\ii\theta},\quad
 0\leq\theta\leq\pi,\quad X_0>1\,,
 \label{eq:J-route}
\end{equation}
such that the square root goes from $e=\ii\sqrt{X_0^2-1}$ at $\theta=0$ to $-e$ at $\theta=\pi$, thus exchanging
$GM(1-e)/\epsilon$ and $GM(1+e)/\epsilon$. Indeed, the function $e(\theta)^2=1-X_0^2e^{2\ii\theta}$ winds once around zero, for $X_0>1$, so that the square root changes sign. This exchanges the two turning points \cite{Kalin:2019rwq,Kalin:2019inp}. Notice that on this reference sheet the two roots are actually complex for $X_0>1$. Reaching the real roots requires a further analytic continuation to the $0<X<1$ physical region.\footnote{For $0<X_0<1$, one can also exchange the two real turning points directly by deforming the semicircle from $X_0$ to $-X_0$ into a path around one of the branch points $X=\pm1$, giving the required sign change $e\to-e$.} \vskip 4pt 

At a given PN order the roots are defined by
\eqref{eq:turning-root-definition}. As discussed in \cite{Kalin:2019rwq,Kalin:2019inp}, the two analytic continuations end at the physical roots $r_-<r_+$. Since at the start it reaches $r_-$, the continued solution has
the bound initial condition $(r_-,0,0)$. Uniqueness of the solution to Hamilton's equations
then identifies the local orbit and source with the bound values. Reaching $r_+$ instead supplies the point of apoapsis. As we shall see next, comparing the resulting Fourier integrands also requires accounting for their different
time and angle at the origins. 

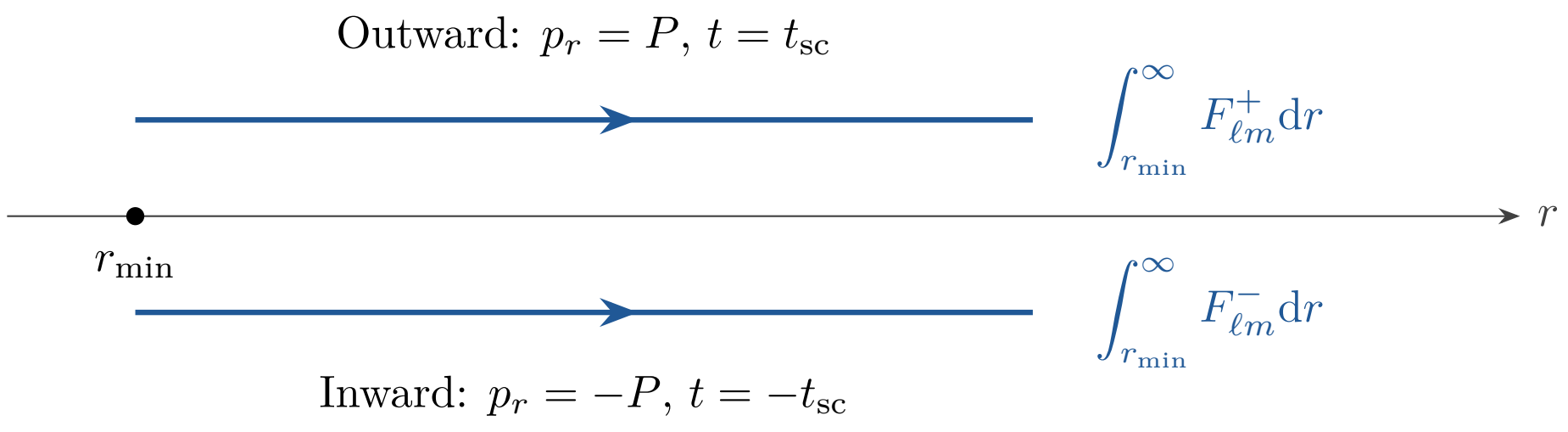
\begin{figure}[t!]
\centering
\begin{tikzpicture}[>=Stealth,font=\small]
\draw[darkgray,->](0,0)--(11.8,0)node[right]{$r$};
\fill(1,0)circle(2pt)node[below=4pt]{$r_{\min}$};
\draw[routea,very thick,arr](1,.75)--(8,.75);
\draw[routea,very thick,arr](1,-.75)--(8,-.75);
\node[routea,anchor=west]at(8.35,.75){$\displaystyle\int_{r_{\min}}^\infty F_{\ell m}^+\dd r$};
\node[routea,anchor=west]at(8.35,-.75){$\displaystyle\int_{r_{\min}}^\infty F_{\ell m}^-\dd r$};
\node at(4.5,1.4){Outward: $p_r=P$, $t=t_{\rm sc}$};
\node at(4.5,-1.4){Inward: $p_r=-P$, $t=-t_{\rm sc}$};
\end{tikzpicture}
\caption{The source integrals in \eqref{eq:scattering-radial}. Both arrows run from $r_{\min}$ to infinity.}
\label{fig:original}
\end{figure}

\paragraph{Two radial passages.}
We are now in position to change variables in \eqref{eq:local-scattering-time}. We start by solving the
energy equation for the {\it outward} momentum branch $P=p_r>0$,
\begin{equation}
 \mathcal H(r,P,J)=E,\qquad
 v_r(r)=\mathcal H_{p_r}(r,P,J).
 \label{eq:ham}
\end{equation}
The integration measure transforms as $\dd t=\dd r/v_r$. We then set the outgoing time and angle to zero at $r_{\min}$, yielding
\begin{equation}
 t_{\rm sc}(r)=\int_{r_{\min}}^{r}\frac{\dd r'}{v_r(r')},\qquad
 \phi_{\rm sc}(r)=\int_{r_{\min}}^{r}
 \frac{\mathcal H_J(r',P(r'),J)}{v_r(r')}\dd r'.
 \label{eq:clocks}
\end{equation}
On the other hand, the {\it inward} branch has $p_r=-P < 0$ with opposite radial velocity, such that $t=-t_{\rm sc}$, $\phi=-\phi_{\rm sc}$ and
$\dd t=-\dd r/v_r$.\vskip 4pt Following the B2B strategy, we now split the time integral around $r_{\rm min}$ and,
reversing the incoming radial limits, we find 
\begin{equation}
 S_{\ell m}=\int_{r_{\min}}^{\infty}(F_{\ell m}^++F_{\ell m}^-)\dd r,
 \qquad F_{\ell m}^\pm=
 \frac{A_{\ell m}(r,\pm P,J)}{v_r}e^{\pm\ii(\omega t_{\rm sc}-m\phi_{\rm sc})}.
 \label{eq:scattering-radial}
\end{equation}
Both terms run over the same real interval in the displayed direction.
The signs label opposite radial momenta and Fourier phases, not
opposite integration limits. We display in Fig.~\ref{fig:original} these two contributions. This is the
scattering integral whose root, source, clock branches and contour
must now be deformed under analytic continuation. For notational convenience, we drop the $\ell m$ labels on the $F_{\ell m}^\pm$ and $S_{\ell m}$ functions below, unless otherwise noted, and distinguish operations on their indices or variables by introducing new functions.  

\subsection{Four events}
The radial form \eqref{eq:scattering-radial} separates the local
integrand from its integration path. We now continue both, keeping
track of the root at which each path starts.
The first route follows the original mode to negative binding energy.
For the second, the reversal of the orbital orientation suggests the use of a 
reflected mode, $m \to -m$, while a continuation in angular momentum from $J$ to $-J$ exchanges the roots. We can understand this by noticing that the reflected amplitude can be connected via a spatial rotation. Since
$Y_{\ell m}(\pi-\vartheta,-\varphi)=(-1)^\ell
Y_{\ell,-m}(\vartheta,\varphi)$, the mass and current projections obey
\begin{equation}
 A_{\ell m}(r,\pm P,J)=(-1)^\ell  A_{\ell,-m}(r,\pm P,-J)\,.
 \label{eq:reflection}
\end{equation}
Rotationally invariant evolution preserves these relations, also after taking 
source derivatives on the multipole moments. Therefore, they hold for the local
waveforms and their analytic continuations.\vskip 4pt 

We call the analytically continued scattering transforms of the direct and reflected routes as $S_{\rm peri}$ and
$S_{\rm apo}$, according to the root reached by the continuation. They
are not the outward and inward pieces of one encounter: \emph{each}
contains both terms of \eqref{eq:scattering-radial}. The proposed B2B
construction subtracts their exterior contributions, leaving the
passage(s) between $r_-$ and $r_+$ associated with a bound orbit.\vskip 4pt We start with the analytic
continuation in the energy of \eqref{eq:scattering-radial}, along the path
\begin{equation}
\begin{aligned}
&\qquad \quad p_\infty(\theta)=s e^{\ii\theta}, \quad 0\leq\theta\leq\pi/2,\\
 \gamma(\theta)&=\sqrt{1+p_\infty^2(\theta)},\quad
 E(\theta)=M\sqrt{1+2\nu[\gamma(\theta)-1]}.
 \label{eq:energy-route}
\end{aligned}
\end{equation}
Here $0<s<1$ fixes the magnitude of $p_\infty$ at the start and the end, while
$J$ and $\omega$ are held fixed. The square
roots are followed from their positive scattering values. Hence, the endpoint
has $p_\infty=\ii s$ and negative binding energy. This continuation acts on the entire radial integral. The root
follows $\mathcal H(r_\star(J,E(\theta)),0,J)=E(\theta)$, the integrand follows the same
momentum and clock branches, and the original interval
$[r_{\min},\infty)$ is transported so that singularities do not cross it.
We denote the resulting paths by $C_{\rm peri}^+$ and
$C_{\rm peri}^-$. The signs retain their meaning from
\eqref{eq:scattering-radial}: they label the continued $p_r=+P$ and
$p_r=-P$ contributions, both oriented away from the turning point. (They do not necessarily remain the same curve in the complex radial plane.) At the endpoint, the root is identified with $r_-$, and the analytically continued
$S_{\rm peri}$ starts at periastron.\vskip 4pt

We now consider the reflected scattering transform $S_{\ell,-m}$, and
apply the same energy continuation, followed by $J$ to $-J$ along a path that exchanges the
turning roots. Similarly to what we did before, following both radial contributions defines
$C_{\rm apo}^\pm$ and their sum $S_{\rm apo}$. The mode label is now
$-m$, the final angular momentum is $-J$, and the new starting root is now
identified with $r_+$. By construction, all four paths descend from \eqref{eq:scattering-radial}:
\begin{equation}
\begin{aligned}
 C_{\rm peri}^\pm &: r_-\longrightarrow\infty,
 &S_{\rm peri}&=\sum_\pm\int_{C_{\rm peri}^\pm}
 F_{\rm peri}^\pm(r)\,\dd r,\\
 C_{\rm apo}^\pm &: r_+\longrightarrow\infty,
 &S_{\rm apo}&=\sum_\pm\int_{C_{\rm apo}^\pm}
 G_{\rm apo}^\pm(r)\,\dd r.
\end{aligned}
\label{eq:continued-encounters}
\end{equation}
Here $F_{\rm peri}^\pm$ is the analytically continued integrand in its original form, while we use  
$G_{\rm apo}^\pm$ to emphasize it is the reflected one.

\subsection{Clocks and angles}
\label{sec:clock-matching}
We now compare $F_{\rm peri}^\pm$ and $G_{\rm apo}^\pm$. Similarly to the B2B map for the radiative fluxes \cite{Cho:2021arx}, the existence of an isotropic-like Hamiltonian (invariant under $J \to -J$) guarantees that both enjoy the same radial measure, $\dd r/{\cal H}_{p_r}$, upon analytic continuation. The key task is to connect the clocks. Let
us introduce the functions $t_{\rm peri}(r),\phi_{\rm peri}(r)$, defined as in
\eqref{eq:clocks}, but starting at $r_-$, such that \begin{equation} \begin{aligned} t_{\rm peri}(r_-)&=\phi_{\rm peri}(r_-)=0\,,\quad t_{\rm peri}(r_+)&= T_r/2, \,\, \phi_{\rm peri}(r_+)=\frac{\Delta\phi}{2}\,. \label{eq:tperi}\end{aligned}\end{equation} The apoapsis encounter instead
has zero clock and angle at $r_+$, and opposite angular momentum. Consequently, since $ \partial_r \phi(r,J) = - \partial_r \phi(r,-J)$, we have\footnote{Due to the $J$-parity invariance  of the Hamiltonian, ${\cal H}(J)={\cal H}(-J)$,  the derivative with respect to $J$ changes sign, and likewise the orbital velocity in \eqref{eq:clocks}.} 
\begin{equation}
\begin{aligned}
 t_{\rm apo}(r)
 &=t_{\rm peri}(r)-\frac{T_r}{2}\,,\quad
 \phi_{\rm apo}(r) =-\phi_{\rm peri}(r)+\frac{\Delta\phi}{2}\,.
\end{aligned}
\label{eq:clockmatch}
\end{equation}
such that the angle reverses sign. Substituting  into
the original scattering integrands we find
\begin{equation}
\begin{aligned}
 F_{\rm peri}^\pm(r)&=\frac{A_{\ell m}(r,\pm P,J)}{v_r}
 e^{\pm\ii[\omega t_{\rm peri}(r)-m\phi_{\rm peri}(r)]},\\
 G_{\rm apo}^\pm(r)&=\frac{A_{\ell,-m}(r,\pm P,-J)}{v_r}
 e^{\pm\ii[\omega t_{\rm apo}(r)+m\phi_{\rm apo}(r)]}.
\end{aligned}
 \label{eq:both-radial-integrands}
\end{equation}
Notice that the plus sign in the angular term comes from the reflected
mode.
Finally, the remaining factors are connected via \eqref{eq:clockmatch}:
\begin{align}
 \omega t_{\rm apo}+m\phi_{\rm apo}
 &=\omega t_{\rm peri}-m\phi_{\rm peri}
       -\tfrac12(\omega T_r-m\Delta\phi),
  \end{align}     
 and using \eqref{eq:reflection} we arrive at the key relationship: 
 \begin{align}
 G_{\rm apo}^\pm
 &=(-1)^\ell e^{\mp i(\omega T_r-m\Delta\phi)/2}F_{\rm peri}^\pm.
 \label{eq:integrand-match-frequency}
\end{align}
It is straightforward to show now that, on the bound spectral frequencies, we have
\begin{equation}
  G_{\rm apo}^\pm= (-1)^{\ell+k} F_{\rm peri}^\pm, \qquad \omega=\omega_{km}.
 \label{eq:integrand-match}
\end{equation}
The factor $(-1)^\ell$ is the source rotation, whereas the $(-1)^k$ is the
half-period time and angle shift. We can therefore summarize the selection rules $U$- and $V$-type harmonics as follows:
\begin{equation}
\begin{array}{c|c|c}
\text{allowed channel}&(-1)^{\ell+k}&\text{continued combination}\\\hline
U:\ \ell+m\text{ even}&(-1)^p&S^U_{\rm peri}-(-1)^pS^U_{\rm apo}\\
V:\ \ell+m\text{ odd}&(-1)^{p+1}&S^V_{\rm peri}+(-1)^pS^V_{\rm apo}
\end{array}
\label{eq:parity-table}
\end{equation}
Substituting \eqref{eq:integrand-match} into
\eqref{eq:continued-encounters}, and using $(-1)^{2(\ell+k)}=1$, yields
\begin{equation}
 S_{\rm peri}-(-1)^{\ell+k} S_{\rm apo}
 =\sum_\pm\left(\int_{C_{\rm peri}^\pm}F_{\rm peri}^\pm\dd r
 -\int_{C_{\rm apo}^\pm}F_{\rm peri}^\pm\dd r\right).
 \label{eq:matched-combination}
\end{equation}
We have now obtained a difference of paths for the \emph{same} integrand
on each radial branch. This is the point at which a contour argument
becomes possible. 

\subsection{Cauchy's input}
\label{sec:path-subtraction}

In \eqref{eq:matched-combination} all final parameters and all branches are now fixed. The remaining operation would consist on deforming the radial integration at {\it fixed} values of $E$ or $J$.  We denote the path connecting $C^\pm_{\rm peri}$ and $C^\pm_{\rm apo}$ in the radial plane by $B^\pm$. It lies over the real interval $[r_-,r_+]$ on the physical bound sheet. Both displayed orientations are $r_-\to r_+$ because the
returning contribution has its radial limits reversed.\vskip 4pt

\begin{figure}[t!]
\centering
\begin{tikzpicture}[>=Stealth,font=\small]
\draw[gray,->](-.4,0)--(12,0)node[right]{$\Re r$};
\draw[gray,->](0,-.4)--(0,3.2)node[above]{$\Im r$};
\coordinate(a)at(1,0);\coordinate(b)at(5,0);\coordinate(R)at(10.5,0);
\draw[routea,very thick,arr](a)..controls(2,3)and(8.8,3)..(R);
\draw[bound,very thick,arr](a)--(b);
\draw[routeb,very thick,arr](b)--(R);
\draw[gray,thick,arr](R)--(11.65,0);
\foreach \x/\lab in {1/{r_-},5/{r_+},10.5/{R_\pm}}{\fill(\x,0)circle(2pt)node[below=4pt]{$\lab$};}
\node[routea]at(5.7,3.1){$C_{\rm peri}^\pm$};
\node[bound]at(3,-.8){$B^\pm$};
\node[routeb]at(7.7,-.8){$C_{\rm apo}^\pm$};
\node[gray,align=center]at(11,1.1){};
\fill[rem](5.3,1.1)circle(2pt);\draw[rem,thick](5.3,1.1)--(6.3,1.1);
\node[rem,below=7pt]at(5.8,1.1){possible obstruction};
\end{tikzpicture}
\caption{Sketched geometry at fixed orbital parameters. The blue arc is a schematic representative of the analytically continued path, which can be joined by running backwards on orange followed by green. A pole or cut would contribute to the integral. The red mark illustrates such a possibility.}
\label{fig:cauchy}
\end{figure}
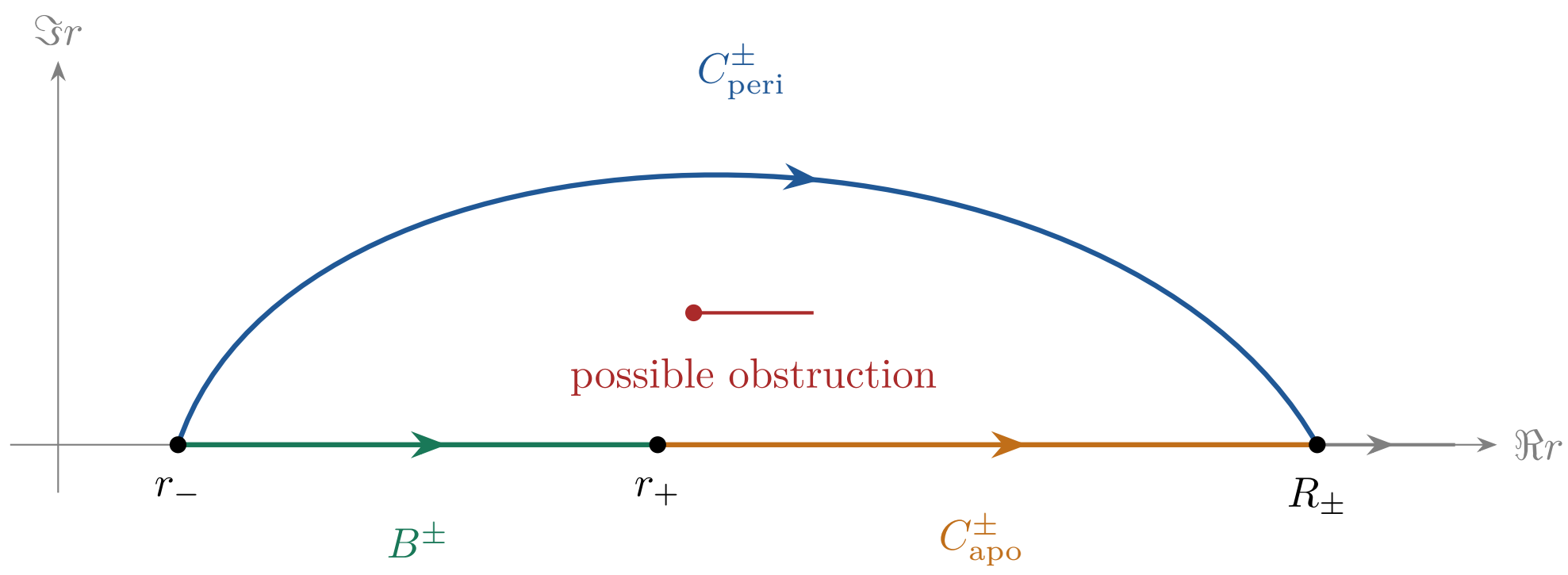

To compare the {\it exterior} parts, assume the two continued paths extending to infinity can
be represented with an identical final portion. Let $R_\pm$ be a point where that shared portion begins. (Notice that $\pm$ refers to the routing of the functions integrated and not to the $r_\pm$ endpoints.) We will eventually take $R_\pm \to \infty$, such that
its exact location is arbitrary.  We display in Fig. \ref{fig:cauchy} a sketched representation in the complex radial plane. 

For the oriented closed path(s) shown in Fig.~\ref{fig:cauchy}, $\mathcal L^\pm \equiv C_{\rm peri}^\pm-C_{\rm apo}^\pm-B^\pm$,
 Cauchy's theorem implies
\begin{align}
 \int_{C_{\rm peri}^\pm}F_{\rm peri}^\pm\dd r
   -\int_{C_{\rm apo}^\pm}F_{\rm peri}^\pm\dd r
 &=\int_{B^\pm}F_{\rm peri}^\pm\dd r
   +{\cal R}_\pm\,, \quad {\cal R}_\pm \equiv \oint_{\mathcal L^\pm}F_{\rm peri}^\pm\dd r.
 \label{eq:closed-difference}
\end{align}
Cauchy's theorem sets the closed integral to zero if
the paths can be deformed into one another. If a pole or a cut obstructs that deformation, the associated contour
contribution remains. As we demonstrate in App.~\ref{app:transport}, no such obstruction exists for a class of local PN sources.\vskip 4pt 

\paragraph{Bound state.} The final task is to identify the radial integral on the $B^\pm$ line with one period of the
bound waveform. In order to do this we start at periastron with $t=\phi=0$. Let us consider first the mass mode. The amplitude $A^U_{\ell m}$ must be evaluated separately on the two
radial momentum branches. On the outward passage, $p_r=P>0$ and the radial velocity is positive. The time and angle are the outward integrals
$t_{\rm peri}(r)$ and $\phi_{\rm peri}(r)$ we already introduced in \eqref{eq:tperi}. Hence, changing the integration variable from $t$ to $r$ in the outward passage contribution to \eqref{eq:bounddef} gives
\begin{equation}
\begin{aligned}
\frac{1}{T_r} \int_0^{T_r/2}e^{i\omega t}U_{\ell m}(t)\,\dd t
 &=
\frac{1}{T_r} \int_{r_-}^{r_+}
 \frac{A_{\ell m}(r,P,J)}{v_r(r)}
 e^{i[\omega t_{\rm peri}(r)-m\phi_{\rm peri}(r)]}\,\dd r\\
 &=\frac{1}{T_r}\int_{r_-}^{r_+}F_{\rm peri}^+(r)\,\dd r.
\end{aligned}
\end{equation}
Therefore, the same integrand, $F_{\rm peri}^+$, is precisely the one entering in the outward half of the bound orbit.

On the return passage, $p_r=-P$, the radius decreases. Since the conservative Hamiltonian
is even in $p_r$, the radial velocity changes sign, and therefore the integral is traversed with the opposite measure,  $dt=-dr/v_r(r)$. The angular velocity, however, does not change sign. Therefore, the time and angle still to be accumulated
before reaching the next periastron is given by
\begin{equation}
 t=T_r-t_{\rm peri}(r),\qquad
 \phi=\Delta\phi-\phi_{\rm peri}(r)\,.
\end{equation}
The synchronization is important: at apoapsis these expressions give $t=T_r/2$ and
$\phi=\Delta\phi/2$, and at the next periastron they give $t=T_r$ and
$\phi=\Delta\phi$, as expected. Substituting these relations into the remaining return contribution yields
\begin{align}
 \frac{1}{T_r} \int_{T_r/2}^{T_r}e^{i\omega t}U_{\ell m}(t)\,\dd t
 &=
 -\frac{1}{T_r} \int_{r_+}^{r_-}
 \frac{A_{\ell m}(r,-P,J)}{v_r(r)}
 e^{i\omega[T_r-t_{\rm peri}(r)]
    -im[\Delta\phi-\phi_{\rm peri}(r)]}\,\dd r
 \\
 &=
  \frac{1}{T_r} e^{i(\omega T_r-m\Delta\phi)}
 \int_{r_-}^{r_+}
 \frac{A_{\ell m}(r,-P,J)}{v_r(r)}
 e^{-i[\omega t_{\rm peri}(r)-m\phi_{\rm peri}(r)]}\,\dd r
 \nonumber\\
 &=
 \frac{1}{T_r} e^{i(\omega T_r-m\Delta\phi)}
 \int_{r_-}^{r_+}F_{\rm peri}^-(r)\,\dd r.\nonumber
\end{align}
with the minus sign reversed  upon changing the radial limits. Both radial integrals therefore run from $r_-$ to $r_+$, even though
the physical return motion runs from $r_+$ to $r_-$.\vskip 4pt 

The overall phase must be now evaluated on the bound frequencies. Indeed, as before, we find that the phase becomes
\begin{equation}
 e^{i(\omega_{km}T_r-m\Delta\phi)}
 =e^{2\pi i k}=1,\qquad k\in\mathbb Z.
\end{equation}
Adding the two passages, with both integrands evaluated at
$\omega=\omega_{km}$, yields
\begin{equation}
\begin{aligned}
  U^b_{\ell m,k} &=\frac{1}{T_r} \int_0^{T_r}e^{i\omega_{km}t}U_{\ell m}(t)\,\dd t =
\frac{1}{T_r} \int_{r_-}^{r_+}
 \bigl(F_{\rm peri}^++F_{\rm peri}^-\bigr)\,\dd r\\
 &= \frac{1}{T_r} \left(\int_{B^+}F_{\rm peri}^+\,\dd r
 +\int_{B^-}F_{\rm peri}^-\,dr\right)\,.
   \label{eq:bound-passage}
 \end{aligned}\end{equation}
 Here $B^\pm$ are both oriented from $r_-$ to $r_+$ on their respective
radial momentum branches. Notice that, crucially, the full  $\Delta\phi$, including the periastron advance, is needed for this equality to hold beyond Newtonian order.  Replacing $U$ by $V$ reproduces the same argument for the current mode.\vskip 4pt  We are therefore led to the following B2B waveform relations
\begin{align}
 U^b_{\ell m,k}=\frac{S^U_{\rm peri}-(-1)^{\ell+k} S^U_{\rm apo}-\mathcal R_U}{T_r},
 \quad
 V^b_{\ell m,k}=\frac{S^V_{\rm peri}-(-1)^{\ell+k} S^V_{\rm apo}-\mathcal R_V}{T_r},
 \label{eq:theorem}
 \end{align}
with the remainders given by 
\begin{align}
 \mathcal R_U=\sum_\pm\oint_{\mathcal L^\pm}F^{\pm,U}_{\rm peri}\dd r,
 &\qquad
 \mathcal R_V=\sum_\pm\oint_{\mathcal L^\pm}F^{\pm,V}_{\rm peri}\dd r.
 \label{eq:channel-remainders}
 \end{align}
 
Our previous analysis isolates what is needed beyond local source
analyticity: control of the continued paths and their closed
difference. We return to this issue in App.~\ref{app:transport}, where we demonstrate that the remainder vanishes for an isotropic local Hamiltonian that is polynomial in momenta and inverse radius, and local multipoles that are
polynomial in momenta and Laurent polynomial in radius. Both may
contain finite powers of radial logarithms. On the common branches,  we show that 
$\mathcal R_{U}^{(n)}=\mathcal R_{V}^{(n)}=0$ at each finite $n$PN order. The superscript $(n)$ denotes truncation through $n$PN, with the
period, harmonic frequency and Fourier phase consistently expanded to that order.
The B2B correspondence is thus established at every PN order:\begin{align}
U^{b(n)}_{\ell m,k}=\frac{1}{T_r}\left(S^{U(n)}_{\rm peri}-(-1)^{\ell+k} S^{U(n)}_{\rm apo}\right)_{\omega^{(n)}_{km}}\,,
 \,
 V^{b(n)}_{\ell m,k}=\frac{1}{T_r}\left(S^{V(n)}_{\rm peri}-(-1)^{\ell+k} S^{V(n)}_{\rm apo}\right)_{\omega^{(n)}_{km}}\,.
 \label{b2bw}
 \end{align}
 
\subsection{Tail effects}
\label{sec:tails}

The previous B2B relations concern the local source. To include propagation,
we consider a single time-dependent multipole ${\cal Q}_{\ell m}$ interacting
with any number of stationary mass monopoles, through terms of the form
$\mathcal M^n{\cal Q}_{\ell m}$ (recall $\mathcal M=E=\Gamma M$).
In this sector, the frequency-space factors introduced in
\eqref{eq:general-tail-organization} allow us to carry the source
relation directly to the radiative waveform. The leading quadrupole
tail illustrates both the factorization and the time convention
needed for this step.\vskip 4pt

For the quadrupole, the leading responses are\footnote{The complex terms in
\eqref{eq:two-leading-tails} carry the retarded boundary condition.
Following the prescription in~\cite{Porto:2012as}, the logarithm is understood as
\begin{equation}
 \ln\!\left[-4\mu_0^2(\omega+i0^+)^2\right]
 =2\ln(2\mu_0|\omega|)
  -i\pi\,\mathrm{sgn}(\omega).
\end{equation}
Multiplication by $iG\mathcal M\omega$ gives both the real
$\pi G\mathcal M\omega\,\mathrm{sgn}(\omega)
=\pi G\mathcal M|\omega|$ term and the imaginary frequency
logarithm in \eqref{eq:two-leading-tails}. They are parts of the
same retarded response. The absolute value and sign function
express its boundary values on the real frequency axis; continuation
uses the logarithm with its inherited prescription, rather than
continuing those two real-axis functions separately.}
\cite{Blanchet:1992br,Porto:2012as}
\begin{align}
 \mathcal D^U_2
 &=1+G\mathcal M\omega
 \left\{\pi\,\mathrm{sgn}(\omega)
 +2i\left[\ln(2\mu_0|\omega|)+\gamma_E-\frac{11}{12}\right]\right\}
 +\cdots,
 \nonumber\\
 \mathcal D^V_2
 &=1+G\mathcal M\omega
 \left\{\pi\,\mathrm{sgn}(\omega)
 +2i\left[\ln(2\mu_0|\omega|)+\gamma_E-\frac76\right]\right\}
 +\cdots.
 \label{eq:two-leading-tails}
\end{align}
Here $\gamma_E$ is Euler's constant and $\mu_0$ is a fixed scale.
The responses are written after removal of the infrared pole and
a choice of radiative-time convention~\cite{Porto:2012as}.
Different finite constants distinguish the mass and current channels.\vskip 4pt

Using that $\mathcal Q_{22}=\Cnorm Q/2$ we have $\widetilde U^{\rm src}_{22}=-\omega^2\widetilde{\mathcal Q}_{22}$
at leading order (see \eqref{eq:Q0}). The mass-tail contribution is therefore
\begin{equation}
 \delta\widetilde U_{22}(\omega)
 =\bigl(\mathcal D^U_2-1\bigr)
   \widetilde U^{\rm src}_{22}(\omega)
 =-\omega^2\bigl(\mathcal D^U_2-1\bigr)
   \widetilde{\mathcal Q}_{22}(\omega).
\end{equation}
Therefore, the propagation changes the amplitude and phase of each Fourier
component without changing its frequency. The same reasoning applies to repeated scattering from the monopole.
Stationarity preserves frequency, while spherical symmetry makes
the response $m$-independent.
Accordingly, within the stationary sector,
\begin{equation}
 \widetilde U_{\ell m}(\omega)
 =\mathcal D^U_\ell(\mathcal M,\omega)
  \widetilde U^{\rm src}_{\ell m}(\omega),\qquad
 \widetilde V_{\ell m}(\omega)
 =\mathcal D^V_\ell(\mathcal M,\omega)
  \widetilde V^{\rm src}_{\ell m}(\omega).
 \label{eq:tail-source-factors}
\end{equation}
where we reintroduced the  ``src'' label to emphasize it carries the local source amenable to the B2B dictionary, while the
$\mathcal D^{U,V}_\ell$'s include tails, tails of tails, and their iterations. In the above expressions we assume that ultraviolet poles cancel against local source contributions, with the associated finite source $\log r$
and propagation $\log|\omega|$ terms retained in the matched waveform
\cite{Goldberger:2009qd,Amalberti:2024jaa}.\vskip 4pt 

Before continuing these expressions, the source and radiation must
be matched in a common convention. The long-range monopole produces an
infrared phase linear in $\omega$, which can be absorbed into the
definition of time \cite{Porto:2012as}. We therefore use the same radiative-time origin and scale $\mu_0$
for scattering and bound waveforms. The matching and higher-order propagation factors are given in~\cite{Goldberger:2009qd,Porto:2012as,Marchand:2016vox,Henry:2021cek}. 

We can then apply the B2B continuation with this time convention held
fixed. Along the energy route \eqref{eq:energy-route},
$\mathcal M=E$ is continued together with the source, while $\omega$
remains independent. Along the subsequent $J$ route, $E$ and
$\omega$ are fixed. The propagation factor is consequently unchanged
on this second leg: it depends neither on $J$ nor on the reflected
label $m$. Both routes retain the same retarded prescription.
In particular, reflecting $m$ does not reverse $\omega$ or
$\mathrm{sgn}(\omega)$.

Let $\mathcal D^U_{\ell,b}$ denote the resulting factor evaluated at
the final bound energy and harmonic frequency $\omega=\omega_{km}$,
with its sign retained in \eqref{eq:two-leading-tails}.
It multiplies both continued scattering terms, while the bound
radiative harmonic is given by
$U^b_{\ell m,k}=\mathcal D^U_{\ell,b}U^{b,\rm src}_{\ell m,k}$ \cite{Morras:2025nbp}.
The source B2B relation therefore gives
\begin{align}
 S^U_{\rm peri}-(-1)^{\ell+k}S^U_{\rm apo}
 &=
 \mathcal D^U_{\ell,b}
 \left[
 S^{U^{\rm src}}_{\rm peri}
 -(-1)^{\ell+k}S^{U^{\rm src}}_{\rm apo}
 \right]
 \nonumber\\
 &=T_rU^b_{\ell m,k}
   +\mathcal D^U_{\ell,b}\mathcal R_{U,\rm src},
 \label{eq:tail-inheritance}
\end{align}
where $\mathcal R_{U,\rm src}$ is the remainder of the local-source
relation. The vanishing source remainder remains zero after
adding propagating tails. The same applies to the current channel with the inclusion of
$\mathcal D^V_{\ell,b}$.\vskip 4pt

The restriction to one time-dependent multipole is essential.
Interactions between two or more such moments mix frequencies
\cite{Porto:2024cwd,Porto:2026fsd}, while conservative nonlocal tail
effects modify the orbital dynamics itself. Neither is incorporated
by the propagation factors in \eqref{eq:tail-source-factors}.
We leave the extension of the B2B map to these sectors to future work.

\section{Applications}
\label{sec:applications}
The B2B correspondence becomes especially transparent in known examples.  We study here both the Newtonian waveform and the 1PN correction, with the latter introducing precession effects. 

\subsection{Newtonian waveforms}

 We begin on the scattering side, and introduce the parameters:
\begin{equation}
 a_s=\frac{GM}{p_\infty^2},\quad
 n_s=\frac{p_\infty^3}{GM},\quad
 e_s=\sqrt{1+j^2p_\infty^2},\quad
 \alpha=\frac{\omega}{n_s}.
 \label{eq:Newton-sc-vars}
\end{equation}
Here $a_s$ is the positive hyperbolic length scale, $n_s$ its inverse
time scale, and $e_s>1$ the scattering eccentricity. At Newtonian order we also have $p_\infty=\sqrt{2\mathcal E}>0$. We use positive frequencies in the scattering Bessel representations below, while negative values follow from the corresponding mode reality relations.

\subsubsection{The $U_{22}$ spectrum}

Projecting the quadrupole moment computed in~\cite{Bini:2021jmj} onto the radiative $U_{22}$ mode,
%\begin{equation}
%S^{U,N}_{22}
%=-\frac{\omega^2}{2}\sqrt{\frac{\pi}{5}}\,
%\bigl(\widehat I_{xx}-\widehat I_{yy}-2i\widehat I_{xy}\bigr),
%\end{equation}
gives us the scattering waveform in the Newtonian regime,\footnote{The Newtonian hyperbolic waveform is
already known in frequency space. For instance, the quadrupole Fourier transforms
are given in~\cite{Grobner:2020fnb}, and explicit strain
polarizations can be found in~\cite{Fontbute:2024amb}. Our representation in terms of the 
sum in \eqref{eq:mass-scatter-full} expresses the $22$ projection in terms of
shifted Bessel functions, using the Bessel identities in~\cite{Bini:2021jmj} and the conventions of
\S\ref{sec:emission}.}
\begin{equation}
 S^{U,N}_{22}
 =-\frac{\Cnorm\mu a_s^2\omega^2}{n_s}
 e^{\pi\alpha/2}
 \sum_{n=-3}^{3}
 c_n(e_s,X)\,
 \ii^nK_{n-\ii\alpha}(\alpha e_s)\,,\quad X = - i  p_\infty j\,.
 \label{eq:mass-scatter-full}
\end{equation}
 The $c_n$ coefficients are (evaluated at $e=e_s$)
\begin{equation}
\begin{array}{c|c@{\qquad}c|c}
 n&c_n(e,X)&n&c_n(e,X)\\\hline
 -3&e(X+1)^2/8
 &1&5e(1-X)(3+X)/8\\
 -2&(X+1)^2(2X-3)/4
 &2&-(X-1)^2(3+2X)/4\\
 -1&5e(X+1)(3-X)/8
 &3&e(X-1)^2/8\\
 0&-5e^2/2&&
\end{array}
\label{eq:Ncoeffs}
\end{equation}

\paragraph{Continue the energy.} At the Newtonian level, the energy route in \eqref{eq:energy-route} becomes
\begin{equation}
 p_\infty(\theta)=\sqrt{\epsilon}\,e^{\ii\theta},
 \qquad 0\leq\theta\leq\frac{\pi}{2},
\end{equation}
while holding $J$, and $\omega$
fixed. Here $\epsilon>0$ fixes the magnitude of the final binding energy. The path starts at $\E=\epsilon/2>0$ and ends at
$\E=-\epsilon/2<0$, where $p_\infty=\ii\sqrt{\epsilon}$. Follow all square roots and Bessel functions continuously
from their scattering values. At the endpoint,
\begin{equation}
 a_s\to-\frac{GM}{\epsilon} = -a_N,\quad
 n_s\to-\ii\Omega_{r,N},\quad
 \alpha\to \ii\frac{\omega}{\Omega_{r,N}},
 \quad
 \Omega_{r,N}=\frac{\epsilon^{3/2}}{GM}.
\end{equation}
The arguments of the $c_n$ coefficients become $X=j\sqrt{\epsilon}$ and $e=\sqrt{1-X^2}$. We first work on the reference range $X>1$, while the continuation to a physical ellipse, with $0<X<1$ and $0<e<1$, follows after combining the two routes.\vskip 4pt

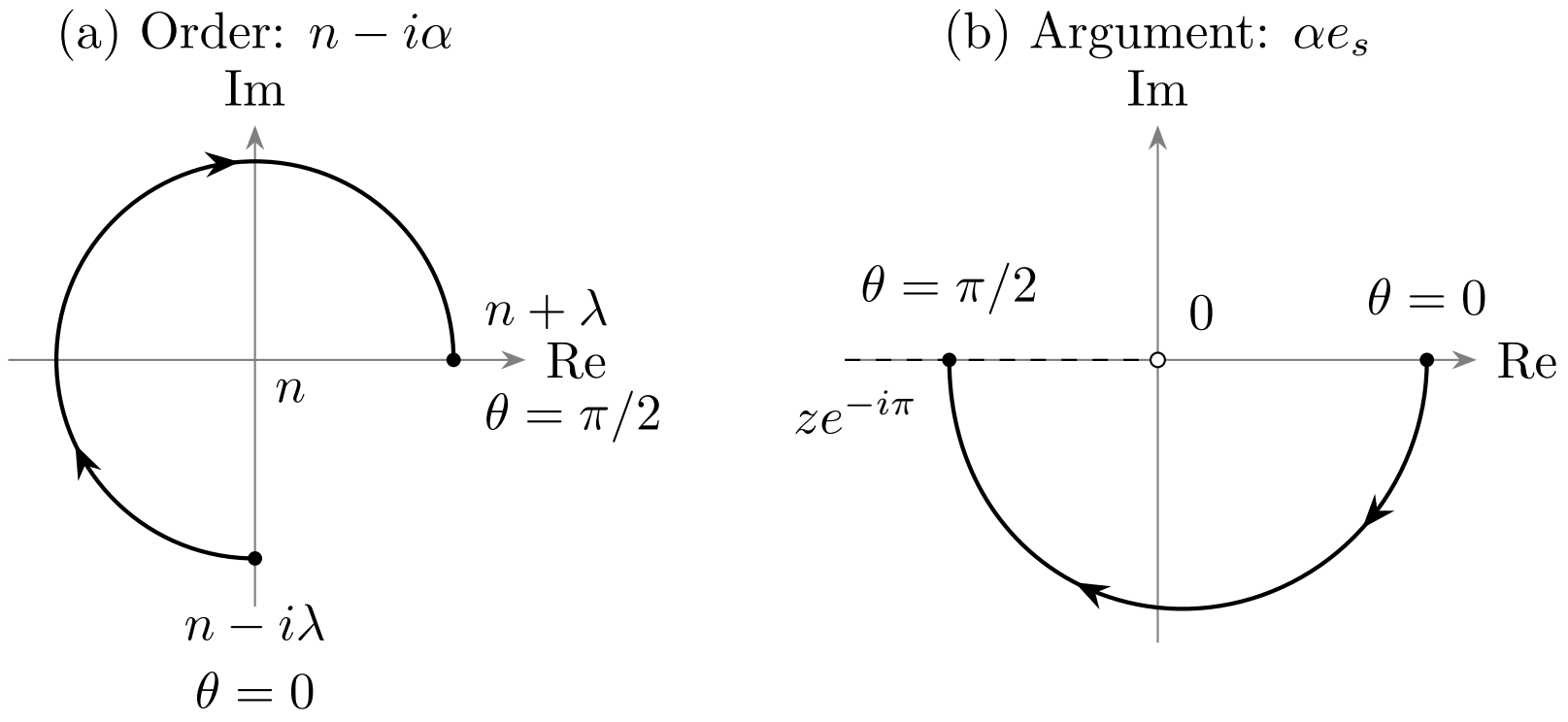
\begin{figure}[t!]
\centering
\begin{tikzpicture}[font=\small,>=Stealth,scale=.82,
  continued/.style={thick,postaction={decorate},decoration={markings,
    mark=at position .24 with {\arrow{Stealth}},
    mark=at position .65 with {\arrow{Stealth}}}}]
  \begin{scope}
    \node at (0,2.7) {(a) Order: $n-\ii\alpha$};
    \draw[->,gray] (-2.05,0)--(2.25,0) node[right,text=black] {$\mathrm{Re}$};
    \draw[->,gray] (0,-2.05)--(0,1.95) node[above,text=black] {$\mathrm{Im}$};
    \node[below right] at (0,0) {$n$};
    \draw[continued] (0,-1.65) arc[start angle=-90,end angle=-360,radius=1.65];
    \fill (0,-1.65) circle (1.7pt)
      node[below=5pt,align=center] {$n-\ii\lambda$\\$\theta=0$};
    \fill (1.65,0) circle (1.7pt)
      node[above right=3pt] {$n+\lambda$}
      node[below right=3pt] {$\theta=\pi/2$};
  \end{scope}
  \begin{scope}[shift={(7.5,0)}]
    \node at (0,2.7) {(b) Argument: $\alpha e_s$};
    \draw[->,gray] (-2.6,0)--(2.65,0) node[right,text=black] {$\mathrm{Re}$};
    \draw[->,gray] (0,-2.35)--(0,1.95) node[above,text=black] {$\mathrm{Im}$};
    \draw[dashed] (-2.6,0)--(0,0);
    \draw[continued] plot coordinates {
      (2.23607,0.00000)
      (2.23384,-0.09655)
      (2.22718,-0.19289)
      (2.21609,-0.28877)
      (2.20061,-0.38400)
      (2.18078,-0.47834)
      (2.15665,-0.57158)
      (2.12829,-0.66352)
      (2.09577,-0.75394)
      (2.05919,-0.84263)
      (2.01864,-0.92941)
      (1.97423,-1.01408)
      (1.92608,-1.09646)
      (1.87433,-1.17637)
      (1.81909,-1.25363)
      (1.76053,-1.32809)
      (1.69880,-1.39960)
      (1.63406,-1.46800)
      (1.56648,-1.53318)
      (1.49624,-1.59499)
      (1.42352,-1.65333)
      (1.34850,-1.70810)
      (1.27139,-1.75920)
      (1.19236,-1.80656)
      (1.11163,-1.85009)
      (1.02940,-1.88975)
      (0.94586,-1.92549)
      (0.86122,-1.95728)
      (0.77570,-1.98508)
      (0.68948,-2.00889)
      (0.60278,-2.02871)
      (0.51580,-2.04455)
      (0.42874,-2.05643)
      (0.34179,-2.06439)
      (0.25515,-2.06847)
      (0.16901,-2.06873)
      (0.08354,-2.06523)
      (-0.00107,-2.05806)
      (-0.08466,-2.04728)
      (-0.16705,-2.03301)
      (-0.24810,-2.01533)
      (-0.32765,-1.99436)
      (-0.40556,-1.97023)
      (-0.48171,-1.94304)
      (-0.55597,-1.91294)
      (-0.62823,-1.88006)
      (-0.69838,-1.84454)
      (-0.76633,-1.80652)
      (-0.83201,-1.76615)
      (-0.89534,-1.72359)
      (-0.95627,-1.67899)
      (-1.01473,-1.63249)
      (-1.07070,-1.58425)
      (-1.12414,-1.53442)
      (-1.17504,-1.48316)
      (-1.22340,-1.43061)
      (-1.26921,-1.37692)
      (-1.31249,-1.32222)
      (-1.35327,-1.26666)
      (-1.39156,-1.21036)
      (-1.42742,-1.15344)
      (-1.46089,-1.09603)
      (-1.49203,-1.03822)
      (-1.52089,-0.98013)
      (-1.54754,-0.92184)
      (-1.57205,-0.86344)
      (-1.59450,-0.80499)
      (-1.61496,-0.74656)
      (-1.63350,-0.68821)
      (-1.65022,-0.62997)
      (-1.66517,-0.57188)
      (-1.67844,-0.51396)
      (-1.69010,-0.45622)
      (-1.70021,-0.39868)
      (-1.70884,-0.34132)
      (-1.71604,-0.28413)
      (-1.72186,-0.22710)
      (-1.72634,-0.17021)
      (-1.72952,-0.11341)
      (-1.73142,-0.05669)
      (-1.73205,-0.00000)
    };
    \draw[fill=white] (0,0) circle (1.7pt) node[above right=3pt] {$0$};
    \fill ({sqrt(5)},0) circle (1.7pt) node[above=5pt] {$\theta=0$};
    \fill ({-sqrt(3)},0) circle (1.7pt)
      node[above=5pt] {$\theta=\pi/2$}
      node[below left=3pt] {$ze^{-\ii\pi}$};
  \end{scope}
\end{tikzpicture}
\caption{Energy continuation of the order and argument of
$K_{n-\ii\alpha}(\alpha e_s)$, with arrows along increasing
$\theta$, shown here for $X^2>3/2$. The order makes a clockwise three-quarter turn about $n$.
The argument starts at $\lambda\sqrt{1+X^2}$ and reaches
$-\lambda\sqrt{X^2-1}=ze^{-\ii\pi}$. The dashed line marks the cut of $K$.}
\label{fig:N-energy-Bessel}
\end{figure}

Although the endpoints are simple, the path of analytic continuation
matters. The Bessel functions depend on both $n-\ii\alpha$ and
$\alpha e_s$, whose paths are shown in Fig.~\ref{fig:N-energy-Bessel}.
For $\omega>0$, we introduce $\lambda=\omega/\Omega_{r,N}$ and use
the endpoint value $X=j\sqrt{\epsilon}$ to parameterize the path:
\begin{equation*}
 \alpha(\theta)=\lambda e^{-3\ii\theta},\qquad
 e_s(\theta)=\sqrt{1+X^2e^{2\ii\theta}}.
\end{equation*}
Here $X$ and $\lambda$ are fixed, while $\theta$ follows the energy
continuation. The square root starts at the positive scattering value
$\sqrt{1+X^2}$. As the energy changes sign at fixed $J$, the
quantity under the square root moves through the upper half-plane
from $1+X^2$ to $1-X^2$.
For $X>1$, the eccentricity therefore reaches
$e=\ii\sqrt{X^2-1}$. (This is the continuation from $e_s$ to $e$, we discuss the exchange $e\to-e$ occurs only in the subsequent $J$ route.)\vskip 4pt 

Along the energy path, $\alpha$ accumulates a $3\pi/2$ phase
clockwise, while $e_s$ accumulates $\pi/2$ counterclockwise.
The net change in the phase of the Bessel argument is therefore
\begin{equation}
 -\frac{3\pi}{2}+\frac{\pi}{2}=-\pi.
\end{equation}
At the endpoint,
\begin{equation}
 \alpha\to\ii\omega/\Omega_{r,N}=\ii\lambda.
\end{equation}
The order therefore reaches $\lambda+n$, and the continued Bessel
factor, still at independent frequency, becomes
\begin{equation}
 K_{\lambda+n}(z e^{-\ii\pi}),
 \qquad z=\lambda\sqrt{X^2-1}>0.
\end{equation}
The argument reaches the negative real axis with accumulated
phase $-\pi$. Since $K_{\lambda+n}$ has a branch cut there,
writing $e^{-\ii\pi}$ records which value is reached by the continuation.\vskip 4pt

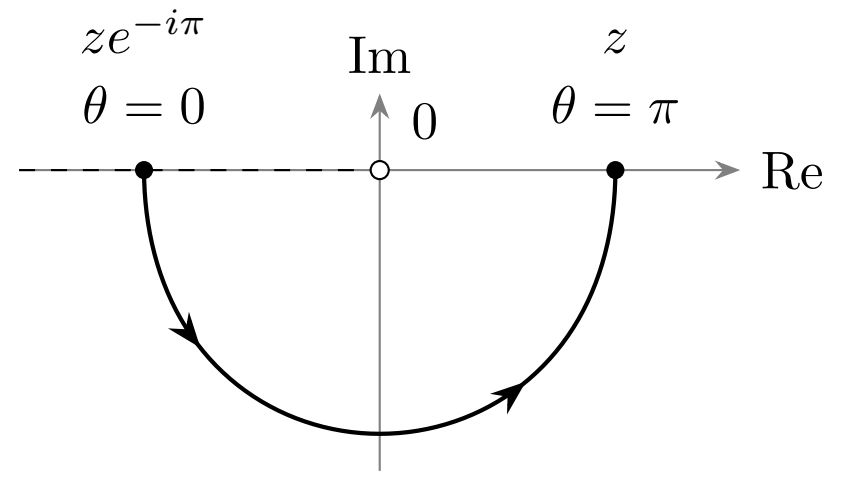
\begin{figure}[t!]
\centering

\begin{tikzpicture}[font=\small,>=Stealth,x=.90cm,y=.78cm]
  \draw[->,gray] (-2.65,0)--(2.65,0) node[right,text=black] {$\mathrm{Re}$};
  \draw[->,gray] (0,-2.55)--(0,.65) node[above,text=black] {$\mathrm{Im}$};
  \draw[dashed] (-2.65,0)--(0,0);
  \draw[thick,postaction={decorate},decoration={markings,
    mark=at position .24 with {\arrow{Stealth}},
    mark=at position .70 with {\arrow{Stealth}}}]
    plot coordinates {
      (-1.73205,0.00000)
      (-1.73086,-0.09066)
      (-1.72730,-0.18113)
      (-1.72135,-0.27124)
      (-1.71301,-0.36079)
      (-1.70226,-0.44962)
      (-1.68908,-0.53756)
      (-1.67347,-0.62445)
      (-1.65541,-0.71014)
      (-1.63488,-0.79449)
      (-1.61189,-0.87736)
      (-1.58643,-0.95864)
      (-1.55852,-1.03819)
      (-1.52815,-1.11591)
      (-1.49536,-1.19170)
      (-1.46017,-1.26544)
      (-1.42261,-1.33706)
      (-1.38272,-1.40646)
      (-1.34056,-1.47355)
      (-1.29617,-1.53825)
      (-1.24962,-1.60049)
      (-1.20098,-1.66018)
      (-1.15031,-1.71725)
      (-1.09770,-1.77165)
      (-1.04323,-1.82329)
      (-0.98698,-1.87213)
      (-0.92905,-1.91810)
      (-0.86953,-1.96114)
      (-0.80853,-2.00121)
      (-0.74613,-2.03826)
      (-0.68246,-2.07224)
      (-0.61761,-2.10311)
      (-0.55170,-2.13083)
      (-0.48483,-2.15537)
      (-0.41714,-2.17670)
      (-0.34872,-2.19479)
      (-0.27970,-2.20963)
      (-0.21020,-2.22119)
      (-0.14033,-2.22945)
      (-0.07023,-2.23441)
      (-0.00000,-2.23607)
      (0.07023,-2.23441)
      (0.14033,-2.22945)
      (0.21020,-2.22119)
      (0.27970,-2.20963)
      (0.34872,-2.19479)
      (0.41714,-2.17670)
      (0.48483,-2.15537)
      (0.55170,-2.13083)
      (0.61761,-2.10311)
      (0.68246,-2.07224)
      (0.74613,-2.03826)
      (0.80853,-2.00121)
      (0.86953,-1.96114)
      (0.92905,-1.91810)
      (0.98698,-1.87213)
      (1.04323,-1.82329)
      (1.09770,-1.77165)
      (1.15031,-1.71725)
      (1.20098,-1.66018)
      (1.24962,-1.60049)
      (1.29617,-1.53825)
      (1.34056,-1.47355)
      (1.38272,-1.40646)
      (1.42261,-1.33706)
      (1.46017,-1.26544)
      (1.49536,-1.19170)
      (1.52815,-1.11591)
      (1.55852,-1.03819)
      (1.58643,-0.95864)
      (1.61189,-0.87736)
      (1.63488,-0.79449)
      (1.65541,-0.71014)
      (1.67347,-0.62445)
      (1.68908,-0.53756)
      (1.70226,-0.44962)
      (1.71301,-0.36079)
      (1.72135,-0.27124)
      (1.72730,-0.18113)
      (1.73086,-0.09066)
      (1.73205,-0.00000)
    };
  \draw[fill=white] (0,0) circle (1.7pt) node[above right=3pt] {$0$};
  \fill ({-sqrt(3)},0) circle (1.7pt)
    node[above=5pt,align=center] {$ze^{-\ii\pi}$\\$\theta=0$};
  \fill ({sqrt(3)},0) circle (1.7pt)
    node[above=5pt,align=center] {$z$\\$\theta=\pi$};
  \node at (0,1.9) {Argument on the $J \to -J$ route};
\end{tikzpicture}
\caption{The additional continuation $J(\theta)=Je^{\ii\theta}$
at fixed energy and frequency. The argument starts at the endpoint
$ze^{-\ii\pi}$ of Fig.~\ref{fig:N-energy-Bessel} and returns to $z$
through the lower half-plane, now counterclockwise. The square root
is followed from $e=\ii\sqrt{X^2-1}$ to $-e$, while the order
$n+\lambda$ stays fixed.}
\label{fig:N-J-Bessel}
\end{figure}

\paragraph{Continue the reflected mode in $J$.} For the second route, start with the reflected scattering mode
$m=-2$. Relative to $U_{22}$, its angular dependence changes from
$e^{-2\ii\phi}$ to $e^{+2\ii\phi}$, and the Fourier transform becomes
\begin{equation*}
 S^{U,N}_{2,-2}
 =-\frac{\Cnorm\mu a_s^2\omega^2}{n_s}
 e^{\pi\alpha/2}
 \sum_{n=-3}^{3}
 c_{-n}(e_s,X)\,
 \ii^n K_{n-\ii\alpha}(\alpha e_s).
\end{equation*}
Hence, the reflection reverses the coefficient index while leaving the frequency and Bessel factors unchanged. We then perform the same energy continuation on this
reflected mode, and subsequently
\begin{equation}
 J(\theta)=J e^{\ii\theta},
 \qquad 0\leq\theta\leq\pi.
\end{equation}
As we discussed earlier, following $e=\sqrt{1-X^2}$ along the root-exchanging path gives $(e,X)\longrightarrow(-e,-X)$~\cite{Cho:2021arx}. During this continuation, the energy and frequency stay fixed,
so $\alpha=\ii\lambda$ and the Bessel order $\lambda+n$ do not change.
For $X>1$, the followed eccentricity makes a counterclockwise
half-turn from $e$ to $-e$. The Bessel argument follows it,
starting from the lower side of the negative real axis selected
by the energy route and returning through the lower half-plane
to the positive real axis: $ze^{-\ii\pi}\ \longrightarrow\ z$. We display in Fig.~\ref{fig:N-J-Bessel} this additional path. In conclusion, the first
route yields $K_{\lambda+n}(ze^{-\ii\pi})$, while the
reflected route reaches $K_{\lambda+n}(z)$.\vskip 4pt

For the reflected mode, we notice that the sum already carries the $c_{-n}$ coefficients. The $J$-route sends their arguments to
$(-e,-X)$, so the explicit coefficients in \eqref{eq:Ncoeffs} transform as
\begin{equation}
 c_{-n}(-e,-X)=(-1)^n c_n(e,X)\,,
\end{equation}
such that, after both continuations, the second route has the same
coefficients as the first but multiplied by a factor of $(-1)^n$. \vskip 4pt
 
\paragraph{Combine the two routes.}  After continuing both routes we set $\omega=p\Omega_{r,N}$, such that 
$\lambda=p$ and $z=p\sqrt{X^2-1}$. The Bessel order becomes $p+n$, and $\ii^n e^{\pi\alpha/2}\to\ii^{p+n}$. Together with the Bessel continuation described above, this gives
\begin{equation}
\begin{aligned}
 S_{\rm peri}^{U,N}
 &=-\frac{\ii\Cnorm\mu GM p^2}{\sqrt{\epsilon}}
 \sum_{n=-3}^{3}
 c_n(e,X)\ii^{p+n}K_{p+n}(ze^{-\ii\pi}),
\\
 S_{\rm apo}^{U,N}
 &=-\frac{\ii\Cnorm\mu GM p^2}{\sqrt{\epsilon}}
 \sum_{n=-3}^{3}
 (-1)^n c_n(e,X)\ii^{p+n}K_{p+n}(z).
 \label{eq:N-continued-routes}
\end{aligned}
\end{equation}
Here $e$ and $X$ in both sums denote their values after the
first energy continuation. The continuation of the
reflected mode is included in the second line.

The remaining step relies on the Bessel identity
\begin{equation}
 K_{p+n}(ze^{-\ii\pi})
 =(-1)^{p+n}K_{p+n}(z)+\ii\pi I_{p+n}(z),
 \qquad
 \ii^{p+n}I_{p+n}(z)=J_{p+n}(\ii z).
 \label{eq:connection}
\end{equation}
From here, and using the B2B combination in \eqref{b2bw}, we find 
\begin{equation}
 \frac{S_{\rm peri}^{U,N}-(-1)^pS_{\rm apo}^{U,N}}{T_{r,N}}
 =
 \frac{\Cnorm\mu\epsilon p^2}{2}
 \sum_{n=-3}^{3}c_n(e,X)J_{p+n}(pe),
 \label{eq:N-normalization}
\end{equation}
after dividing by the period, $T_{r,N}=2\pi GM/\epsilon^{3/2}$, and using the key identity: 
\begin{align}
 &c_n \ii^{p+n}
 \left[K_{p+n}(ze^{-\ii\pi})
       -(-1)^{p+n}K_{p+n}(z)\right]
 =\ii\pi c_n \ii^{p+n}I_{p+n}(z)
 =\ii\pi c_n J_{p+n}(pe).
 \label{eq:N-term-map}
\end{align}
The common $K_\lambda$ part cancels, and the remaining $I_\lambda$ function becomes the bound $J_\lambda$
function.\vskip 4pt

For nonzero frequency, and $0<e<1$, the Newtonian harmonic given in~\cite{Morras:2025nbp} reads\footnote{The Newtonian zero-frequency harmonic vanishes and is not obtained by setting $p=0$ in this expression.}
\begin{align}
 U^{b,N}_{22,p}&=-\Cnorm\mu\epsilon \,\frac{2}{e^2}\left[
 \left(-1+\frac{e^2}{2}+pX^3\right)J_p(pe)
 +eX(pX-1)J_p'(pe)\right].
 \label{eq:HN0}
 \end{align}
where the prime differentiates the argument. To compare with
\eqref{eq:N-normalization}, we use the identities
\begin{equation}
\begin{aligned}
 J_{p+n-1}(pe)+J_{p+n+1}(pe)
 &=\frac{2(p+n)}{pe}J_{p+n}(pe),\\
 J_{p-1}(pe)-J_{p+1}(pe)&=2J'_p(pe).
 \label{eq:recurrence}
\end{aligned}
\end{equation}
Repeated use of the first identity reduces the shifted orders in the
sum to $J_p$ and $J_{p\pm1}$; the second then expresses the result
in terms of $J_p$ and $J'_p$. Substituting the coefficients
\eqref{eq:Ncoeffs} and using $X^2+e^2=1$ gives
\begin{equation}
- \frac{2}{e^2}\left[
 \left(-1+\frac{e^2}{2}+pX^3\right)J_p(pe)
 +eX(pX-1)J_p'(pe)\right]=\frac{p^2}{2}\sum_{n=-3}^{3}c_n(e,X)J_{p+n}(pe)\,,\label{eq:HN}
\end{equation}
such that the B2B expression in \eqref{eq:N-normalization} reproduces the bound
harmonic:
\begin{equation}
U^{b,N}_{22,p}= \frac{1}{T_{r,N}}\left(S_{\rm peri}^{U,N}-(-1)^pS_{\rm apo}^{U,N}\right)\,.
\end{equation}
\vskip 4pt

Technically speaking, in order to complete the match there is yet a final analytic continuation, from $X>1$ to $0<X<1$, to be done. This can be achieved by taking a detour below $X=1$ so that
$e=+\ii\sqrt{X^2-1}$ reaches $e=+\sqrt{1-X^2}$ without passing
through zero. See App.~\ref{app:transport} for more details.  

\subsubsection{The $V_{21}$ spectrum}

Projecting the Newtonian current quadrupole onto the $21$ mode
and applying the Fourier-transform identities of~\cite{Bini:2021jmj}, we obtain
\begin{equation}
 S^{V,N}_{21}
 =-\frac{2C_{21}J a_s\omega^2}{n_s}
 e^{\pi\alpha/2}
 \sum_{n=-2}^{2}
 c_n^V(e_s,X)\,
 \ii^nK_{n-\ii\alpha}(\alpha e_s)\,,
 \label{current:eq:scatterN}
\end{equation}
with  $C_{21}=\frac{2\ii\Delta}{3}\sqrt{\frac{\pi}{5}}$, $\Delta \equiv \frac{m_1-m_2}{M}$.  The $c^V_n$ coefficients are (evaluated also at $e=e_s$)
\begin{equation}
\begin{array}{c|c@{\qquad}c|c}
 n&c_n^V(e,X)&n&c_n^V(e,X)\\\hline
 -2&-e(X+1)/4
 &1&(-X+e^2+1)/2\\
 -1&(X+e^2+1)/2
 &2&e(X-1)/4\\
 0&-3e/2&&
\end{array}
\label{current:eq:cn}
\end{equation}

\paragraph{The combined routes.}
We use the same energy path and reflected $J$ path as for
$U_{22}$, keeping $\omega$ independent until both continuations
are complete. The energy route brings the Bessel argument to
its branch with phase $-\pi$. The subsequent root-exchanging
$J$ route takes $(e,X)$ to $(-e,-X)$ and brings that argument
back to the positive real axis, on the reference region $X>1$.\vskip 4pt

As before, reflection also
reverses the index in the coefficient sum, which combines with the sign flip from the $J$ to $-J$ continuation. There is, however, an overall minus sign due to parity, such that the coefficients transform as 
\begin{equation}
 c_{-n}^V(-e,-X)=- (-1)^{n}c_n^V(e,X)\,,
 \label{current:eq:coefficient-parity}
\end{equation}
which produces an additional minus sign with respect to the $U_{22}$ term (see the table in \eqref{eq:parity-table}).
Once again we set $\omega=\omega_p=p\Omega_{r,N}$. The overall coefficient then transforms as: 
\begin{equation}
 -\frac{2C_{21}J a_s\omega^2}{n_s}
 \to
 \frac{2\ii C_{21}J a_N\omega^2}{\Omega_{r,N}}.
\end{equation}
Using the same $z=p\sqrt{X^2-1}$  variable as before, the two continued transforms are
\begin{equation}
\begin{aligned}
 S_{\rm peri}^{V,N}
 &=\frac{2\ii C_{21}J a_N\omega_p^2}{\Omega_{r,N}}
 \sum_{n=-2}^{2}
 c_n^V(e,X)\ii^{p+n}
 K_{p+n}(ze^{-\ii\pi}),\\
 S_{\rm apo}^{V,N}
 &=\frac{2\ii C_{21}J a_N\omega_p^2}{\Omega_{r,N}}
 \sum_{n=-2}^{2}
 (-1)^{n+1}c_n^V(e,X)\ii^{p+n}
 K_{p+n}(z).
 \label{current:eq:banks}
\end{aligned}
\end{equation}

The Bessel formula in \eqref{eq:connection} is unchanged. To cancel its common $K$ contribution, we multiply the second route by $(-1)^{p+1}$ as dictated by the B2B map for the $V_{21}$ mode. The cancellation then proceeds term by term:
\begin{align}
 c_n^V\ii^{p+n}
 \left[
 K_{p+n}(ze^{-\ii\pi})
 -(-1)^{p+n}K_{p+n}(z)
 \right] =\ii\pi c_n^V\ii^{p+n}I_{p+n}(z)
 =\ii\pi c_n^V J_{p+n}(pe).
\end{align}
Including the common prefactor, and dividing by
$T_{r,N}=2\pi/\Omega_{r,N}$, gives
\begin{equation}
 \frac{
 S_{\rm peri}^{V,N}-(-1)^{p+1}S_{\rm apo}^{V,N}
 }{T_{r,N}}
 =
 -C_{21}J a_N\omega_p^2
 \sum_{n=-2}^{2}c_n^V(e,X)J_{p+n}(pe).
 \label{current:eq:continued-bound}
\end{equation}

The leading current harmonic can be found in~\cite{Morras:2025nbp}. Expressed in our normalization, it becomes
\begin{equation}
 V^{b,N}_{21,p}
 =
 -\frac{C_{21}J a_N\omega_p^2}{p}
 \left[
 J'_p(pe)+\frac{X}{e}J_p(pe)
 \right].
 \label{current:eq:boundN}
\end{equation}
Similar  Bessel identities reduce the terms obtained
from scattering to the precise combination:
\begin{equation}
 \sum_{n=-2}^{2}c_n^V(e,X)J_{p+n}(pe)
 =
 \frac{1}{p}
 \left[
 J'_p(pe)+\frac{X}{e}J_p(pe)
 \right].
\end{equation}
Consequently,
\begin{equation}
 V^{b,N}_{21,p}
 =
 \frac{1}{T_{r,N}}
 \left[
 S_{\rm peri}^{V,N}
 -(-1)^{p+1}S_{\rm apo}^{V,N}
 \right].
 \label{current:eq:exact}
\end{equation}

Finally, the same continuation from $X>1$ to $0<X<1$, gives the physical elliptic harmonic.

\subsection{Mass harmonic to 1PN}
\label{sec:1pn}
At 1PN order, the scattering coefficients, orbital time scale and eccentricities receive corrections. The angular motion is also
corrected, giving periastron advance in the bound regime and
changing both the harmonic frequency and the angular phase.
Keeping these changes together gives the bound spectrum. We illustrate
this with the instantaneous mass harmonic at 1PN order and compare with the results in~\cite{Morras:2025nbp}.\vskip 4pt 

On the scattering side the Newtonian reference
variables are given by $X=-\ii \sqrt{2\E} j$ and $e^2=1-X^2$.
The corrected time eccentricity $e_t$ enters the Fourier phase,
so the Bessel argument is now $\alpha e_t$, with $\alpha = \omega/n_s$ as before, and $e_t=e-2\E\delta e_t$. The required clock and angular corrections are
\begin{align}
 n_s&=\frac{(2\E)^{3/2}}{GM}(1-2\E n_1),\qquad
 \Korb_s=1-2\E k_1,\qquad
 n_1=\frac{\nu-15}{8},\quad k_1=\frac3{X^2},
 \label{eq:1pn-frequencies}\\
 \delta e_t&=\frac{-2(1-\nu)-(7\nu-17)X^2/4}{2e}.
 \label{eq:et-main}
\end{align}
Since $X^2=-2\E j^2$, the angular coefficient is
$\Korb_s=1+3/j^2$ on the scattering orbit (which at this order coincides directly with the bound value for $\Korb$). The scattering spectrum reads\footnote{The full expression in~\eqref{eq:1pn-sc-spectrum} follows from the local quadrupole and hyperbolic motion obtained in~\cite{Mishra:2015bqa,Cho:2018upo}. See
App.~\ref{app:1pn} for more details. The PM
expanded values agree with the scattering coefficients through
$G^3$ in \cite{Bini:2024rsy,Bini:2024ijq}, see \S\ref{sec:PM-bound}.}
\begin{equation}
 S_{22}^{U}=-\frac{\Cnorm\mu a_{s,N}^2\omega^2}{n_s}
 \left[\sum_{n=-3}^{3}(c_n-2\E\delta c_n)\mathcal B_n
       +2\E k_1\sum_{n=-3}^{3}c_n\mathcal A_n
       +\Ord(\E^2)\right],
 \label{eq:1pn-sc-spectrum}
\end{equation}
with $a_{s,N}=GM/(2\E)$. We keep the same $c_n$ notation for the Newtonian values in \eqref{eq:Ncoeffs}, such that the corrections $\delta c_n$ include the orbital and source contributions listed in
App.~\ref{app:1pn}. The first sum uses
\begin{equation}
 \mathcal B_n =e^{\pi\alpha/2}\ii^nK_{n-\ii\alpha}(\alpha e_t).
 \label{eq:sc-Bmaster}
\end{equation}
This keeps the normalization used at Newtonian order, with
corrected coefficients, order and argument.  The second term is the precession contribution:\footnote{In this normalization, the anomaly
integral without the logarithm is equal to $2\mathcal B_n$. The factor of 2 cancels the $1/2$ in the mode projection
\eqref{eq:Q0}, yielding \eqref{eq:1pn-sc-spectrum}.} 
\begin{align}
 \mathcal A_n &=\int \dd v\,
 e^{(n-\ii\alpha)v+\ii \alpha e_t\sinh v}\log R_0(v), \label{eq:sc-Amaster}
\\
 R_0(v)&=\frac{e-\cosh v-X\sinh v}{e\cosh v-1},
 \qquad \log R_0(0)=0.\nonumber
\end{align}
The integration coordinate $v$ is the hyperbolic anomaly. Here
$R_0=e^{\ii\phi_N}$, with $\phi_N$ the Newtonian angle measured
from periastron, with $e$ the Newtonian eccentricity.\footnote{At 1PN order, the $R$ function parameterizing the angular evolution uses the angular eccentricity $e_\phi$ in place of $e$, and $\sqrt{1-e_\phi^2}$ in place of $X$. The orbital angle becomes $\phi=-\ii\Korb_s\log R$, so its contribution to the mode splits as
$e^{-2\ii\phi}=R^{-2}\exp[4\E k_1\log R]$ through 1PN, see \S\ref{sec:angterm}. On the one hand, the correction to $R^{-2}$ is included in the shifted $\delta c_n$ coefficients. On the other hand, the precession is already
of 1PN order, and therefore $R$ can be replaced by $R_0$, up to higher order corrections. Its Fourier transform, including the $1/2$ in
\eqref{eq:Q0}, gives the coefficient $2\E k_1$ of
$\mathcal A_n$ in \eqref{eq:1pn-sc-spectrum}.}

\paragraph{Continue the two routes.}
We keep $J$ and $\omega$ fixed and follow
$p_\infty(\theta)=s e^{\ii\theta}$ along \eqref{eq:energy-route}. Through 1PN we have
\begin{equation}
 s=\sqrt{\epsilon}\left[1-\frac{1+\nu}{8}\epsilon
 +\Ord(\epsilon^2)\right].
 \label{eq:sE}
\end{equation}
Since \eqref{eq:1pn-sc-spectrum} is already written in terms of $\E$,
we continue its energy dependence directly: $2\E$ reaches
$-\epsilon$ through the upper half-plane, such that the power $(2\E)^{3/2}$ reaches $-\ii\epsilon^{3/2}$. On the other hand,
$X$ reaches $j\sqrt{\epsilon}$, and each factor $-2\E$
in the corrections becomes $+\epsilon$. The continued orbital quantities are therefore
\begin{equation}
 \begin{gathered}
 a_{s,N}\to-a_N,\qquad n_s\to-\ii\Omega_r,\qquad
 \alpha\to\ii\lambda,\\
 e_t\to e+\epsilon\delta e_t,\qquad
 \Korb_s\to\Korb=1+\epsilon k_1,\\
 a_N=\frac{GM}{\epsilon},\qquad
 \lambda=\frac{\omega}{\Omega_r},\qquad
 \Omega_r=\Omega_{r,N}(1+\epsilon n_1).
 \end{gathered}
 \label{eq:1pn-endpoint-dictionary}
\end{equation}
%On a bound cycle the angular anomaly advances by $2\pi$, so the bound coefficient becomes $\Korb=\Korb_s$, as we mentioned before.\vskip 4pt 

We now follow the Bessel argument along the same parameter paths
as at Newtonian order. On the reference range, away from the
eccentricity zeros, the 1PN factors multiplying $e$ and $n_s$
stay close to one and introduce no additional winding.
The energy route therefore still brings $\alpha e_t$ to the
negative real axis from below.

For the reflected mode, this energy continuation is followed by
$J(\theta)=J e^{\ii\theta}$, $0\leq\theta\leq\pi$, at fixed
energy and frequency. We evaluate $e_t=e+\epsilon\delta e_t$
continuously along this path. At its endpoint, the value of $e$ changes sign \cite{Cho:2021arx}. The numerator of $\delta e_t$ in \eqref{eq:et-main} is therefore
unchanged, while its denominator $2e$ changes sign, and therefore $e_t$
reaches $-e_t$ through 1PN.\footnote{The same reasoning applies to $e_r$ and $e_\phi$. See
App.~\ref{app:1pn} for more details.} Since $n_s$ depends only on energy
at this order, $\alpha$ remains fixed, and the phase of the
Bessel argument increases by $\pi$, as in the Newtonian example.
Introducing $z_t=-\ii\lambda e_t>0$, with the (complex) $e_t$ evaluated at the energy endpoint, the two steps give
\begin{equation}
 \alpha e_t\ \xrightarrow{\ \text{energy}\ }\ z_t e^{-\ii\pi}
 \ \xrightarrow{\ \text{reflected }J\ }\ z_t,
 \qquad \ii z_t=\lambda e_t.
 \label{eq:1pn-zt-routes}
\end{equation}

The reflected mode again reverses the summation index.
Evaluating its coefficients at the endpoint of the $J$ path gives
\begin{equation}
 c_{-n}(-e,-X)=(-1)^nc_n(e,X),\qquad
 \delta c_{-n}(-e,-X)=(-1)^n\delta c_n(e,X).
 \label{eq:1pn-coefficient-parity}
\end{equation}
Hence, in terms of the two continued values of
\eqref{eq:sc-Bmaster}, the scattering spectra have the
same form as their Newtonian counterparts:
\begin{equation}
\begin{gathered}
\begin{aligned}
 S_{\rm peri}^{U}
 &=-\frac{\ii\Cnorm\mu a_N^2\omega^2}{\Omega_r}
 \sum_{n=-3}^{3}(c_n+\epsilon\delta c_n)
 \mathcal B_n^{\rm peri}+\cdots,\\
 S_{\rm apo}^{U}
 &=-\frac{\ii\Cnorm\mu a_N^2\omega^2}{\Omega_r}
 \sum_{n=-3}^{3}(-1)^n(c_n+\epsilon\delta c_n)
 \mathcal B_n^{\rm apo}+\cdots.
\end{aligned}\\[6pt]
 \mathcal B_n^{\rm peri}
 =e^{\ii\pi(\lambda+n)/2}K_{\lambda+n}(z_t e^{-\ii\pi}),
 \qquad
 \mathcal B_n^{\rm apo}
 =e^{\ii\pi(\lambda+n)/2}K_{\lambda+n}(z_t).
\end{gathered}
\label{eq:1pn-B-routes}
\end{equation}
The dots denote the precession contribution proportional to
$\mathcal A_n$ in \eqref{eq:1pn-sc-spectrum}, which we combine
with these terms below. The same identities used in the Newtonian
case give~\cite[Eqs.~10.34.2 and 10.27.6]{DLMF}
\begin{equation}
 K_{\lambda+n}(z_t e^{-\ii\pi})
 =e^{\ii\pi(\lambda+n)}K_{\lambda+n}(z_t)
       +\ii\pi I_{\lambda+n}(z_t),\quad  
 \ii^{\lambda+n}I_{\lambda+n}(z_t)
 =J_{\lambda+n}(\ii z_t).
 \label{eq:1pn-B-connection}
\end{equation}
The change from Newtonian order lies in the corrected eccentricity
and frequency. After completing both continuations, we set
\begin{equation}
 \omega_p=[p+2(\Korb-1)]\Omega_r,\quad
 \lambda=p+2\epsilon k_1,\quad i z_t \to p e +2\epsilon k_1 e +  p \epsilon \delta e_t +{\cal O}(\epsilon^2)
 \label{eq:lambda1pn}
\end{equation}
through relative 1PN order. Once again, the half-period clock and angle shifts
 give (for $m=2$):
\begin{equation}
 e^{-\ii\omega_pT_r/2+\ii\Delta\phi}
 =e^{-\ii\pi\lambda+2\ii\pi\Korb}
 =(-1)^p\,,
 \label{eq:1pn-phase}
\end{equation}
such that the subtraction remains
$S_{\rm peri}-(-1)^pS_{\rm apo}$. Since the Bessel identity in \eqref{eq:1pn-B-connection}
contains $e^{\ii\pi(\lambda+n)}$, with noninteger $\lambda$,
the ordinary Bessel terms alone leave a contribution
proportional to $K_{\lambda+n}$. As we shall, see the angular
term is needed to remove this unwanted piece. 

\paragraph{The angular correction.}
We now apply the same two continuations to the angular
frequency function in \eqref{eq:sc-Amaster}. On the original
scattering encounter, $v$ runs from $-\infty$ to $+\infty$.
Introducing $w=e^v$, such that $0<w<\infty$ and
$\dd v=\dd w/w$, we have
\begin{equation}
 \mathcal A_n
 =\int_0^\infty\frac{\dd w}{w}\,
 w^{n-\ii\alpha}
 e^{\frac{\ii\alpha e_t}{2}(w-w^{-1})}
 \log R_0(w), \qquad R_0(w)=\frac{w-\beta}{1-\beta w},
 \qquad \beta=\frac{e}{1+X}.
 \label{eq:1pn-angular-w-input}
\end{equation}
As in \eqref{eq:1pn-B-routes}, we construct
$\mathcal A_n^{\rm peri}$ and $\mathcal A_n^{\rm apo}$, through the energy route and the
energy route followed by the reflected $J$ path,
respectively. The angular term enters the same B2B subtraction as the
ordinary Bessels. However, the relative coefficient changes sign between
the modes: after the energy continuation, the expansion
of $e^{-2\ii\phi}$ gives $-2\epsilon k_1\log R_0$,
whereas $e^{+2\ii\phi}$ gives $+2\epsilon k_1\log R_0$.
Using the normalization of \eqref{eq:1pn-sc-spectrum}
and the coefficient parity in \eqref{eq:1pn-coefficient-parity},
the $n$th angular term in the bracket becomes
\begin{equation}
 \begin{aligned}
 -\epsilon k_1c_n\mathcal A_n^{\rm peri}
 -(-1)^p\bigl[\epsilon k_1(-1)^nc_n
                  \mathcal A_n^{\rm apo}\bigr]=-\epsilon k_1c_n
       \bigl[\mathcal A_n^{\rm peri}
              +(-1)^{p+n}\mathcal A_n^{\rm apo}\bigr]\,.
 \end{aligned}
 \label{eq:logr0}
\end{equation}

In order to perform the analytic continuation, we notice that $\alpha\to\ii\lambda$ after the energy route, and it remains fixed
on the subsequent $J$ path. Hence, the power in
\eqref{eq:1pn-angular-w-input} becomes
$w^{n-\ii\alpha}\to w^{n+\lambda}$ on both routes.
The two continued Fourier kernels are therefore
\begin{equation}
 w^{n-\ii\alpha}
 e^{\frac{\ii\alpha e_t}{2}(w-w^{-1})}
 \ \longrightarrow\
 \begin{cases}
 w^{n+\lambda}e^{-\frac{\ii z_t}{2}(w-w^{-1})},
       &\text{energy route},\\[3pt]
 w^{n+\lambda}e^{+\frac{\ii z_t}{2}(w-w^{-1})},
       &\text{energy and $J$ routes}.
 \end{cases}
 \label{eq:1pn-angular-continued-kernels}
\end{equation}
These are the same kernels underlying the two Bessel functions
in \eqref{eq:1pn-B-routes}. The additional factor is the angular logarithm.\vskip 4pt

The angular parameters follow the same root exchange as
the time eccentricity. Through 1PN, with 
$e_\phi=e+\epsilon\delta e_\phi$, we have
\begin{equation}
 \sqrt{1-e_\phi^2}
 =X-\epsilon\frac{e\,\delta e_\phi}{X}+\Ord(\epsilon^2)
 \ \longrightarrow\
 -\left(X-\epsilon\frac{e\,\delta e_\phi}{X}\right)
 +\Ord(\epsilon^2).
 \label{eq:1pn-angular-root-branch}
\end{equation}
The logarithmic insertion already carries one power of
$\epsilon$, so it needs only the Newtonian angular
function $R_0$. Using $e,X$ for the values obtained after
the energy route, the $J$ path sends
$(e,X)$ to $(-e,-X)$ and therefore
\begin{equation} \beta=e/(1+X) \ \longrightarrow\ -e/(1-X)=-\beta^{-1}\,, \quad \text{$J$ route}.\end{equation}

We must analytically continue the logarithm together with the Fourier kernel. The two
continued values of $\mathcal A_n$ are fixed by the
parameter paths in \eqref{eq:1pn-zt-routes}. We notice, however, that the two
kernels in \eqref{eq:1pn-angular-continued-kernels} have
opposite signs in the exponential. In the integral representation, the substitution
$w\mapsto e^{-\ii\pi}w$ in the reflected mode puts them in the same form:
\begin{equation}
 \left.
 w^{\lambda+n}e^{\frac{\ii z_t}{2}(w-w^{-1})}
 \right|_{w\mapsto e^{-\ii\pi}w}
 =e^{-\ii\pi(\lambda+n)}
 w^{\lambda+n}e^{-\frac{\ii z_t}{2}(w-w^{-1})}.
 \label{eq:1pn-angular-kernel-shift}
\end{equation}
The measure $\dd w/w$ is unchanged, and the contour and
logarithms are carried with this change of variable. To follow this change in the logarithm, we rewrite it as
\begin{equation}
 \begin{aligned}
 \log R_0(w)&=\log w+
       \log\frac{1-\beta/w}{1-\beta w}.
 \end{aligned}
 \label{eq:1pn-R-factor}
\end{equation}
Substituting the continued $\beta$ at fixed $w$ gives
\begin{equation}
 \begin{aligned}
 \left.R_0(w)\right|_{\beta\to-\beta^{-1}}
 &=\frac1w\frac{1+\beta w}{1+\beta/w},\\
 \left.\log R_0(w)\right|_{\beta\to-\beta^{-1}}
 &=-\log w-\log\frac{1+\beta/w}{1+\beta w}\,,
 \end{aligned}
 \label{eq:1pn-reflected-log-main}
\end{equation}
where the branches are followed along both parameter paths with the normalization $\log R_0(1)=0$ inherited from \eqref{eq:sc-Amaster}. The argument branch is fixed by \eqref{eq:1pn-zt-routes}, and therefore
$\log(e^{-\ii\pi}w)=\log w-\ii\pi$. Applying this substitution to the reflected angular
function gives\footnote{The constant $\ii\pi$ carries the angular-origin
shift in \eqref{eq:clockmatch}. At Newtonian order, the
energy-continued angle is zero at periastron and reaches
$\pi$ at apoapsis. The energy-reflected continuation is zero
at apoapsis instead, and reverses the direction of angular motion: $\phi_{\rm apo}=\pi-\phi_{\rm peri}$. The factor of
$\pi$ ensures that the reflected angle vanishes at its
starting root. Since $\log R_0=\ii\phi_N$, this same relation
gives $\ii\pi-\log R_0$. The logarithmic continuation
therefore carries the normalization fixed
in \eqref{eq:clocks} and inherited from the condition $\phi_N(r_{\rm min})=0$. In fact, the same normalization is also apparent in \eqref{eq:1pn-angular-kernel-shift}, with the factor of $e^{-\ii\pi\lambda}= e^{-\ii \omega T_r/2}$ carrying the time shift instead.}
\begin{equation}
 \begin{aligned}
 \left.R_0(e^{-\ii\pi}w)\right|_{\beta\to-\beta^{-1}}
 &=-\frac1{R_0(w)},\\
 \left.\log R_0(e^{-\ii\pi}w)\right|_{\beta\to-\beta^{-1}}
 &=\ii\pi-\log R_0(w).
 \end{aligned}
 \label{eq:1pn-logshift}
\end{equation}
We now insert \eqref{eq:1pn-logshift} into the angular
combination in \eqref{eq:logr0}. The constant $\ii\pi$
multiplies the reflected integral without a logarithm,
which is $2\mathcal B_n^{\rm apo}$ in the normalization
of \eqref{eq:sc-Bmaster}. The remaining insertions are
$+\log R_0(w)$ on the first path and $-\log R_0(w)$ on
the transformed reflected path. Since the angular term
already carries $\epsilon$, we need its value only at
$\lambda=p$. The outside phase in
\eqref{eq:1pn-angular-kernel-shift} then cancels the B2B
weight $(-1)^{p+n}$, leaving a difference of integrals
with the same kernel and logarithm. Denoting their paths
by $\mathrm E$ and $\mathrm{EJ}$, respectively, we obtain
\begin{equation}
 \begin{aligned}
 \left[\mathcal A_n^{\rm peri}
       +(-1)^{p+n}\mathcal A_n^{\rm apo}\right]_{\lambda=p}
 &=2\ii\pi(-1)^{p+n}
        \left.\mathcal B_n^{\rm apo}\right|_{\lambda=p}\\
 &+\left[\left(\int_{\rm E}-\int_{\rm EJ}\right)
 \frac{\dd w}{w}\,w^{\lambda+n}
 e^{-\frac{\ii z_t}{2}(w-w^{-1})}\log R_0(w)
 \right]_{\lambda=p}.
 \end{aligned}
\label{eq:1pn-angular-combined-paths}
\end{equation}
To evaluate this difference, we separate $\log R_0$ as in
\eqref{eq:1pn-R-factor}. The factor $\log w$ can be
inserted by differentiating the power of $w$:
\begin{equation}
 \frac{\partial}{\partial\lambda}w^{\lambda+n}
 =w^{\lambda+n}\log w,
 \label{eq:1pn-angular-log-insertion}
\end{equation}
with $z_t$ and $\beta$ fixed. The first integral without the logarithm produces 
$2\mathcal B_n^{\rm peri}$. The second is equal to
$2e^{\ii\pi(\lambda+n)}\mathcal B_n^{\rm apo}$, since the
phase multiplying the transformed integral in
\eqref{eq:1pn-angular-kernel-shift} must be restored to
recover $2\mathcal B_n^{\rm apo}$. Their difference
follows from \eqref{eq:1pn-B-routes} and
\eqref{eq:1pn-B-connection}, such that
\begin{equation*}
 2\left[\mathcal B_n^{\rm peri}
 -e^{\ii\pi(\lambda+n)}\mathcal B_n^{\rm apo}\right]
 =2\pi\ii J_{\lambda+n}(\ii z_t).
\end{equation*}
Differentiating this equality and setting $\lambda=p$ later gives (see App.~\ref{app:1pn} for more details)
\begin{equation}
 2\left.\partial_\lambda\left[
 \mathcal B_n^{\rm peri}
 -e^{\ii\pi(\lambda+n)}\mathcal B_n^{\rm apo}\right]
 \right|_{\lambda=p}
 =2\pi\ii\left.\partial_\lambda
 J_{\lambda+n}(\ii z_t)\right|_{\lambda=p}.
 \label{eq:1pn-angular-aligned-derivative}
\end{equation}
\vskip 4pt

The remaining logarithmic ratio multiplies the same Bessel
kernel as before. As we have shown, e.g., \eqref{eq:1pn-B-connection}, the difference in the paths without the logarithm would ultimately give a factor of $2\pi\ii J_{\lambda+n}(\ii z_t)$. We therefore use the same contour deformation underlying the Bessel
identities~\cite[\S10.9(ii)]{DLMF}, now carrying the
logarithmic ratio along the way. We show in Fig.~\ref{fig:1pn-angular-paths} the contours
at the continued values for $z_t$ and $\beta$.\vskip 4pt

In the integral representation, the energy continuation of the Bessel argument
in \eqref{eq:1pn-zt-routes} also requires turning the ends towards
directions of exponential decay. The finite arcs then connect
these ends to the real segments. As a result, the two paths in the
left and middle panels Fig.~\ref{fig:1pn-angular-paths} represent the analytically continued Bessels shown in 
\eqref{eq:1pn-B-routes}. Both run from $0$ to infinity, approaching these ends in the
lower imaginary direction, where the exponential decay as:
\begin{equation}
 \left.e^{-\frac{\ii z_t}{2}(w-w^{-1})}\right|_{w=-\ii|w|}
 =e^{-\frac{z_t}{2}(|w|+|w|^{-1})}\,, \qquad z_t>0.
\end{equation}
The addition of the $\log R_0$ term does not change this picture. After separating
the factor of $\log w$, it introduces at most logarithmic growth, still
dominated by the exponential decay. The only remaining issue is the existence of finite cuts in the logarithmic ratio in \eqref{eq:1pn-R-factor}.\vskip 4pt

On the $X>1$ branch, the cuts join $0$ to $\beta$ and
$\beta^{-1}$ to infinity, as drawn in red on the picture to the right of~Fig.~\ref{fig:1pn-angular-paths}. Following the
branches from \eqref{eq:1pn-logshift} gives the same
logarithm on the common gray portions. At $\lambda=p$,
factors of $w^{p+n}$ also agree, so these portions cancel out.
What remains is the finite loop on the right. It encloses
the inner cut and leaves the outer cut outside. Moving its
upper arc inwards and its lower arc outwards to $|w|=1$
crosses neither cut, so the integral is unchanged. On this 
circle (in green) the logarithm has a convergent Laurent series,
\begin{figure}[t!]
\centering
\begin{tikzpicture}[x=.57cm,y=.57cm,font=\small]
 % Each panel is a contour representative at fixed continued parameters.
 \foreach \x in {0,6.5,13}{
  \begin{scope}[shift={(\x,0)}]
   \draw[gray!25](-2.65,0)--(2.65,0);
   \draw[rem,thick](0,0)--(0,.7);
   \fill[rem](0,.7) circle(1.2pt);
   \draw[rem,thick,->](0,-1.4286)--(0,-3.15);
   \fill[rem](0,-1.4286) circle(1.2pt);
   \fill(0,0) circle(1pt);
  \end{scope}
 }
 \begin{scope}
  \node at(0,3.35){$\int_{\rm E}$ path};
%  \node at(0,2.9){$\int_{\rm E}$};
  \node[anchor=north east] at(-.08,.05){$0$};
  \draw[gray!80,thick,arr](0,-.04)--(0,-.55);
  \draw[routea,thick](0,-.55) arc(-90:0:.55);
  \draw[routea,thick,arr](.55,0)--(2.4,0);
  \draw[routea,thick,arr](2.4,0) arc(0:180:2.4);
  \draw[gray!80,thick,arr](-2.4,0) arc(180:260:2.4);
  \draw[gray!80,thick,->]
   ({2.4*cos(260)},{2.4*sin(260)})
   --({2.4*cos(260)},-3.15);
  \node[anchor=north east] at(-.2,-3.2){$-\ii\infty$};
 \end{scope}
 \node at(3.25,.4){$-$};
 \begin{scope}[shift={(6.5,0)}]
  \node at(0,3.35){$\int_{\rm EJ}$ path};
 % \node at(0,2.9){$\int_{\rm EJ}$};
  \node[anchor=north east] at(-.08,.05){$0$};
  \draw[gray!80,thick,arr](0,-.04)--(0,-.55);
  \draw[routeb,thick](0,-.55) arc(270:180:.55);
  \draw[routeb,thick,arr](-.55,0)--(-2.4,0);
  \draw[gray!80,thick,arr](-2.4,0) arc(180:260:2.4);
  \draw[gray!80,thick,->]
   ({2.4*cos(260)},{2.4*sin(260)})
   --({2.4*cos(260)},-3.15);
  \node[anchor=north east] at(-.2,-3.2){$-\ii\infty$};
 \end{scope}
 \node at(9.75,.4){$=$};
 \begin{scope}[shift={(13,0)}]
  \node at(0,3.35){Closed loop};
%  \node at(0,2.9){$\lambda=p$};
  \draw[routea,thick,arr](.55,0)--(2.4,0);
  \draw[routea,thick,arr](2.4,0) arc(0:180:2.4);
  \draw[routeb,thick,arr](-2.4,0)--(-.55,0);
  \draw[routeb,thick](-.55,0) arc(180:270:.55);
  \draw[routea,thick](0,-.55) arc(270:360:.55);
  \draw[bound,thick,densely dashed](0,0) circle(1);
  \node[rem,anchor=west] at(.08,.7){$\beta$};
  \node[rem,anchor=west] at(.08,-1.43){$\beta^{-1}$};
  \node[bound,anchor=west] at(.85,-.85){$|w|=1$};
  \draw[gray!65,->](1.35,1.5)--(.67,.75);
  \draw[gray!65,->](-.13,-.63)--(-.13,-.96);
 \end{scope}
\end{tikzpicture}
\caption{In the left and middle panels we show the integral contours in the $w$-plane for the two analytically continued Bessel functions in
\eqref{eq:1pn-B-routes}, after the reflected substitution in
\eqref{eq:1pn-angular-kernel-shift}, at fixed $z_t>0$.
The logarithmic ratio adds the two red cuts on the right panel, shown for
$\beta=\ii|\beta|$ on the reference range $(X>1$).
The common gray portions cancel at $\lambda=p$. The subtraction of the orange
path then closes the loop on the right.
Its upper and lower arcs can be moved to the (green) dashed unit circle
without crossing the logarithmic cuts (in red). See App.~\ref{app:transport} for a similar argument.}
\label{fig:1pn-angular-paths}
\end{figure}
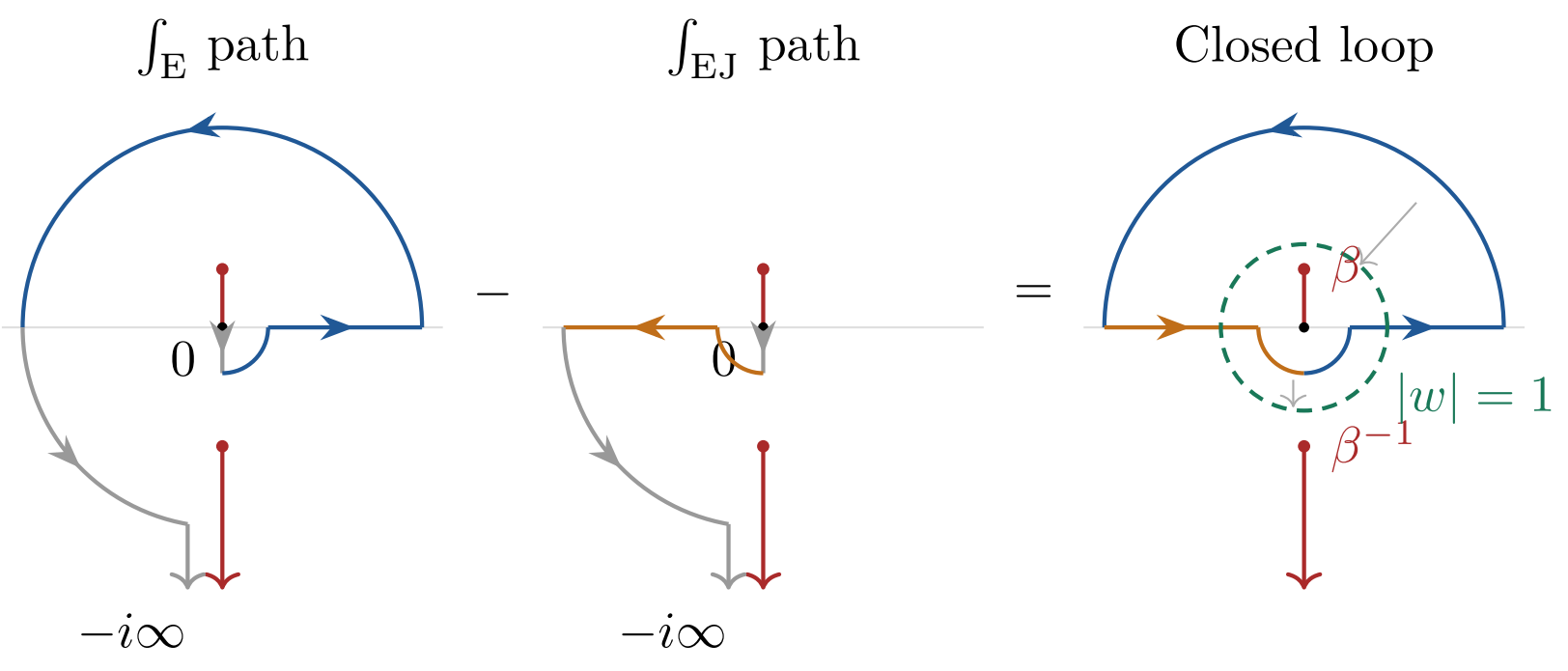
\begin{equation}
 \log\frac{1-\beta/w}{1-\beta w}
 =\sum_{s=1}^{\infty}\frac{\beta^s}{s}(w^s-w^{-s}),
 \qquad |\beta|<|w|<|\beta|^{-1}.
 \label{eq:periodic-log}
\end{equation}
Setting $w=e^{\ii u}$ gives $\dd w/w=\ii\dd u$ and,
with $\ii z_t=pe$ at this order, the Fourier factor becomes
$e^{\ii(p+n)u-\ii pe\sin u}$. Each factor of $w^{\pm s}$
shifts the resulting Bessel order by $\pm s$, yielding
$2\pi\ii J_{p+n\pm s}(pe)$.\footnote{The reader will immediately notice that, without the logarithmic insertion, the same contour integral gives $2\pi\ii J_{p+n}(pe)$, reproducing the identities in \eqref{eq:1pn-B-connection}.} 

Combining this periodic integral with the constant from
\eqref{eq:1pn-logshift} and the order derivative in
\eqref{eq:1pn-angular-aligned-derivative}, we obtain
\begin{equation}
 \begin{aligned}
 &\left[\mathcal A_n^{\rm peri}
       +(-1)^{p+n}\mathcal A_n^{\rm apo}\right]_{\lambda=p}
 =2\ii\pi(-1)^{p+n}
       \left.\mathcal B_n^{\rm apo}\right|_{\lambda=p}\\
 &+2\pi\ii\Biggl[
 \left.\frac{\partial J_{\lambda+n}(pe)}{\partial\lambda}
       \right|_{\lambda=p}
 -\sum_{s=1}^{\infty}\frac{\beta^s}{s}
       \bigl[J_{p+n-s}(pe)-J_{p+n+s}(pe)\bigr]\Biggr],
 \end{aligned}
 \label{eq:1pn-angular-integral-connection}
\end{equation}
where the minus sign comes from reversing the order of the two
Bessel terms generated by $w^s-w^{-s}$ in \eqref{eq:periodic-log}.\vskip 4pt

We now combine the ordinary and angular contributions at
the bound frequency in \eqref{eq:lambda1pn}. The angular
term already carries $-\epsilon k_1$, so its frequency
functions are needed only at $\lambda=p$. The ordinary Bessel term, however,
must be evaluated at $\lambda=p+2\epsilon k_1$. Using \eqref{eq:1pn-B-routes} and
\eqref{eq:1pn-B-connection}, we first write
\begin{equation*}
 \mathcal B_n^{\rm peri}
 -(-1)^{p+n}\mathcal B_n^{\rm apo}
 =
 \pi\ii J_{\lambda+n}(\ii z_t)
 +\bigl[e^{\ii\pi(\lambda+n)}-(-1)^{p+n}\bigr]
       \mathcal B_n^{\rm apo}.
\end{equation*}
At Newtonian order, $\lambda=p$, and the coefficient of
$\mathcal B_n^{\rm apo}$ vanishes. The 1PN frequency shift
instead gives
$e^{\ii\pi(\lambda+n)}
=(-1)^{p+n}[1+2\pi\ii\epsilon k_1]+\Ord(\epsilon^2)$, such that we end up with a factor of 
$2\pi\ii\epsilon k_1(-1)^{p+n}\mathcal B_n^{\rm apo}|_{\lambda=p}$.
The $\ii \pi$ factor in \eqref{eq:1pn-angular-integral-connection},
multiplied by $-\epsilon k_1$, precisely cancels this term.  Their combined coefficient, with $\lambda=p+2\epsilon k_1$, becomes
\begin{equation}
 \begin{aligned}
 &e^{\ii\pi(\lambda+n)}-(-1)^{p+n}
       -2\pi\ii\epsilon k_1(-1)^{p+n}\\
 &\quad=(-1)^{p+n}
 \bigl[1+2\pi\ii\epsilon k_1+\Ord(\epsilon^2)
       -1-2\pi\ii\epsilon k_1\bigr]
 =\Ord(\epsilon^2)\,.
 \end{aligned}
 \label{eq:1pn-log-match}
\end{equation}
Hence, no $K$ contribution remains at 1PN order.\vskip 4pt

The same frequency shift also changes the order of the
remaining $J$ function. Expanding its order while keeping
its argument fixed gives
\begin{equation*}
 J_{\lambda+n}(\ii z_t)
 =
 J_{p+n}(\ii z_t)
 +2\epsilon k_1
 \left.\frac{\partial J_{\lambda+n}(pe)}
                  {\partial\lambda}\right|_{\lambda=p}
 +\Ord(\epsilon^2)\,.
\end{equation*}
In the derivative term we may use the Newtonian argument
$pe$, since this term already multiplies $\epsilon$.
The first term keeps the corrected argument $\ii z_t$
from \eqref{eq:lambda1pn}. The order derivative in
\eqref{eq:1pn-angular-integral-connection}, multiplied by
$-\epsilon k_1$, also cancels this term. The Bessel order returns to $p+n$, while
the clock and frequency corrections remain in the argument.
Together with the shifted Bessel sum from the angular
logarithm, this gives
\begin{equation}
 \begin{aligned}
 &\Bigl[\mathcal B_n^{\rm peri}
       -(-1)^{p+n}\mathcal B_n^{\rm apo}
       -\epsilon k_1\bigl(\mathcal A_n^{\rm peri}
              +(-1)^{p+n}\mathcal A_n^{\rm apo}\bigr)
       \Bigr]_{\lambda=p+2\epsilon k_1}\\
 &\quad=\pi\ii\Biggl\{
 J_{p+n}(\ii z_t)
 +2\epsilon k_1\sum_{s=1}^{\infty}\frac{\beta^s}{s}
       \bigl[J_{p+n-s}(pe)-J_{p+n+s}(pe)\bigr]
 \Biggr\}+\Ord(\epsilon^2).
 \end{aligned}
 \label{eq:1pn-order-shift}
\end{equation}
The second term is the periodic angular contribution that
survives these cancellations. The argument must still be expanded at the corrected
frequency and time eccentricity:
\begin{equation}
 \ii z_t=\lambda e_t
 =pe+\epsilon(p\delta e_t+2k_1e)+\Ord(\epsilon^2).
 \label{eq:1pn-secular-cancellation}
\end{equation}
Multiplying \eqref{eq:1pn-order-shift} by $c_n$
therefore supplies
$(p\delta e_t+2k_1e)c_nJ'_{p+n}(pe)$ at 1PN, together
with the angular sum. The polynomial correction in
\eqref{eq:1pn-sc-spectrum} adds
$\delta c_nJ_{p+n}(pe)$. These are the three contributions
that survive the B2B analytic continuation.\vskip 4pt

To put the angular sum in the form that appears in the literature \cite{Morras:2025nbp}, we use the explicit form of the
 $c_n$  coefficients and use integration by parts on its
periodic Fourier representation. This produces an angular
function, which we denote as $\widetilde f_p(e)$ (see below), and finite ordinary Bessel
terms.  Expanding the common prefactor and dividing  by $T_r$ gives the 1PN correction in  
\begin{equation}
 U^b_{22,p}=U^{b,N}_{22,p}+\epsilon U^{b,(1)}_{22,p}+\cdots\,,
\end{equation}
which takes the form
\begin{align}
 \left.U^{b,(1)}_{22,p}\right|_{\rm B2B}
 &=\frac23(n_1+k_1)U^{b,N}_{22,p}
   +\delta e_t\partial_eU^{b,N}_{22,p}\nonumber\\
 &\quad-\frac{\Cnorm\mu\epsilon}{X^2}
 \left[
 \frac{A_p^{\rm B2B}J_p(pe)+B_p^{\rm B2B}eJ'_p(pe)}{e^2}
 +p\widetilde f_p(e)\right].
 \label{eq:1pn-continued-complete}
\end{align}
The first two terms display the amplitude and eccentricity
changes when the result is written in the corrected orbital
variables. The polynomials $A_p^{\rm B2B}$ and $B_p^{\rm B2B}$
include the finite terms from the angular reduction. We provide more details in App.~\ref{app:1pn}.

\paragraph{Comparison with bound harmonic.} The bound expression in \cite{Morras:2025nbp} uses $e_t$ and
$y^2=(GM\Omega_\phi)^{2/3}/(1-e_t^2)$. Its conversion to our
energy variables is
\begin{equation}
 (GM\Omega_\phi)^{2/3}
 =\epsilon\left[1+\frac23\epsilon(n_1+k_1)\right]
 +\Ord(\epsilon^3),\qquad
 y^2=\frac{\epsilon}{X^2}+\Ord(\epsilon^2). \label{eq:Omegaphi}
\end{equation}
Together with $e_t=e+\epsilon\delta e_t$, this gives
\begin{align}
 \left.U^{b,(1)}_{22,p}\right|_{\text{Ref.~\cite{Morras:2025nbp}}}
 &=\frac23(n_1+k_1)U^{b,N}_{22,p}
 +\delta e_t\partial_eU^{b,N}_{22,p}
 -\frac{\Cnorm\mu\epsilon}{X^2}N_1(e,p,\nu),
 \label{eq:publishedH1}\\
 N_1(e,p,\nu)
 &=\frac{A_p^{\mathrm{1PN}}J_p(pe)+B_p^{\mathrm{1PN}}eJ'_p(pe)}{e^2}
       +p\widetilde f_p(e).
 \label{eq:pubN1}
\end{align}
Here $A_p^{\mathrm{1PN}}(e,\nu)$ and $B_p^{\mathrm{1PN}}(e,\nu)$ are the
full 1PN polynomials of~\cite{Morras:2025nbp}, given in App.~\ref{app:1pn}. The precession function is exactly the one obtained above from
the continued logarithm, and the remaining finite pieces satisfy
\begin{equation}
 A_p^{\rm B2B}=A_p^{\mathrm{1PN}},\qquad
 B_p^{\rm B2B}=B_p^{\mathrm{1PN}}.
 \label{eq:1pn-polynomial-match}
\end{equation}
With the amplitude and eccentricity conversions identical on
both sides, the complete coefficients agree:
\begin{equation}
 \left.U^{b,(1)}_{22,p}\right|_{\rm B2B}
 =\left.U^{b,(1)}_{22,p}\right|_{\text{Ref.\cite{Morras:2025nbp}}}\,,
  \label{eq:1pn-analytic-match}
\end{equation}
such that the B2B correspondence  recovers the instantaneous bound harmonic through relative 1PN order. See App.~\ref{app:1pn} for more details. 

\section{From PM data to bound waveforms}
\label{sec:PM-bound}
The Newtonian and 1PN examples retain all powers of $G$ at their
respective PN orders. A finite PM calculation instead supplies a limited
number of powers of $G$. As we discuss, the PM-truncated data therefore does not reproduce the bound motion.  
We use our 1PN example to show how additional PN structure can constrain the missing information. Finally, using the results of~\cite{Bini:2024rsy,Bini:2026dvn},
we explore how PM scattering data, combined with PN information
on the frequency dependence, can guide a bootstrap of bound
waveforms while retaining the full velocity dependence from 
the conservative sector. 

\subsection{PM truncation}
\label{sec:PM}
For scattering there are two time scales: $b/p_\infty$, the duration of the encounter,
and $1/n_s \simeq GM/p_\infty^3$, the gravitational time scale. Accordingly, the waveform depends on the variables:
\begin{equation}
 u=\frac{\omega b}{p_\infty},\qquad
 \alpha=u\frac{GM}{bp_\infty^2},\qquad
 \alpha e_s=u\sqrt{1+\left(\frac{GM}{bp_\infty^2}\right)^2}\,,
 \label{eq:PM-N-scales}
\end{equation}
where $b = J/P_\infty$ is the impact parameter. Notice that the product $\alpha e_s$ becomes $u$, and is therefore independent of $G$, at leading PM order.\vskip 4pt  Expanding in $G$ at fixed $J,p_\infty,\omega$ keeps $u$ fixed and expands both the Bessel
order and its argument, as well as the polynomial coefficients.  At fixed $u$, the PM expansion replaces the exact Bessel function
$K_{n-\ii\alpha}(\alpha e_s)$ by its Taylor series: the order is
expanded around $n$, and the argument around $u$. The exponential
$e^{\pi\alpha/2}$ and polynomial coefficients are also expanded. A finite PM truncation retains a finite set of terms
in this series, losing the full dependence on the gravitational
deflection already contained in the exact Newtonian result.\footnote{Similarly, the Newtonian scattering angle includes infinite powers of $1/j$, even though the Newtonian interaction is linear in $G/r$ \cite{Kalin:2019rwq,Kalin:2019inp}.} A direct application of the B2B continuation to this truncation therefore cannot, in general, recover the complete bound harmonic. \vskip 4pt 

To illustrate this point let us consider the $22$ Newtonian mode. The first three terms are (with $p_\infty = \sqrt{2\E}$)
\cite{Bini:2024rsy,Bini:2024ijq}
\begin{align}
 \left.S^{U,N}_{22}\right|_{G+G^2+G^3}
 =-\frac{2\Cnorm\nu GM^2}{p_\infty}
 \Biggl\{&\left(1+\frac{\pi\alpha}{2}\right)
 \bigl[(2u+1)K_0(u)+2(u+1)K_1(u)\bigr]\nonumber\\
 &+\left(\frac{GM}{bp_\infty^2}\right)^2F_N^{(3)}(u)\Biggr\}.
 \label{eq:PM-N-scattering}
\end{align}
The factor of $F_N^{(3)}$ enters at $\Ord(G^3)$. Its explicit expression, including the
Bessel-order derivatives produced by the expansion, is given in
App.~\ref{app:PM}.\vskip 4pt 

 We use the same analytic routes as for the Newtonian calculation before, holding
$\omega$ independent until both continuations are performed. Similarly to our previous analysis, the two routes give
\begin{equation}
\renewcommand{\arraystretch}{1.35}
\begin{array}{c|ccc}
 &p_\infty&u&GM/(bp_\infty^2)\\\hline
 \mathrm{peri}&\ii\sqrt\epsilon&xe^{-\ii\pi}&-\ii/X\\
 \mathrm{apo}&\ii\sqrt\epsilon&x&+\ii/X
\end{array}
\label{eq:PMroute-table}
\end{equation}
where, as before, $X=j\sqrt\epsilon$ after analytic continuation ($\epsilon>0$), and we defined
 $x=\frac{\omega J}{\mu\epsilon}$, such that $\lambda=\frac{\omega}{\Omega_{r,N}}=\frac{x}{X}$\,. 
The first row follows the energy continuation: $u\propto J/p_\infty^2$ makes a clockwise half-turn. The second row also continues the reflected mode in $J$, bringing $u$ back to the positive axis.\footnote{Recall that for the full Newtonian expression, we have  $K_{n-\ii\alpha}(\alpha e_s)\to K_{\lambda+n}(ze^{-\ii\pi})$. The reflected $J$ route then gives $K_{\lambda+n}(z)$, and both the order and the argument are kept unexpanded.} 
 The variable $x$ is related to the exact argument of the Newtonian Bessel function (post energy route) via 
 \begin{equation}
z = \lambda \sqrt{X^2-1} = x\sqrt{1- \frac{1}{X^2}} \, \xrightarrow[X\gg 1] \, x-\frac{x}{2X^2}+\Ord(X^{-4})\,.
 \end{equation}
 We can then use the same Bessel relations, as before, but this time evaluated on $x$:
\begin{equation}
 K_0(xe^{-\ii\pi})=K_0(x)+\ii\pi I_0(x),\qquad
 K_1(xe^{-\ii\pi})=-K_1(x)+\ii\pi I_1(x).
 \label{eq:PM-K01}
\end{equation}
Moreover, the polynomial factors have $u=-x$ on the first route and $u=x$ on the second.\vskip 4pt
For the exact Newtonian result,  $\omega=p\Omega_{r,N}$ gives
\begin{equation}
 \alpha\to \ii p, \quad \lambda\to p,\qquad
 \frac{S_{\rm peri}^{U,N}-(-1)^pS_{\rm apo}^{U,N}}{T_{r,N}}
 =U^{b,N}_{22,p}.
 \label{eq:nonuniform}
\end{equation}
 By contrast, a finite PM expression
contains only finitely many terms in the $\alpha$ expansion. 
At fixed $\epsilon,p$, the harmonic frequency itself becomes
$\omega_p=p\epsilon^{3/2}/(GM)$, with $\alpha=\ii p$, that is no longer
small. For instance, powers of $\alpha$ from the factor of 
$e^{\pi\alpha/2}$ become powers of $\ii p$, and all of them would be needed to
recover the correct $e^{\ii\pi p/2}$ value. Likewise, the derivatives of $K_q$ generated
by the order expansion must recombine into the finite shift from
$n$ to $n+p$. Keeping the full $u$ dependence of a few PM coefficients
does not restore these terms.\vskip 4pt The inability of the B2B correspondence to reconstruct the correct Newtonian waveform from PM-truncated scattering data can be readily seen after performing the analytic continuation. For instance, at $p=2,X=4$, for the $22$ mode we find:
\begin{equation}
\begin{array}{c|c}
 \text{scattering input}&
 (S_{\rm peri}^{U,N}-(-1)^pS_{\rm apo}^{U,N})/(T_{r,N}U^{b,N}_{22,p})\\\hline
 G&-1.56549\\
 G+G^2&-1.56549-4.91812\ii\\
 G+G^2+G^3&6.84777-4.91812\ii\\
 \text{exact Newtonian}&1
\end{array}
\label{eq:PM-N-truncation}
\end{equation}
These truncations do not approximate the exact result at this harmonic. The discrepancy is manifest even before proceeding
to physical ellipses $(0<X<1)$. It already shows why a finite PM truncation does
not reconstruct the bound harmonic. Although the continuation of each
 coefficient is well defined, the missing terms are
not controlled after imposing $x=pX$. This is a limitation of the
PM expansion, not a failure of the B2B map.\vskip 4pt 

There is, however, still useful information in the PM expansion.
Before imposing $\omega=\omega_p$, we can take the large-$X$ limit
at fixed $x$ to extract features of the waveform that may extend
to the bound domain. In this limit, the ratio
$x/X=\omega/\Omega_{r,N}$ becomes small. The bound-harmonic condition
is not imposed at this stage. The fixed-$x$ expansion supplies
coefficients that any proposed completion must reproduce.
This suggests a reconstruction strategy: use these coefficients
to constrain functions of $X$ and frequency that can subsequently
be evaluated at a bound harmonic. A natural approach is therefore to use PN information to supply the class
within which to bootstrap the answer.

\subsection{Bootstrap approach}
\label{sec:PM-1pn-bootstrap}
We now use the PN structure of the waveform to supply the frequency
dependence missing from a finite PM expansion. The 1PN example makes
this strategy explicit: we use the Bessels and logarithmic integrals
identified in \S\ref{sec:1pn} as the frequency functions of the ansatz.
Conservative dynamics supplies the orbital parameters, while
PM data determine the weights of these frequency functions.
We then compare the reconstructed harmonic
with the known bound result.

\subsubsection{Extended B2B relation}
The first step is to specify which continued expression we expand.
At a bound harmonic the B2B subtraction carries $(-1)^p$.
At independent values of $\omega$, the change of time and angle origin from
periastron to apoapsis instead gives, for instance for the $m=2$ mode, 
\begin{equation}
 e^{\tfrac{\ii}{2} (\omega T_r-2\Delta\phi)}
 =e^{\ii\pi[\lambda-2(\Korb-1)]},\qquad
 \lambda=\frac{\omega}{\Omega_r}.
 \label{eq:PM-half-period-factor}
\end{equation}
We therefore introduce the frequency-dependent combination
\begin{equation}
 \left. U^{b}_{\ell m}\right|_{\rm B2B} (\lambda) \equiv \frac{1}{T_r}\left(S_{\rm peri}^{U}(\omega)
 -(-1)^\ell e^{\ii \frac{\Theta_m(\lambda)}{2}}S_{\rm apo}^{U}(\omega)\right)\,, \label{eq:PM-continuous-choice}
\end{equation}
 with
 \begin{equation}
  \Theta_m(\lambda) \equiv \omega T_r-m\Delta\phi = 2\pi[\lambda-m\Korb] \,,
\end{equation}
which agrees with the B2B correspondence on the harmonic condition $\lambda=p+m(\Korb-1)$.
Away from this spectrum, it defines our chosen extension of the B2B map.\footnote{The sign of the phase matters away from the harmonics. For the Newtonian case, we have $e^{\ii\Theta_2/2}=e^{\ii\pi\lambda}$, and the combination in \eqref{eq:PM-continuous-choice}  produces the anticipated factor of $J_{\lambda+n}(\lambda e)$ (see below). Choosing $e^{-\ii\pi\lambda}$ instead would replace each Bessel function by
\begin{equation*}
e^{\ii\pi\lambda}
\left[\cos(\pi\lambda)J_{\lambda+n}(\lambda e)
-\sin(\pi\lambda)Y_{\lambda+n}(\lambda e)\right].
\end{equation*}
Although both choices agree at $\lambda=p\in\mathbb Z$, for fixed $0<e<1$ and large positive $\lambda$, the first choice decays while the second grows exponentially away from the harmonic condition. Requiring a bounded extension at positive real frequency
 selects the choice in \eqref{eq:PM-continuous-choice}.}\vskip 4pt 

The exact Newtonian result now helps us identify what is retained through $G^3$. Before selecting a harmonic, we have (with $\Korb_N=1$)
\begin{equation}
\frac{1}
 {\Cnorm\mu\epsilon}  \left. U^{b,N}_{22}\right|_{\rm B2B} (\lambda)
 =\frac{\lambda^2}{2}
 \sum_{n=-3}^{3}c_n(e,X)J_{\lambda+n}(\lambda e).
 \label{eq:PM-HN-definition}
\end{equation}
At $\lambda=p$, this is the function in \eqref{eq:HN}. We keep the frequency independent and expand at fixed $x=\lambda X$. For the waveform the large-$X$ expansion we find
\begin{equation}
 - \frac{1}{\Cnorm\mu\epsilon}\left.U^{b,N}_{22}\right|_{\rm B2B} = -\frac{x^2}{2X^2}
 \sum_{n=-3}^{3}c_n(e,X)
 J_{n+\tfrac{x}{X}}\!\left(\frac{xe}{X}\right)
 =h_0(x)+\frac{h_1(x)}{X}+\frac{h_2(x)}{X^2}
 +\Ord(X^{-3}).
 \label{eq:PM-newton-main}
\end{equation}
The coefficients are
\begin{equation}
\begin{aligned}
 h_0(x)={}&(1-2x)I_0(x)+2(1-x)I_1(x),\\
 h_1(x)={}&x\bigl[(2x-1)K_0(x)+2(1-x)K_1(x)\bigr]
 +\frac{\ii\pi x}{2}h_0(x),\\
 h_2(x)={}&\frac{x^2}{2}
 \left[
 (1-2x)\left.\partial_q^2 I_q(x)\right|_{q=0}
 +2(1-x)\left.\partial_q^2 I_q(x)\right|_{q=1}
 \right]\\
 &-2x(1-x)K_0(x)
 +(x^2-3x+2)I_0(x)
 +\left(x^2-\frac52x+2\right)I_1(x)\\
 &+\frac{\ii\pi x}{2}h_1(x)
 +\frac{\pi^2x^2}{8}h_0(x).
\end{aligned}
\label{eq:hs}
\end{equation}
The derivatives act on the Bessel order holding $x$ fixed. In particular,
$\left.\partial_q I_q(x)\right|_{q=0}=-K_0(x)$, which explains
why $K$ functions appear in these expansion coefficients even
after the common $K$ contribution in the two routes cancelled.\vskip 4pt 

The same three functions can then be obtained from the scattering
expansion in \eqref{eq:PM-N-scattering}. After applying the B2B map at leading order in $G$, the two Bessel continuations
in \eqref{eq:PM-K01} give
\begin{equation}
\begin{aligned}
 S_{{\rm apo},G}^{U,N}
 &=-\frac{2\ii\Cnorm\mu GM}{\sqrt\epsilon}
 \bigl[(2x-1)K_0(x)+2(1-x)K_1(x)\bigr],\\
 S_{{\rm peri},G}^{U,N}
 &=S_{{\rm apo},G}^{U,N}-\Cnorm\mu\epsilon T_{r,N}h_0(x).
 \label{eq:PM-N-tree-matching}
\end{aligned}
\end{equation}
 At $\Ord(G^2)$, the factor of $\pi\alpha/2$ in the scattering expression
becomes $\ii\pi x/(2X)$ on both routes. The phase, however,
multiplies only the apoapsis route and expands as
\begin{equation} e^{\ii\pi x/X}=1+\ii\pi x/X-\pi^2x^2/(2X^2)+\cdots\,.\end{equation}
Keeping both contributions yields
\begin{equation}
 -\frac{S_{{\rm peri},G^2}^{U,N}-S_{{\rm apo},G^2}^{U,N}
 -\dfrac{\ii\pi x}{X}S_{\rm apo,G}^{U,N}}
 {\Cnorm\mu\epsilon T_{r,N}}
 =\frac{h_1(x)}{X}.
 \label{eq:PM-N-loop-matching}
\end{equation}
Notice that the expansion of the clock phase is crucial to recover the correct result.\vskip 4pt 

At $\Ord(G^3)$, we also continue the Bessel-order derivatives in
$F_N^{(3)}$. Combining this term with the clock corrections multiplying the
loop and tree gives (see App.~\ref{app:PM} for more details)
\begin{equation}
 -\frac{S_{{\rm peri},G^3}^{U,N}-S_{{\rm apo},G^3}^{U,N}
 -\dfrac{\ii\pi x}{X}S_{{\rm apo},G^2}^{U,N}
 +\dfrac{\pi^2x^2}{2X^2}S_{{\rm apo},G}^{U,N}}
 {\Cnorm\mu\epsilon T_{r,N}}
 =\frac{h_2(x)}{X^2}.
 \label{eq:PM-N-third-matching}
\end{equation}

At Newtonian order, these manipulations show that the two-route analytic continuation
of the first three PM terms determines the three coefficients in the
large-$X$ expansion of the waveform combination defined by the extended
B2B map in \eqref{eq:PM-continuous-choice}, keeping the frequency as an
independent parameter---therefore without yet selecting a bound harmonic.\vskip 4pt At these orders and fixed $x$, the analytic continuation and the PM expansion therefore commute through the extended B2B correspondence. Expanding the frequency-dependent phase plays an essential role. The comparison holds at fixed $x$, where the neglected terms are
$\Ord(X^{-3})$. Selecting a bound harmonic instead requires $x=pX$,
so powers of $x/X$ are no longer suppressed. This again illustrates why all orders in $\alpha$ are needed to recover
the exact bound harmonic. The coefficients obtained above nevertheless
constrain the frequency dependence to be reconstructed. The harmonic
condition can then be imposed once that dependence has been completed,
or a controlled approximation has been established.

\subsubsection{The 1PN reconstruction}
\label{sec:PM-reconstruction}
We now use the PM coefficients to determine the weights of a
bound-frequency ansatz at 1PN. The PN calculation supplies the
allowed Bessel and angular functions, while conservative dynamics
fixes the changes in the clock, eccentricities and normalization.
We leave the remaining polynomial weights entirely free. Matching the
large-$X$ expansion of this ansatz to the continued scattering
coefficients will determine their values, such that the resulting 
functions can then be evaluated at a bound harmonic.

 We separate
the two mass-ratio contributions as
\begin{equation}
 U^b_{22,p}=-\Cnorm M\nu\epsilon
 \left[H_N(e,p)+\epsilon\bigl(H_1^{(0)}(e,p)
                    +\nu H_1^{(1)}(e,p)\bigr)+\Ord(\epsilon^2)\right],
 \label{eq:PM-reduced-normalization}
\end{equation}
where
\begin{equation}
 H_N(e,p)=-\frac{U^{b,N}_{22,p}}{\Cnorm M\nu\epsilon}
 =-\frac{p^2}{2}\sum_{n=-3}^{3}c_n(e,X)J_{p+n}(pe)
 \label{eq:PM-HN-normalization}
\end{equation}
is the (normalized) Newtonian result already established in \eqref{eq:HN}. At fixed values of the eccentricity, the
$H_1^{(0)}$ and $H_1^{(1)}$ contribute at overall orders $\nu$ and $\nu^2$. Since the angular correction in the scattering waveform is proportional to $k_1=3/X^2$, which is independent of $\nu$, determining 
$H_1^{(1)}$ requires only the finite Bessel sum. The probe coefficient $H_1^{(0)}$ also contains the
logarithmic frequency function. Its shape is supplied, while PM
matching determines the polynomial correction and the overall
angular normalization. We give the $H_1^{(1)}$ reconstruction below and
the probe matching in App.~\ref{app:PM}.\vskip 4pt

\paragraph{PN constraints.}
The ansatz uses the ordinary Bessel functions identified in
\eqref{eq:1pn-sc-spectrum}. Separating the supplied amplitude
and eccentricity shifts, we write the frequency function on the
continued side of \eqref{eq:PM-continuous-choice}, with
$\lambda$ still independent of the harmonic condition:
\begin{equation}
 \begin{aligned}
 H_1^{(1)}(e,\lambda)={}&\frac{H_N(e,\lambda)}{12}
 +\frac{2-7X^2/4}{2e}\,\partial_eH_N(e,\lambda)\\
 &+\frac{A^{(1)}(e,\lambda)J_\lambda(\lambda e)
       +B^{(1)}(e,\lambda)eJ'_\lambda(\lambda e)}{X^2e^2}.
 \end{aligned}
 \label{eq:nu-regular-completion}
\end{equation}
Both fixed terms follow by keeping the Newtonian waveform and expanding its orbital variables through 1PN.
The first comes from the overall amplitude factor $(GM\Omega_\phi)^{2/3}/\epsilon$, see \eqref{eq:Omegaphi}.
At fixed binding energy, the fractional shift in $\Omega_\phi$ proportional to $\nu$ is $\epsilon\nu/8$,
so the $2/3$ power gives $\epsilon\nu/12$ multiplying $H_N$.
The second comes from replacing $e$ by $e_t=e+\epsilon\,\delta e_t$ inside $H_N$.
By \eqref{eq:et-main}, the coefficient of $\nu$ in $\delta e_t$ is $(2-7X^2/4)/(2e)$.
It multiplies $\partial_e H_N$, taken at fixed $\lambda$ with $X^2=1-e^2$.\vskip 4pt

The remaining information is in $A^{(1)}$ and $B^{(1)}$.
The seven allowed Bessel shifts reduce to $J_\lambda,J'_\lambda$
using \eqref{eq:recurrence}. We supply the resulting polynomial
degrees and parity as PN input, with ten unknown constants
$a_i,b_i$, $i=0,\ldots,4$:
\begin{align}
 A^{(1)}&=X^2(a_0+a_1e^2)
       +\lambda X^3(a_2+a_3e^2)+\lambda^2X^6a_4,
       \label{eq:nu-class}\\
 B^{(1)}&=X(b_0+b_1e^2)
       +\lambda X^2(b_2+b_3e^2)+\lambda^2X^5b_4.\nonumber
\end{align}
 The PM waveform will determine those constants by
requiring the large-$X$ expansions to agree at independent
values of $x$.\vskip 4pt 

\paragraph{The scattering information.}
For the ${\cal O}(\nu^2)$ contribution, the two modes through
$G^3$ can be written together as
\cite{Bini:2024rsy,Bini:2024ijq}
\begin{equation}
 \begin{aligned}
 \left.S^U_{2,2\varsigma}\right|_{1\rm PN,\nu^2}
 =\nu^2\Cnorm GM^2p_\infty\Biggl[
 &-\frac47P_1^\varsigma(u)
 -\frac{\pi}{14}\frac{GM}{bp_\infty^2}uP_2^\varsigma(u)\\
 &+\frac1{14}\left(\frac{GM}{bp_\infty^2}\right)^2P_3^\varsigma(u)
 +\Ord\!\left(\left(\frac{GM}{bp_\infty^2}\right)^3\right)\Biggr].
 \end{aligned}
 \label{eq:nu-scattering}
\end{equation}
Here $\varsigma=\pm 1$ labels $m=\pm 2$. The functions $P_i^\varsigma$
include the associated Bessel functions, e.g.,
\begin{align}
 P_1^\varsigma(u)&=(6u^2+3\varsigma u+2)K_0(u)
             +[\varsigma(6u^2+4)+6u]K_1(u),\nonumber\\
 P_2^\varsigma(u)&=(24u^2+26\varsigma u+15)K_0(u)
             +[\varsigma(24u^2+30)+38u]K_1(u).
 \label{eq:nu-tree-loop}
\end{align}
The expression for $P_3^\varsigma$ is given in App.~\ref{app:PM}.\vskip 4pt

\paragraph{The two-route extended map.}
Since \eqref{eq:nu-scattering} is already at 1PN order, the analytic continuation is similar to the Newtonian case (see  \eqref{eq:PMroute-table}): the energy route acts on the $m=2$ expression with $u=xe^{-\ii\pi}$,
while the reflected $J$ route acts on $m=-2$ with $u=x$.
The factors $GM/(bp_\infty^2)$ become $-\ii/X$ and $+\ii/X$,
respectively. Furthermore, the polynomial factors are evaluated on $u=-x$ for the energy route and $u=x$ for the reflected mode. For example, \eqref{eq:PM-K01} gives
\begin{equation}
 P_1^+(xe^{-\ii\pi})=P_1^-(x) + \ii\pi\bigl[(6x^2-3x+2)I_0(x)
                    +(6x^2-6x+4)I_1(x)\bigr].
 \label{eq:PM-tree-discontinuity}
\end{equation}
The common $K$ functions cancel in the tree contribution.
The same continuation applies to $P_2^\varsigma$ and
$P_3^\varsigma$, including the products generated by the
relative phase in \eqref{eq:PM-loop-phase} at each PM
order.\vskip 4pt

There are two further contributions at 1PN. Firstly, the
Newtonian scattering waveform changes when expressed in the
bound-endpoint variables. Using \eqref{eq:sE} and
$b=\Gamma J/(\mu p_\infty)$ gives, on each route, 
\begin{equation}
 u\ \to\
 x\left[1+\frac{1-\nu}{4}\epsilon\right]e^{-\ii\pi}
 \quad\hbox{or}\quad u \to 
 x\left[1+\frac{1-\nu}{4}\epsilon\right].
 \label{eq:PM-shifted-argument}
\end{equation}
Secondly, the period and clock in \eqref{eq:PM-continuous-choice}
must be expanded consistently:
\begin{equation}
 \begin{gathered}
 \lambda=\frac{x}{X}(1-\epsilon n_1)+\Ord(\epsilon^2),\qquad
 \frac1{T_r}=\frac{1+\epsilon n_1+\Ord(\epsilon^2)}{T_{r,N}},\\
 e^{\ii\pi[\lambda-2(\Korb-1)]}
 =e^{\ii\pi x/X}
 \left[1-\ii\pi\epsilon\left(n_1\frac{x}{X}+2k_1\right)\right]
 +\Ord(\epsilon^2).
 \end{gathered}
 \label{eq:PM-loop-phase}
\end{equation}
The explicit 1PN scattering terms, the induced Newtonian
changes and the clock products must be combined before
extracting a PM coefficient. We collect in App.~\ref{app:PM}
the full expressions that result from implementing these operations.

\paragraph{Matching the PM data.}
The continued PM waveform and the ansatz are now expressed in
the same variables. To compare them, we expand the ansatz at fixed
$x$, leaving all ten weights unknown. Matching the
leading coefficient to the continued tree gives 
\begin{equation}
 \begin{aligned}
 a_1+xa_3-x^2a_4=\frac{12x^2+22x-17}{42},\quad
 b_1+xb_3-x^2b_4=\frac{6x^2+8x+25}{21}.
 \end{aligned}
 \label{eq:PM-tree-weights}
\end{equation}
where we have subtracted on both sides the induced shift on the parameters of the Newtonian waveforms.\vskip 4pt

The next coefficient tests the frequency dependence retained in
the ansatz. The tree fixes six weights, but these multiply
Bessels whose order is still $\lambda=x/X$. Expanding their
order and accompanying phase produces a definite $1/X$ term.
For example, 
\begin{equation}
 J_{x/X}\!\left(\frac{xe}{X}\right)
 =I_0(x)+\frac{x}{X}
       \left[\frac{\ii\pi}{2}I_0(x)-K_0(x)\right]
       +\Ord(X^{-2})\,,
 \label{eq:PM-loop-from-order}
\end{equation}
where $e=\ii\sqrt{X^2-1}$ in the reference sheet with $X>1$ (prior to analytic continuation to $0<X<1$).\vskip 4pt 

The remaining coefficients, $a_0,a_2,b_0,b_2$, first appear at
$1/X^2$ in the fixed-$x$ expansion.\footnote{The $1/X$ term is therefore
already determined by the coefficients fixed from the tree
waveform. We can test it directly against the $G^2$
results, and we find that the two expressions agree without extra adjustments.} To determine these four parameters, we therefore need the
$1/X^2$ contribution and the scattering waveform at $G^3$. We expand the ansatz one order further
and compare it with the continued $G^3$ scattering contribution. Both sides include the Bessel corrections
generated by the tree coefficients and the known orbital contributions. After accounting for all these terms, matching the
coefficients of $I_0(x)$ and $I_1(x)$ yields
\begin{equation}
 -a_0-xa_2-x^2a_4=-b_0-xb_2-x^2b_4
 =\frac{6x^2-73x+67}{21}.
 \label{eq:nu-last}
\end{equation}
The constant term fixes $a_0=b_0=-67/21$, and the term linear in $x$ fixes
$a_2=b_2=73/21$. The quadratic term gives
$a_4=b_4=-2/7$, in agreement with the values already obtained
from the waveform at $\Ord(G^1)$.

Hence, the tree fixes six coefficients, the loop tests the
frequency dependence they predict, and $G^3$ determines the
four remaining coefficients while providing a further check.
All ten coefficients are fixed within the PN-informed
function space.\vskip 4pt

Once the bootstrapped waveform is self-consistently determined for generic values of $\omega$, the remaining step is to select the bound harmonic: 
\begin{equation}
 \lambda=p+2\epsilon k_1+\Ord(\epsilon^2),\qquad
 x=pX+\epsilon(n_1pX+6/X)+\Ord(\epsilon^2),
 \label{eq:PMfreqshift}
\end{equation}
The reconstructed ${\cal O}(\nu^2)$ contribution becomes
\begin{equation}
U^{b}_{22, {\rm 1PN,\nu^2}} = -\Cnorm M\nu^2\epsilon^2 H_1^{(1)}(e,\lambda=p).
 \label{eq:nu-ansatz}
\end{equation}
The reconstructed polynomials agree coefficient by coefficient
with the contributions to the bound result quoted in App.~\ref{app:1pn}. Together with the supplied orbital corrections,
they give the complete $H_1^{(1)}$ contribution at 1PN order. We discuss the probe term in 
App.~\ref{app:PM}. Evaluating the total expression
at the bound harmonic reproduces the result of
\cite{Morras:2025nbp}.\vskip 4pt

The above manipulations illustrate what a bootstrap  approach can add to the B2B map within a truncated PM regime. 
Analytic continuation determines the large-$X$ information in the
PM data. The specified PN structure relates this information to a
function that can then be evaluated on a bound harmonic. 

\subsection{Relativistic clock and angle}
\label{sec:all-velocity}
The preceding construction suggests how relativistic scattering might
extend the bound waveform beyond the supplied PN orders. In the
conservative B2B dictionary, the coefficient of powers of $1/j$
retains full energy dependence. For waveforms, conservative
information instead enters the clock and angle of the frequency functions in the PN calculation. The conservative factors can therefore be kept unexpanded at a given PM order.\vskip 4pt

\subsubsection{The clock in the loop} The one-loop waveform contains a term proportional to the tree
\cite{Heissenberg:2025fcr}:
\begin{equation}
 S^U_{1\text{-loop}}
 =\frac{\pi \gamma(3-2\gamma^2)}{2(\gamma^2-1)^{3/2}} GE\omega\,
       S^U_{\rm tree}+S^{U,\rm gen}_{1\text{-loop}}\,\,, \label{eq:allv-loop-split}
\end{equation}
with $S^{U,\rm gen}_{1\text{-loop}}$ denoting what remains after this part is subtracted.\vskip 4pt 

We can understand the origin of this term as follows. Using the same variables as before, $\alpha=\omega/n_s$ and $z_t=\alpha e_t$, they enter in the orbital Fourier phase through the relation
\begin{equation}
 \omega t(v)=z_t\sinh v-\alpha v.
 \label{eq:allv-old-clock}
\end{equation}
 These are the quantities that became the Bessel order and argument in
\eqref{eq:sc-Bmaster}. The motion at 1PM order admits the same hyperbolic
anomaly, $v$, and the same form as \eqref{eq:allv-old-clock}, but with coefficients that depend on $\gamma$, resuming  velocity dependence \cite{Kalin:2019rwq,Kalin:2019inp}. We denote these parameters as: $\alpha_{\mathrm{1PM}}$ and $z_{t,\mathrm{1PM}}$. The first is fixed by the scattering time:\footnote{Since $v$ grows as $\log r$ at large separations, the term $-\alpha_{\mathrm{1PM}}v/\omega$ in \eqref{eq:allv-old-clock} is related to the time delay. At Newtonian order, $\alpha_{\mathrm{1PM}}\to {GM\omega\over p_\infty^3} \to {\omega \over n_{s,N}}$, recovering \eqref{eq:Newton-sc-vars}. See App.~\ref{app:PM} for more details.}
\begin{equation}
 \alpha_{\mathrm{1PM}}
 =\frac{\gamma(3-2\gamma^2)}{(\gamma^2-1)^{3/2}} GE\omega\,.
 \label{eq:allv-delay-coefficient}
\end{equation}
We can therefore immediately identify the factor in \eqref{eq:allv-loop-split} with $\pi\alpha_{\rm 1PM}/2$. The coefficient of $\sinh v$ also comes from the same conservative solution. To make the PM counting explicit, we expand in $G$ at fixed $b$, energy and frequency. In terms of
$u=\omega b/p_\infty$, the 1PM conservative value gives
\begin{equation}
 z_{t,\mathrm{1PM}}
 =u\left(\gamma-\frac{\nu(\gamma^2-1)}{\Gamma^2}\right)
   +\Ord(G^2).
 \label{eq:allv-flight-time-main}
\end{equation}
Notice that $z_t$ starts at order $G^0$, whereas $\alpha_{\mathrm{1PM}}$ starts at order $G$. For notational convenience, we keep  the label ``1PM''  to denote
the conservative dynamics used to construct the solution, rather
than the powers of $G$ inside the variable.\footnote{The same distinction is already present at Newtonian order.
Although $\alpha=GM\omega/p_\infty^3$ is proportional to $G$,
the scattering eccentricity grows as $1/G$ at fixed $b$.
Their product is
\begin{equation}
 z_{t,N}=\alpha e_s
 =u\sqrt{1+\left(\frac{GM}{bp_\infty^2}\right)^2}
 =u+\Ord(G^2).
\end{equation}
The leading Bessel argument therefore remains finite as
$G\to0$.}
See App.~\ref{app:PM} for more details.\vskip 4pt  

Following the structure of the 1PN waveform in \eqref{eq:1pn-sc-spectrum}, we can now replace these PM coefficients into the frequency-dependent part. Replacing the clock gives
\begin{align}
\frac12\int \dd v\,e^{nv}e^{\ii\omega t(v)} =\frac12\int \dd v\,
 e^{(n-\ii\alpha_{\mathrm{1PM}})v
      +\ii z_{t,\mathrm{1PM}}\sinh v}\,,
 \label{eq:allv-reference-integral}
\end{align}
which evaluates to the same Bessel function in \eqref{eq:sc-Bmaster}, now dressed by PM data:
\begin{equation}
 \mathcal B_n(\alpha_{\mathrm{1PM}},z_{t,\mathrm{1PM}})
 =\ii^n e^{\pi\alpha_{\mathrm{1PM}}/2}
   K_{n-\ii\alpha_{\mathrm{1PM}}}(z_{t,\mathrm{1PM}}).
 \label{eq:allv-reference-Bessel}
\end{equation}
In this fashion, the PN expression supplies the frequency structure, and the conservative
clock fixes the relativistic parameters to a given PM order.\vskip 4pt 

The term proportional to the tree in~\eqref{eq:allv-loop-split}
now appears as the linear term in the expansion
\begin{equation}
 e^{\pi\alpha_{\rm 1PM}/2}
 =1+\frac{\pi\alpha_{\rm 1PM}}{2}+\Ord(G^2)
 =1+\frac{\pi\gamma(3-2\gamma^2)}{2p_\infty^3}GE\omega
 +\Ord(G^2).
 \label{eq:allv-clock-tree}
\end{equation}
Hence, the conservative clock accounts for this part of the
one-loop waveform, carrying its full velocity dependence. Moreover, since the exponential prefactor and the shifted order in
\eqref{eq:allv-reference-Bessel} arise together from
$e^{-\ii\alpha_{\rm 1PM}v}$ in the Fourier integral, we can now expand the clock's influence beyond one-loop order. Let us consider, for instance, the probe limit.
At leading order in $G$, with energy and impact parameter
fixed, \eqref{eq:allv-flight-time-main} gives $z_t=\gamma u$, such that
the leading Fourier phase is given by $\omega \uret = u\gamma\sinh v = \left(\frac{\omega b}{p_\infty}\right) \gamma\sinh v$. Therefore,
\begin{equation}
 v=\mathrm{arcsinh}\frac{\tau}{\gamma}
 =\log\left(\frac{\tau+\sqrt{\tau^2+\gamma^2}}{\gamma}\right),
 \qquad \tau=\frac{p_\infty\uret}{b}.
 \label{eq:allv-encounter-anomaly}
\end{equation}
The clock factor $e^{-\ii\alpha_{\rm 1PM}v}$ is expanded
inside the integral in \eqref{eq:allv-reference-integral}.
Since $\alpha_{\rm 1PM}$ is proportional to $G$, its successive
terms insert also powers of $v$ at successive PM orders.
Writing the integral in terms of $\tau$ turns these powers of the anomaly into powers of
the logarithm in \eqref{eq:allv-encounter-anomaly}. Since $\partial_\tau e^{\ii u\tau}=\ii u e^{\ii u\tau}$,
integration by parts moves the frequency factors generated
by the clock onto the remaining integrand. For each Bessel, the term with $k$ insertions becomes
\begin{equation}
 \begin{aligned}
 &\frac{(-\ii\alpha_{\rm 1PM})^k}{k!}
 \int\dd\tau\,e^{\ii u\tau}
 \frac{v^k e^{nv}}{\sqrt{\tau^2+\gamma^2}}=
 \frac{1}{k!}\left(\frac{\alpha_{\rm 1PM}}{u}\right)^k
 \int\dd\tau\,e^{\ii u\tau}
 \frac{\dd^k}{\dd\tau^k}
 \left[\frac{v^k e^{nv}}{\sqrt{\tau^2+\gamma^2}}\right],
 \end{aligned}
 \label{eq:allv-clock-insertions}
\end{equation}
where $v$ is the function of $\tau$ in
\eqref{eq:allv-encounter-anomaly}, and the derivatives act
on the entire bracket. Derivatives acting on $v^k$ lower the
logarithmic power, while those acting on the other factors
leave that power unchanged. Expanding through the stated PN order then puts these terms
in the Fourier-integral families used
in~\cite{Geralico:2026kbm,Geralico:2026efi}, whose coefficients
can be compared after the same integral reductions. In particular, we can compare the powers of the logarithm in~\eqref{eq:allv-encounter-anomaly}. At $G^5$, four clock insertions acting on the tree fix the
coefficient of the fourth power of the logarithm.
The third power also receives three insertions acting on
the loop (after subtracting its tree-generated clock
contribution to avoid double counting). The lower powers similarly involve the remaining scattering
coefficients at $G^3$ and $G^4$. Therefore, parts of the higher-PM
Fourier integrand can be determined through relativistic clock insertions. Keeping the full $\gamma$ dependence throughout the process, but expanding through relative 2PN order for the comparison, we find perfect agreement with the scattering results in \cite{Geralico:2026kbm,Geralico:2026efi} for the coefficients of the highest logarithmic power through $G^5$, and of the next power through $G^3$--$G^5$, as well as of all positive powers at $G^5$. See App.~\ref{app:PM} for more details.\vskip 4pt

\subsubsection{The angular term}\label{sec:angterm}

The clock accounts for $e^{\ii\omega t}$ in the Fourier
integral. Each $m=2\varsigma$ mode, with $\varsigma=\pm1$,
also carries an angular phase, $e^{-2\ii\varsigma\phi}$. At 1PN, expanding
this factor gave the logarithmic term in
\eqref{eq:1pn-sc-spectrum}. To keep its frequency dependence,
we proceed similarly to the clock. The scattering angular anomaly
is described by~\cite{Cho:2018upo}
\begin{equation}
 R(v)=\frac{e_\phi-\cosh v-\sqrt{1-e_\phi^2}\sinh v}
             {e_\phi\cosh v-1},\qquad \log R(0)=0\,,
 \label{eq:allv-R-main}
\end{equation}
which in the Newtonian limit becomes the function $R_0=e^{\ii\phi_N}$ used in \S\ref{sec:1pn}. Through 1PN, the angle becomes $\phi=-\ii\Korb_s\log R$, such that
the angular factor separates into two terms, $R^{-2\varsigma}$ and $e^{-2\varsigma(\Korb_s-1)\log R}$. The latter can be expanded as (for the $\varsigma=1$ mode)
\begin{equation}
 e^{-2(\Korb_s-1)\log R}
 =1-2(\Korb_s-1)\log R
   +2(\Korb_s-1)^2(\log R)^2+\cdots.
 \label{eq:allv-periastron-expansion}
\end{equation}
 The coefficient has the conservative PM expansion
\cite{Kalin:2019inp}
\begin{equation}
 \Korb_s-1=\frac{3(5\gamma^2-1)}{4\Gamma j^2}+\Ord(j^{-4}),
 \qquad \Gamma=\sqrt{1+2\nu(\gamma-1)}.
 \label{eq:allv-periastron-full}
\end{equation}
Hence, by keeping the full $\gamma$ dependence, the structure of the 1PN waveform already carries the radiative input for 
higher-order corrections.\vskip 4pt

We can now perform the same type of comparison as before. We treat products with the clock in the same fashion as in \eqref{eq:allv-clock-insertions}. In the probe limit, at the Newtonian accuracy needed for the comparison, the leading PM encounter gives
$\phi_N=\arctan\tau$. The square term therefore produces
$(\arctan\tau)^2$ inside the Fourier integral. Expanding through $G^5$ at relative 2PN, the coefficient of $(\arctan\tau)^2$ agrees with the instantaneous probe result~\cite{Geralico:2026kbm,Geralico:2026efi}.
See App.~\ref{app:PM}  for more details.

\subsubsection{The bound contributions}
\label{sec:allv-bound}
The same two continuations carry the clock and precession
contributions into the bound frequency integrals. We first
follow $p_\infty\to\ii s$ at fixed $J,\omega$, carrying
the branches of $\gamma$ and $\Gamma$ along the energy
route. The clock coefficient becomes
\begin{equation}
 \alpha_{\rm 1PM}\longrightarrow
 \ii\omega\frac{T_{r,\rm 1PM}}{2\pi},\qquad
 \frac{T_{r,\rm 1PM}}{2\pi}
 =\frac{GE\gamma(3-2\gamma^2)}{(1-\gamma^2)^{3/2}}.
 \label{eq:allv-clock}
\end{equation}
Continuing \eqref{eq:allv-periastron-full} similarly gives
the $j^{-2}$ contribution to $\Korb-1$ at bound energy.
Both conservative coefficients retain their full $\gamma$
dependence. Their effect on the waveform follows through
the frequency functions, as in the 1PN calculation. 

Using $\lambda=\omega T_{r,\rm 1PM}/(2\pi)$, the
Bessel order becomes $\lambda+n$ (with the
harmonic frequency evaluated using the same period\footnote{The additional 2PN
clock contribution is separated in the comparison with
the full bound waveform.}). The two argument paths are those of \eqref{eq:1pn-zt-routes}: the energy route reaches
$z_t e^{-\ii\pi}$, and the reflected $J$ route returns to
$z_t$, and once again we can use the identities in \eqref{eq:1pn-B-connection} to relate the scattering and bound frequency functions. The angular factor follows these same paths. In the
reflected integral, \eqref{eq:1pn-logshift} sends
$\log R$ to $\ii\pi-\log R$, so the precession factor becomes
\begin{equation}
 e^{2(\Korb-1)[\ii\pi-\log R(v)]}
 =e^{2\pi\ii(\Korb-1)}e^{-2(\Korb-1)\log R(v)}.
 \label{eq:allv-angle-continuation}
\end{equation}
The constant factor $e^{2\pi\ii(\Korb-1)}$ combines with
the relative phase $e^{\ii\pi[\lambda-2(\Korb-1)]}$ in
\eqref{eq:PM-continuous-choice} to give $e^{\ii\pi\lambda}$.
This is the phase accompanying the Bessel subtraction.
Using $T_{r,\rm 1PM}$ also in the harmonic condition,
$\lambda-2(\Korb-1)=p$, the Fourier and angular phase
closes over a radial cycle, giving the cancellation
established in \S\ref{sec:1pn}. In the ordinary Bessel part, the continued clock prefactor
converts the surviving $I$ function according to
$\ii^{\lambda+n}I_{\lambda+n}(z_t)=J_{\lambda+n}(\ii z_t)$.
The identities in \eqref{eq:recurrence} reduce the finite
integer shifts to $J_\lambda,J'_\lambda$. The terms containing $\log R$ require the
angular reduction developed in \S\ref{sec:1pn}, which
also contributes to the Bessel coefficients.\vskip 4pt

To compare the resulting angular terms with the bound
waveform, we express them in the time eccentricity $e_t$
used in \eqref{eq:publishedH1}. The clock already uses
$e_t$, whereas $R$ contains $e_\phi$. Using
$e_\phi=e+\epsilon\delta e_\phi$ and
$e_t=e+\epsilon\delta e_t$, we expand about $R_0$ in \eqref{eq:sc-Amaster},
\begin{equation}
 \begin{aligned}
 \log R(v)
% &=\log R_0(v;e) +\epsilon\delta e_\phi\,\partial_e\log R_0(v;e)   +\Ord(\epsilon^2)\\
 &=\left[
   \log R_0(v;e)
   +\epsilon(\delta e_\phi-\delta e_t)
      \partial_e\log R_0(v;e)
   \right]_{e=e_t}
   +\Ord(\epsilon^2).
 \end{aligned}
 \label{eq:allv-bound-angle-expansion}
\end{equation}
The eccentricity correction first contributes at 2PN
when multiplied by precession. Its derivative of $\log R_0$,
together with the Newtonian polynomial, gives ordinary
Bessel terms. The terms that retain an angular logarithm
therefore follow from $R_0(v;e_t)$. The harmonic condition also shifts the frequency in the
clock factor $e^{\lambda(v-e_t\sinh v)}$. Combining it
with precession gives
\begin{equation}
 \begin{aligned}
 e^{\lambda(v-e_t\sinh v)}
 e^{-2(\Korb-1)\log R_0(v;e_t)}
 &=e^{p(v-e_t\sinh v)}
   e^{2(\Korb-1)\mathcal L(w)},\\
 \mathcal L(w)
 &\equiv v-e_t\sinh v-\log R_0(v;e_t).
 \end{aligned}
 \label{eq:allv-bound-periodic-angle}
\end{equation}
The frequency shift supplies $v-e_t\sinh v$, while
precession subtracts $\log R_0$. Over one radial cycle,
$v$ and $\log R_0$ each increase by $2\pi\ii$, whereas
$\sinh v$ returns to its initial value. The
$\mathcal L$ function is therefore periodic.\vskip 4pt

The same clock and angle enter the overall $\omega^2$
in \eqref{eq:1pn-sc-spectrum}. Using
$\omega_p=[p+2(\Korb-1)]\Omega_r$, and keeping
$e^{p(v-e_t\sinh v)}$ as the common Fourier factor, the Newtonian part transforms as
\begin{equation}
 \sum_{n=-3}^{3}c_n w^n\ \longrightarrow\
 \left(\frac{T_{r,N}}{T_{r,\rm 1PM}}\right)^2
 \left[1+\frac{2(\Korb-1)}p\right]^2
 e^{2(\Korb-1)\mathcal L(w)}
 \sum_{n=-3}^{3}c_n w^n\,,
 \label{eq:allv-bound-joint-factor}
\end{equation}
The clock and periastron coefficients keep their full
energy dependence here, while the polynomial and eccentricity
corrections remain at their known PN orders. At 2PN, leading
precession also multiplies the 1PN polynomial. Its coefficients
are already fixed by the bootstrap, so these angular
contributions require no new waveform weights.\vskip 4pt

For the comparison, we now expand the conservative factors
through 2PN.
The clock gives $T_{r,N}/T_{r,\rm 1PM}=1+\epsilon n_1+\cdots$, with
$n_1=(\nu-15)/8$. Since this ratio is squared in
\eqref{eq:allv-bound-joint-factor}, its first correction
is $2\epsilon n_1$. Likewise, the angular exponential
starts with $2\epsilon k_1\mathcal L$, where
$k_1=3/X_t^2$ and $X_t=\sqrt{1-e_t^2}$. Their product
therefore contributes $4\epsilon^2n_1k_1\mathcal L$. The exponential also supplies
$2\epsilon^2k_1^2\mathcal L^2$ from the square of the leading
precession. Expanding the energy dependence in
\eqref{eq:allv-periastron-full} gives a further linear
term, $2\epsilon^2k_{2,\gamma}\mathcal L$, with
$k_{2,\gamma}=3(2\nu-5)/(4X_t^2)$. The mixed clock-precession term and the precession square
multiply the same Newtonian polynomial in
\eqref{eq:allv-bound-joint-factor}. Reducing this polynomial
by integration by parts in the radial-cycle integral gives
the following contributions proportional to $\mathcal L$
and $\mathcal L^2$, respectively:
\begin{align}
 \left.\delta U^b_{22,p}\right|_{\mathrm{clock}\times\mathrm{prec},{\cal L}}
 ={}&-\frac{4\Cnorm\mu\epsilon^3 n_1k_1}{2\pi\ii}
 \oint_{|w|=1}\frac{\dd w}{w}\,
 w^p e^{-pe_t(w-w^{-1})/2}\nonumber\\
 &\quad\times\left[
 \frac{(1+X_t)(2pX_t^2-1-X_t)}{e_t^2}
 -\frac{2X_t(pX_t-1)}{e_t}w\right]\mathcal L(w),
 \label{eq:allv-bound-clock-precession-waveform}
\end{align}
and
 \begin{align}
 \left.\delta U^b_{22,p}\right|_{\mathrm{prec}^2,\mathcal L^2}
 ={}&-\frac{2\Cnorm\mu\epsilon^3 k_1^2}{2\pi\ii}
 \oint_{|w|=1}\frac{\dd w}{w}\,
 w^p e^{-pe_t(w-w^{-1})/2}\nonumber\\
 &\quad\times\left[
 \frac{(1+X_t)(2pX_t^2-1-X_t)}{e_t^2}
 -\frac{2X_t(pX_t-1)}{e_t}w\right]\mathcal L^2(w).
 \label{eq:allv-bound-precession-square-waveform}
\end{align}
To identify these terms in the bound harmonic
of~\cite{Tessmer:2010ii}, we also include leading precession
acting on the known 1PN polynomial.\footnote{The pure clock
contribution enters the ordinary Bessel coefficients together
with orbital and radiative corrections at 2PN. Isolating it
in the bound harmonic requires additional waveform information.} The Fourier identities
of~\cite{Boetzel:2017zza} then put both expressions in the
same form. Separating the additional 2PN clock correction
and the $j^{-4}$ periastron term gives agreement
for the two displayed contributions, as well as the periastron
energy correction, for $p\geq1$, $0<e_t<1$ and arbitrary
$\nu$. See App.~\ref{app:PM} for more details. Completing the bound harmonic requires the remaining
radiative frequency dependence. Further PM data, such as
those in~\cite{Geralico:2026efi}, can constrain it through
the bootstrap illustrated at 1PN, while the conservative
factors can remain unexpanded.

\section{Conclusions and Outlook}\label{sec:conclusions}

Following the basic ideas of the B2B correspondence---both in the conservative and radiative sectors---we have extended the B2B dictionary to frequency-domain
waveforms. For local conservative dynamics and source moments
satisfying the conditions in \S\ref{sec:theorem},
the mass-type relation takes the form
\begin{equation}
 U^{b}_{\ell m,k}({\cal E}^-,J)
 =\frac{1}{T_r}
 \left[
 S^{U}_{\ell m}({\cal E}^-,J,\omega)
 -(-1)^{\ell+k}
 S^{U}_{\ell,-m}({\cal E}^-,-J,\omega)
 \right]_{\omega=\omega_{km}},
 \label{eq:conclusion-map}
\end{equation}
and similarly for $V^{b}_{\ell m,k}$. The basic idea follows the original B2B construction \cite{Kalin:2019rwq,Kalin:2019inp,Cho:2021arx}. Starting from $U_{\ell m}=A_{\ell m}e^{-\ii m\phi}$ and changing
$\dd t$ to $\dd r$ gives
\begin{equation}
 S_{\ell m}(J,\omega)
 =\sum_\pm\int_{r_{\min}(J)}^\infty F^\pm_{\ell m}\,\dd r,
 \qquad
 F^\pm_{\ell m}
 =\frac{A_{\ell m}(r,\pm P,J)}{v_r}
   e^{\pm\ii(\omega t-m\phi)}.
\end{equation}
The incoming radial limits have been reversed, so both terms
are oriented towards infinity. The time and angle vanish at
$r_{\min}(J)$. We then analytically continue the energy and follow the root-exchanging
$J\to-J$ path at fixed $\omega$, yielding
$r_{\min}(J,{\cal E}^-)=r_-$ and $r_{\min}(-J,{\cal E}^-)=r_+$.
Since the Hamiltonian is even in $J$, the radial velocity is
unchanged while the angular velocity reverses sign, hence
\begin{equation}
 \begin{aligned}
 t(-J,r)-t(J,r)
 &=-\int_{r_{\min}(J)}^{r_{\min}(-J)}
       \frac{\dd r'}{v_r}=-\frac{T_r}{2},\\
 \phi(-J,r)+\phi(J,r)
 &=\int_{r_{\min}(J)}^{r_{\min}(-J)}
       \frac{\dot\phi}{v_r}\,\dd r'=\frac{\Delta\phi}{2}.
 \end{aligned}
\end{equation}
Together with the reflected-source parity, these shifts imply
\begin{equation}
 F^\pm_{\ell,-m}(-J,r;\omega)
 =(-1)^\ell e^{\mp\ii\Theta_m/2}
 F^\pm_{\ell m}(J,r;\omega)\,, \quad \Theta_m = \omega T_r-m\Delta\phi\,.
\end{equation}
Only now select $\omega=\omega_{km}$, for which
$\Theta_m \to 2\pi k$. Both phases become $(-1)^k$, and the subtraction reduces to
the original B2B endpoint construction 
\begin{equation}
\begin{aligned}
&S_{\ell m}(J)
 -(-1)^{\ell+k}S_{\ell,-m}(-J)=\sum_\pm
 \left[\int_{r_{\min}(J)}^\infty-\int_{r_{\min}(-J)}^\infty\right]
 F^\pm_{\ell m}\,\dd r
 =\sum_\pm\int_{r_-}^{r_+}F^\pm_{\ell m}\,\dd r\\
&\quad=\int_0^{T_r}e^{\ii\omega_{km}t}U_{\ell m}(t)\,\dd t
 =T_rU^b_{\ell m,k}.
\end{aligned}
\end{equation}
Because of subtleties of the analytic continuation of the waveforms, the B2B construction cannot be implemented as a simple connection between $J^{-n}$ coefficients in a PM expansion of a given observable, as it does, for instance, in the original map between periastron advance and scattering angle \cite{Kalin:2019rwq}.\footnote{Some of the subtleties we encountered here, however, are already present in the connection between the total radiated energy between hyperbolic and elliptic motion. In particular, the path of root-exchanging analytic continuation, yielding $e \to -e$, becomes important when applied to the nontrivial dependence on the eccentricity entering the full scattering result, e.g., the Clausen $\text{Cl}_2(z)$ function at 3PN order \cite{Cho:2021arx}.} 

Therefore, the schematic manipulations above need to be carefully established as  proper analytic continuations. However, the B2B logic remains essentially unaltered. The scattering expressions follow the energy continuation
and the root-exchanging $J$ path at independent frequency, with the two encounters starting at $r_-$ and $r_+$,
respectively. Only at the end is the bound harmonic selected. The relative clock origin fixes the harmonic phase, source
rotation fixes the parity factor, and subtraction of the common exterior passages leaves one radial period. 
No large-angular-momentum limit is required. Propagation through a stationary mass monopole preserves the B2B relation: tails and their monopole iterations multiply the same
frequency-domain combination within each parity channel. Analogous relations follow for the
current-type harmonics.\vskip 4pt 

We have explicitly shown that the Newtonian mass and current modes are indeed related through the B2B correspondence. At 1PN, the frequency shift and the angular term, including the periastron advance, must be continued together. The combination gives the complete instantaneous mass harmonic. While the B2B correspondence applies to these waveforms at a given PN order, a finite PM truncation presents a different problem. The latter expands the frequency dependence needed at a bound harmonic,
so the harmonic condition cannot in general be imposed on
the truncation. The Newtonian example makes this
distinction explicit: the first three PM terms reproduce
the fixed-$x$ expansion, but setting $x=pX$ removes the suppression of powers of $x/X$. Therefore, we no longer control the bound waveform on a given harmonic (with $p>0$). To tackle this problem we introduced a bootstrap-type approach, in which we retain the complete frequency functions from the PN domain but use PM scattering data to determine their respective weights. At 1PN, the supplied Bessel and angular structure
is sufficient to reconstruct the bound harmonic. In the $\nu^2$ sector, $G^1$ and $G^3$ fix the polynomial weights, while
$G^2$ and the independent bound expression test the
completion. The test-body construction similarly determines
the polynomial part while retaining the known angular function. The clock and angle suggest how to retain some relativistic
information beyond this 1PN reconstruction. Conservative dynamics
supplies the full energy dependence that dresses the frequency functions.\vskip 4pt This is already
visible in the elementary integral (with the prescription in \S\ref{sec:1pn})
\begin{equation}
 \frac{1}{2} \int_{-\infty}^{\infty}\dd v\,e^{nv}e^{\ii\omega t(v)}
 =\ii^n e^{\pi\alpha/2}K_{n-\ii\alpha}(z_t),
 \qquad \omega t(v)=z_t\sinh v-\alpha v.
 \label{eq:conclusion-clock-structure}
\end{equation}
Keeping the Bessel function intact retains the conservative
energy dependence of $\alpha,z_t$, while the established
continuation guides it into the bound problem. Here the
object being dressed is a frequency function, extending the
role of the energy-dependent $1/j$ coefficients in the
conservative B2B dictionary. The clock identified in the loop makes this mechanism particularly
simple. Its coefficient is fixed by the scattering time delay,
and the same function of energy enters both the Bessel order
and the exponential. Expanding $e^{\pi\alpha/2}$
gives the loop term $\pi\alpha S^U_{\rm tree}/2$, while retaining
the exponential together with the shifted order fixes further
contributions. After analytic continuation, the same clock
determines the phase and the radial period. In this sense it acts as an energy-dependent
form factor, correlating terms that appear separately in the
PN and PM expansions.\vskip 4pt  

We tested these clock-generated contributions against
probe scattering data through $G^5$. At $G^5$, the clock
insertions on the tree waveform determine the
coefficient of the fourth power of the logarithm in
\eqref{eq:allv-encounter-anomaly}. The cubic coefficient also
requires the one-loop waveform. To determine the quadratic
and linear coefficients, we needed the scattering
contributions at $G^3$ and $G^4$, respectively. After subtracting
the clock terms already generated at lower orders, these
coefficients agree with the probe scattering result through
relative 2PN order. The periastron advance
gives a further test: its quadratic contribution reproduces
 the term proportional to $\phi^2_N=(\arctan\tau)^2$ at ${\cal O}(G^5)$ and relative 2PN order.\vskip 4pt The B2B correspondence allows us to carry this scattering information
into the bound waveform. The clock and angle determine the contributions to $T_r$ and $\Korb$, retaining
their full energy dependence \cite{Kalin:2019rwq, Kalin:2019inp}.
Their effect on the waveform follows by combining the change
in radial frequency with the precession of the orbital phase,
as in~\eqref{eq:allv-bound-joint-factor}. Acting on the already
known Newtonian and 1PN frequency coefficients, these factors
generate the squared angular logarithm, the mixed
clock--precession term, and the first energy correction to
the periastron contribution. We find exact agreement with
the corresponding terms of the 2PN bound harmonic at finite
eccentricity and arbitrary mass ratio. Extending this comparison to the ordinary Bessel terms requires
additional waveform information: their coefficients combine
the higher clock corrections with further orbital and radiative
contributions. The bootstrap provides a way to determine the
remaining frequency dependence from scattering, while preserving
the conservative factors. Preliminary work along these lines reproduces the full instantaneous 2PN bound
harmonic at finite eccentricity. We will present this reconstruction elsewhere.\vskip 4pt
Keeping the conservative factors unexpanded retains the
higher-PN contributions they generate. However, to recover the full
$\gamma$-dependence of the waveform, the frequency class must
also be extended beyond the PN regime. The exact-velocity tree already contains functions
with two distinct velocity-dependent scales
\cite{DeAngelis:2026kew}. For example, its probe $22$ projection
contains the function
\begin{equation}
 \mathfrak F(u,\gamma)=\int_{-\infty}^{\infty}
 \frac{e^{\ii u\tau}\,\dd\tau}
 {\sqrt{1+\tau^2}\,(\tau^2+\gamma^2)}\,.
 \label{eq:allv-extra-master}
\end{equation}
At $\gamma=1$ the two denominator factors combine into the
usual Bessel kernel. Expanding in $p_\infty^2=\gamma^2-1$
generates higher powers of that denominator, and hence shifted
Bessels. At~any finite PN order these can be reduced using the identities we used throughout the paper. 
The unexpanded function, however, retains both scales. A finite Bessel pair with polynomial frequency
weights cannot represent this dependence, even with coefficients
that are promoted to functions of~$\gamma$. Therefore, an enlarged function space is needed to encapsulate the full velocity dependence.\vskip 4pt 

The B2B continuation would also act on the extended function space. Following $p_\infty\to\ii s$ at fixed
$J,\omega$ sends $u$ to $u e^{-\ii\pi}$, where $s^2=1-\gamma^2$, with
$0<\gamma<1$ at the endpoint. The reflected $J$ route brings the argument back to $u$.
For instance, taking the expression in \eqref{eq:allv-extra-master}, at leading PM order in the probe limit, the B2B analytic continuation yields (see App.~\ref{app:PM})
\begin{equation}
 -\frac{\mathfrak F(u e^{-\ii\pi},\gamma)
             -\mathfrak F(u ,\gamma)}{2\pi\ii}
 =\frac1\gamma\int_0^u\dd u'\,
       \sinh[\gamma(u-u')]I_0(u').
 \label{eq:conclusion-master-continuation}
\end{equation}
Expanding its right-hand side in powers of
$\gamma^2-1$ produces, $uI_1(u)$, then
$(\gamma^2-1)u^2I_2(u)/3$, and so on with higher Bessel shifts, which are the functions expected in the PN expansion. The remaining step would be to complete its finite-$X$ dependence, including the clock and precession, and determine its weight from scattering. This approach would extend the logic of the 1PN bootstrap:
the supplied frequency structure relates the continued PM
coefficients to an expression that can be later evaluated at the
bound harmonic.\vskip 4pt 

To conclude, let us emphasize that our B2B extension to waveforms so far applies to local source terms, including 
their {\it radiative dressing} by tail-type terms~\cite{Galley:2015kus,Blanchet:1992br,Goldberger:2009qd,Porto:2012as,Marchand:2016vox}. More general hereditary interactions \cite{Porto:2024cwd,Porto:2026fsd} and (nonlogarithmic) nonlocal conservative dynamics~\cite{Dlapa:2024cje,Bini:2024tft,Dlapa:2025biy,Dlapa:2026oyq} require extensions beyond the
construction developed here. Moreover, we have restricted our applications to nonspinning bodies. However, following \cite{Kalin:2019inp,Liu:2021zxr}, it is straightforward to generalize our constructions to the case of nonprecessing spins. Nonperturbative regimes may likewise
offer insight into a broader B2B framework. In~particular, the
compatibility of monopole tails with the frequency-space correspondence
motivates exploring the waveform map in the Schwarzschild/Kerr probe
limit.\footnote{Preliminary results suggest a relation between the
analytically continued solutions, with numerical tests supporting the
map in the cases examined. We will return to a more in depth analysis of the nonperturbative regime elsewhere.} Higher self-force orders could then supply
further constraints on its extension.
We will return to these questions elsewhere.
\paragraph{Acknowledgments} 
RAP would like to thank Matias Zaldarriaga for discussions on seemingly unrelated topics that nonetheless inspired revisiting the extension of the B2B dictionary to waveforms. GK acknowledges OpenAI for providing access to ChatGPT and Codex through the ChatGPT for Academic Researchers program.  {\it AI disclaimer:} The scientific ideas, conceptual framework, and research strategy are entirely our own and build on the B2B methodology established in our previous papers \cite{Kalin:2019rwq,Kalin:2019inp,Cho:2021arx}, much before the AIpocalypse. We employed the assistance of OpenAI Codex mostly to assess the rigor of mathematical arguments and to perform algebraic cross-checks, all based on our original work. OpenAI ChatGPT also provided assistance with manuscript preparation. The results and interpretations are entirely our own and we take full responsibility.  

\addtocontents{toc}{\protect\setcounter{tocdepth}{1}}
\appendix
\section{B2B for local PN sources}
\label{app:transport}
The source and clock matching in \S\ref{sec:theorem} puts the two
continued scattering integrals in the same form. To obtain a bound
harmonic, their difference must then reduce to one radial period,
without an additional contribution from a singularity or from infinity.
This is the step that sets the remainder \eqref{eq:channel-remainders}
to zero. We give the contour argument for the local PN class below,
first at $X>1$ and then on physical ellipses.

\subsection{The local PN class}
In this section we use reduced units $GM=\mu=1$ and write
$H=(\mathcal H-M)/\mu=\E$. We denote with $(r,P_r)$
the reduced radial distance and momentum. We assume
\begin{equation}
 H(r,\mathbf P^2;\eta)=\frac{\mathbf P^2}{2}-\frac1r+
 \sum_{n\ge1}\eta^n\sum_{q=0}^{q_n^H}
 H_{nq}(1/r,\mathbf P^2)[\log(r)]^q,
 \qquad \mathbf P^2=P_r^2+\frac{j^2}{r^2}\,,
 \label{eq:PN-isotropic}
\end{equation}
Here $\mathbf P$ is the reduced relative momentum and $\eta$
counts PN orders. The logarithm is real on
physical phase space, and we assume that any (fixed) logarithmic scale has been absorbed into the coefficients.
Each $H_{nq}$ is a finite polynomial, with finite $q_n^H$.
For $q>0$ we require $H_{nq}(0,\mathbf P^2)=0$, so the logarithmic
potential terms vanish at infinity.\vskip 4pt 

The polynomial sector is the isotropic local form used in B2B and WEFT
\cite{Kalin:2019rwq,Kalin:2019inp,Kalin:2020mvi}, with the radial logarithms induced both from potential and radiation modes \cite{Foffa:2019yfl,Dlapa:2021npj}. We assume the associated divergences (and renormalization scale) have already canceled out between near and far zones \cite{Porto:2017dgs}. However, we exclude the (nonlogarithmic)
nonlocal contributions from the tail Hamiltonian \cite{Damour:2014jta,Galley:2015kus,Dlapa:2024cje,Bini:2024tft,Dlapa:2025biy,Dlapa:2026oyq}.\vskip 4pt

In the same gauge, and under the same conditions, the mass and current source multipoles have the form
\begin{equation}
 \begin{aligned}
 I_{\ell m}
 &=e^{-\ii m\phi}
 \sum_{n\ge0}\eta^n\sum_{q=0}^{q_n}
 I_{\ell m}^{(n,q)}(r,P_r,j)\,[\log(r)]^q,\\
 J_{\ell m}
 &=e^{-\ii m\phi}
 \sum_{n\ge0}\eta^n\sum_{q=0}^{q_n}
 J_{\ell m}^{(n,q)}(r,P_r,j)\,[\log(r)]^q.
 \end{aligned}
 \label{eq:PN-source-class}
\end{equation}
The coefficients are Laurent polynomials in $r$ and polynomials
in momenta. They obey the reflection rule in \eqref{eq:reflection}.
By construction, the only finite singularity is at $r=0$, including
its logarithmic branch point.\footnote{This representation therefore includes the logarithms in mass and current moments, e.g., ~\cite{Mishra:2015bqa,Henry:2021cek,Amalberti:2024jaa}.} For any local source function, $F(r,P_r,\phi,j)$, Hamilton's equations preserve this structure. 
Differentiation can raise the pole order at $r=0$ or the
logarithmic degree, but introduces no new singularities at fixed
PN order.\vskip 4pt 

We first work on a compact reference range $X=j\sqrt{\epsilon}>1$.
The first route is the energy continuation \eqref{eq:energy-route}.
For the reflected mode it is followed by the outer path
$j(\theta)=j_0e^{\ii\theta}$, $0\leq\theta\leq\pi$.
At the endpoint the frequency is set to $\omega=\omega_{km}$,
including PN corrections. The continuation to physical ellipses
is discussed in \S\ref{app:PN-physical} below. Let $[\eta^n]$ denote the coefficient of $\eta^n$ after the source,
measure, phase and harmonic frequency have all been expanded together.
Then for \eqref{eq:PN-isotropic} and \eqref{eq:PN-source-class},
the energy route and the root-exchanging angular-momentum routes
constructed below give
\begin{equation}
 [\eta^n]\mathcal R_U=[\eta^n]\mathcal R_V=0
 \qquad(n\geq0)\,,
 \label{eq:PN-zero}
\end{equation}
at the final bound harmonic. Equivalently,
\begin{equation}
 S^U_{\rm peri}-(-1)^{\ell+k}S^U_{\rm apo}
 =T_rU^b_{\ell m,k}\,, \quad S^V_{\rm peri}-(-1)^{\ell+k}S^V_{\rm apo}
 =T_rV^b_{\ell m,k}\,.
\end{equation}

\subsection{The paths}
The contour argument requires the integrand to be regular in the
region swept as one path is moved into the other. A singularity
enclosed by both paths does not obstruct the deformation.
Cauchy's theorem then equates the two integrals. For the local
PN class, the relevant restrictions come from $r=0$ and from infinity.
A turning point is a nonzero separation where the radial motion
reverses, and therefore it causes no obstruction. Near a simple
root, $r_\star$, both $P_r$ and $\dot r$ are proportional to
$\sqrt{r-r_\star}$, so $\dd r/\dot r$ is regular in this
square-root coordinate. Away from the roots, solving $H=\E$
order by order introduces only inverse powers of $r$ and its
logarithm. The new coefficient of $\mathbf P^2$ enters linearly, because
the Newtonian Hamiltonian contains $\mathbf P^2/2$.\vskip 4pt

To study the analytic properties after the energy continuation,
it is useful to introduce a (complex) radial anomaly, $v$, via
\begin{equation}
 r(v)=\frac{r_++r_-}{2}
      -\frac{r_+-r_-}{2}\cosh v.
\end{equation}
The turning points occur at $v=0$ and $v=\ii\pi$. On a physical bound orbit, $v=\ii u$ reduces this expression
to the usual cosine parameterization, with $u$ running from
$0$ to $\pi$ between periastron and apoapsis.
Including the PN corrections in $r_\pm$ shifts the radial
turning points without changing their anomaly values. At Newtonian order this gives $r_N(v)=a_N(1-e\cosh v)$, with $e=\frac{r_+-r_-}{r_++r_-}$ and  $a_N=\epsilon^{-1}$. Introducing $w=e^v$, it can be written as
\begin{equation}
 r_N(w)=\frac{a_N(1+X)}{2}
          (1-\beta/w)(1-\beta w),
 \quad \beta=\frac{e}{1+X}, \quad e^2+X^2=1\,.
 \label{eq:PN-log-radius}
\end{equation} 
The two zeros of this expression, $w=\beta$ and $w=\beta^{-1}$,
are the points where the continued separation vanishes. For $X>1$,
$e=\ii\sqrt{X^2-1}$, so $\beta$ lies above the real axis inside
$|w|=1$, while $\beta^{-1}$ lies below it outside the circle.
The ends $w=\{0,\infty\}$ correspond instead to infinite separation.
Expanding the local PN terms in this coordinate can increase the
powers of these singularities, i.e., $1/(r_N+\delta r_{\rm PN}) =1/r_N - \delta r_{\rm PN}/r_N^2 +\cdots$, but introduces no new locations. The logarithms do not change this geometry. Place their cuts from $0$ to $\beta$ and from $\beta^{-1}$ downwards to infinity. A loop around the inner cut encloses one zero and one pole of
$r_N$, so $\log r_N$ has no net change on that loop. It is therefore
single-valued on the region needed for the deformation, as are its
PN corrections. The followed starting root remains in $\Re r>0$
during the energy and $J$ continuations. At the fixed
reference endpoint, the half-cycle joining $r_-$ to $r_+$ also
lies in $\Re r>0$. These two paths therefore reach the same
logarithm at $r_+$ without winding around zero. Small PN
perturbations preserve this branch.\vskip 4pt

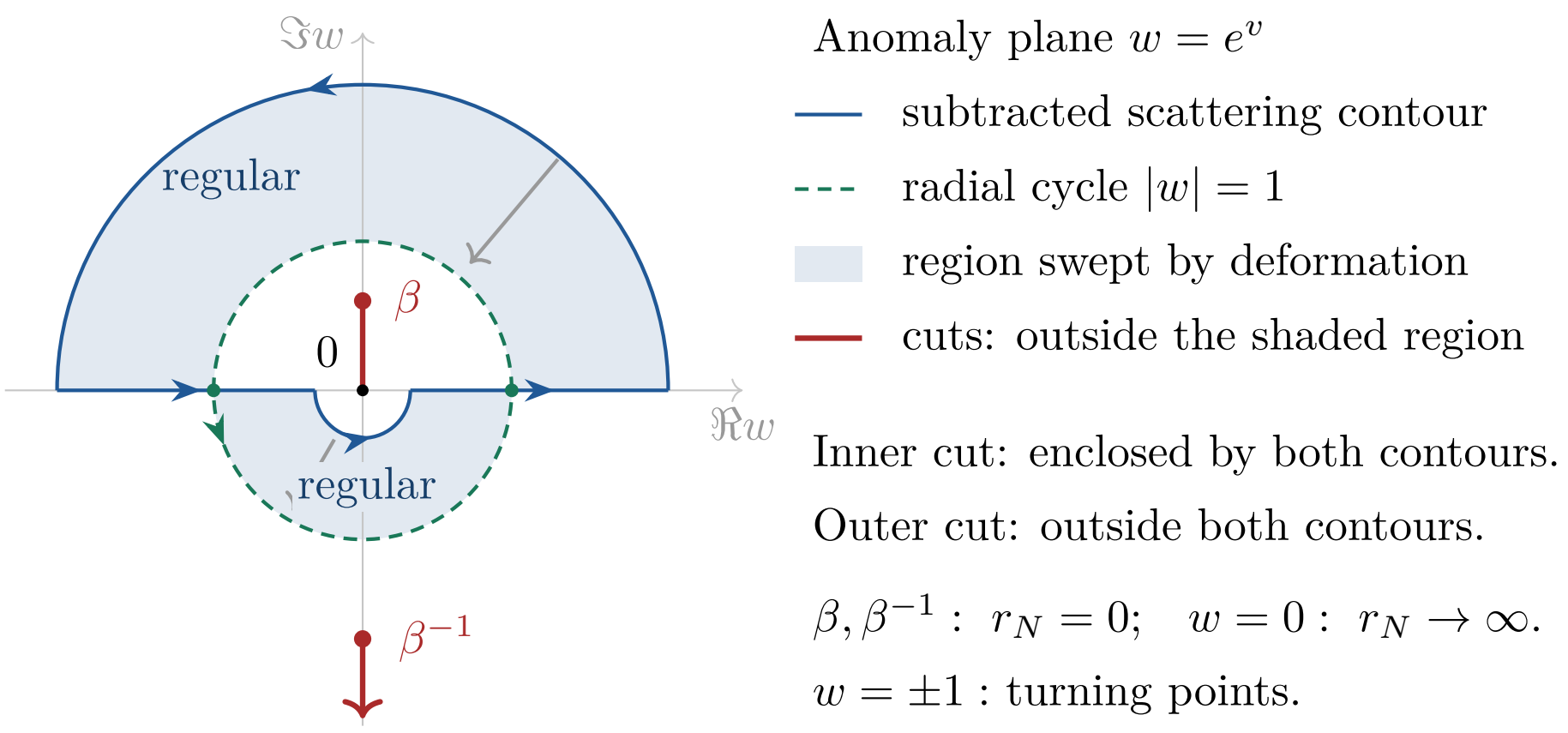
\begin{figure}[t!]
\centering
\begin{tikzpicture}[x=1.15cm,y=1.15cm,font=\small]
 % Only these two half-annuli are swept by the deformation.
 \fill[routea!13](1,0)--(2.05,0) arc(0:180:2.05)
   --(-1,0) arc(180:0:1)--cycle;
 \fill[routea!13](-.32,0)--(-1,0) arc(180:360:1)
   --(.32,0) arc(360:180:.32)--cycle;
 \draw[->,gray!45](-2.4,0)--(2.55,0)
   node[below,text=gray!80]{$\Re w$};
 \draw[->,gray!45](0,-2.25)--(0,2.4)
   node[left,text=gray!80]{$\Im w$};
 \draw[routea,thick,arr](.32,0)--(2.05,0);
 \draw[routea,thick,arr](2.05,0) arc(0:180:2.05);
 \draw[routea,thick,arr](-2.05,0)--(-.32,0);
 \draw[routea,thick,arr](-.32,0) arc(180:360:.32);
 \draw[bound,thick,densely dashed,arr](1,0) arc(0:360:1);
 \draw[rem,very thick](0,0)--(0,.6);
 \draw[rem,very thick,->](0,-1.667)--(0,-2.2);
 \fill[rem](0,.6) circle(1.9pt)
   node[right=3pt]{$\beta$};
 \fill[rem](0,-1.667) circle(1.9pt)
   node[right=4pt]{$\beta^{-1}$};
 \fill(0,0) circle(1.3pt) node[above left=2pt]{$0$};
 \fill[bound](1,0) circle(1.5pt);
 \fill[bound](-1,0) circle(1.5pt);
 \draw[->,gray!80,thick](1.31,1.55)--(.72,.85);
 \draw[->,gray!80,thick](-.19,-.33)--(-.47,-.81);
 \node[routea!70!black]at(-.88,1.42){regular};
 \node[routea!70!black,fill=routea!13,inner sep=1pt]
   at(.03,-.65){regular};
 \node[anchor=west]at(2.9,2.35){Anomaly plane $w=e^v$};
 \draw[routea,thick](2.9,1.85)--(3.35,1.85);
 \node[anchor=west]at(3.5,1.85){subtracted scattering contour};
 \draw[bound,thick,densely dashed](2.9,1.35)--(3.35,1.35);
 \node[anchor=west]at(3.5,1.35){radial cycle $|w|=1$};
 \fill[routea!13](2.9,.73) rectangle(3.35,.97);
 \node[anchor=west]at(3.5,.85){region swept by deformation};
 \draw[rem,very thick](2.9,.35)--(3.35,.35);
 \node[anchor=west]at(3.5,.35){cuts: outside the shaded region};
 \node[anchor=north west,align=left]at(2.9,-.18)
 {Inner cut: enclosed by both contours.\\[3pt]
  Outer cut: outside both contours.};
 \node[anchor=north west,align=left]at(2.9,-1.25)
 {$\beta,\beta^{-1}:\ r_N=0$;\quad $w=0:\ r_N\to\infty$.\\[3pt]
  $w=\pm1:$ turning points.};
\end{tikzpicture}
\caption{Contour deformation in the $w=e^v$ plane on the complex
reference range $X>1$, with finite outer and inner cutoffs.
The upper blue arc moves inwards and
the lower blue arc outwards to the dashed unit (green) circle, sweeping only
the shaded regions. Neither region contains a singularity or a cut.
The inner cut from $0$ to $\beta$ stays enclosed by both contours, while
the outer cut stays outside both. Hence, the contours give equal
integrals even though their common interior is not regular.
After continuation to a physical ellipse, the green circle represents
the real radial passage $r_-\to r_+\to r_-$.}
\label{fig:PN-contour}
\end{figure}

We illustrate the continued paths in Fig.~\ref{fig:PN-contour},
drawn in the $w$ plane. The energy and reflected $J$
continuations have already been performed. We now compare
the resulting contours at fixed orbital parameters, as in
\eqref{eq:matched-combination}. The central portion of the first encounter follows real
values of $v$, and therefore lies on the positive real
$w$ axis. For the reflected encounter, exchanging $r_-$
and $r_+$ reverses the coefficient of $\cosh v$ in the radius.
The reflected encounter at real anomaly $v$ therefore has
the radius of the first at $v+\ii\pi$.
Representing both paths in the figure's common $w$ plane
places the reflected segment at $e^{v+\ii\pi}=-e^v$, on
the negative real axis. The clocks and angular branches
follow the parameter paths throughout this identification.
The turning points themselves are at $w=\pm1$. Each encounter has two distant ends, one on either side
of its turning point. In the anomaly coordinate these are
$\Re v\to-\infty$ and $\Re v\to+\infty$, which become
$w\to0$ and $|w|\to\infty$, respectively. Both limits
describe infinite separation and infinite time.\vskip 4pt

Both scattering integrals extend to infinity, whereas the
bound integral covers a finite radial cycle. We must therefore
show that their distant parts cancel, leaving no additional
term. We first keep the endpoints at a finite distance.
The continued paths reach their distant ends in directions
with $\Im t>0$, where the Fourier factor decays for positive
frequency,
$\lvert e^{\ii\omega t}\rvert=e^{-\omega\Im t}$.
The leading Newtonian time behaves as
\begin{equation}
 t_N\sim\frac{e_s}{2n_s}e^v
 \quad(\Re v\to+\infty),\qquad
 t_N\sim-\frac{e_s}{2n_s}e^{-v}
 \quad(\Re v\to-\infty),
\end{equation}
with $e_s/n_s$ evaluated at its continued value. For negative frequency the decaying ends instead lie
where $\Im t<0$. Changing $\Im v$ rotates these times while keeping that
coefficient fixed. Along the resulting decaying directions,
the Fourier exponential dominates the powers of the
separation and logarithms that occur at each fixed PN order.
The distant endpoints can therefore be taken to infinity,
provided the connections to the rest of the paths are
retained.\vskip 4pt

To see what these connections contribute, consider the
large-$w$ end. The upper part is inherited from the
energy continuation. Along that path, the phase of
$e_s/n_s$ decreases by $\pi$, so the distant end must
turn counterclockwise by $\pi$ to keep the leading time
positive imaginary. Retaining its connection to the
positive real segment, and subtracting the common distant
portion of the reflected encounter, leaves the upper arc behind.
The corresponding cancellation at the other end,
$w\to0$, leaves the lower inner connection. The Fig.~\ref{fig:PN-contour} shows this remaining contour,
after the common distant portions have been subtracted.
The blue arcs are the surviving connections, drawn as
semicircles. Their integrals have not been discarded:
they remain part of the contour that is to be deformed
to the bound radial cycle.\vskip 4pt  In order to do so, we move the upper blue arc inwards
and the lower blue arc outwards until both reach $|w|=1$.
These movements sweep only the shaded regions, which
contain no singularity or cut: $\beta$ remains inside
both contours and $\beta^{-1}$ outside. Cauchy's theorem
therefore leaves the integral unchanged.\footnote{Once around $|w|=1$, $e^{\ii\omega t-\ii m\phi}$ is multiplied by $e^{\ii(\omega T_r-m\Delta\phi)}$, so the integrand is single-valued only at the harmonic value $\omega=\omega_{km}$, see \eqref{eq:periodphase}. At fixed PN order, $\log w$ terms recombine into periodic functions, such as the $\mathcal L(w)$ function in \eqref{eq:allv-bound-periodic-angle}.} Since $|w|=1$ traces one radial cycle, the remaining integral is precisely
the continued bound integral. Under the stated path and branch conditions, this establishes \eqref{eq:PN-zero}
on the reference sheet. For the latter, we have $v=\ii u$,
so that $w=e^{\ii u}$ lies on the unit circle:
\begin{equation}\label{eq:circle}
 r_N(w)=a_N\left[1-\frac e2\left(w+\frac1w\right)\right]
       =a_N(1-e\cos u),\qquad 0\leq u\leq2\pi.
\end{equation}
For a physical ellipse, the upper semicircle traces the
real interval from $r_-$ to $r_+$, and the lower semicircle
traces the return. The figure, instead, shows the complex reference orbit, with $X>1$,
where $e$ is imaginary. We must therefore continue this result 
to the physical domain $0<e<1$, as described below.

\begin{figure}[t!]
\centering
\begin{tikzpicture}[font=\small,>=Stealth]
 \begin{scope}[x=3.8cm,y=2.5cm]
  \draw[->,gray](0,0)--(1.12,0)
    node[right,text=black]{$\Re e$};
  \draw[->,gray](0,-.12)--(0,1.22)
    node[above,text=black]{$\Im e$};

  \draw[blue!65!black,thick,
    postaction={decorate},decoration={markings,
    mark=at position .5 with {\arrow{Stealth}}}]
    (0,1)--(.8,0);

  \fill[blue!65!black](0,1) circle(1.7pt)
    node[right=5pt,text=black]{$e=\ii$};
  \fill[blue!65!black](.8,0) circle(1.7pt)
    node[below left=6pt,text=black]{$e=0.8$};
  \draw[red!65!black,fill=white](0,0) circle(2pt)
    node[below left=3pt,text=black]{$e=0$};
  \draw(1,.025)--(1,-.025)
    node[below=4pt]{$1$};

  \node[anchor=west]at(.1,1.55){eccentricity path};
 \end{scope}

 \begin{scope}[shift={(5.4cm,0.976cm)},x=4.3cm,y=5cm]
  \draw[->,gray](.48,0)--(1.53,0)
    node[right,text=black]{$\Re X$};
  \draw[->,gray](1,-.27)--(1,.10);
  \node[anchor=south]at(1,.10){$\Im X$};

  \draw[blue!65!black,thick,
    postaction={decorate},decoration={markings,
    mark=at position .55 with {\arrow{Stealth}}}]
    plot[domain=0:1,samples=81,variable=\t]
    ({sqrt((sqrt((1-.64*\t*\t+(1-\t)*(1-\t))^2
                      +(1.6*\t*(1-\t))^2)
                  +1-.64*\t*\t+(1-\t)*(1-\t))/2)},
     {(-.8*\t*(1-\t))/sqrt(
        (sqrt((1-.64*\t*\t+(1-\t)*(1-\t))^2
                      +(1.6*\t*(1-\t))^2)
                  +1-.64*\t*\t+(1-\t)*(1-\t))/2)});

  \fill[blue!65!black]({sqrt(2)},0) circle(1.7pt)
    node[above=5pt,text=black]{$\sqrt2$};
  \fill[blue!65!black](.6,0) circle(1.7pt)
    node[above=5pt,text=black]{$0.6$};
  \draw[red!65!black,fill=white](1,0) circle(2pt)
    node[above left=3pt,text=black]{$1$};

  \node[anchor=west]at(.54,.5798)
    {\hspace{1.2cm} $J$ path};
  \node[align=center]at(1,.4048)
    {$X=\sqrt{1-e^2}$\\[2pt]
     $J\propto X$ at fixed $\E<0$};
 \end{scope}
\end{tikzpicture}
\caption{The same continuation in the eccentricity and angular-momentum
planes, shown for the example $e:\ii\to0.8$.
The straight segment on the left maps to the curve on the right,
which passes below $X=1$. The open red points mark the circular
orbit, where the turning points coincide. The integration contour $|w|=1$ stays fixed throughout this change
of orbital parameters.}
\label{fig:PN-physical-path}
\end{figure}
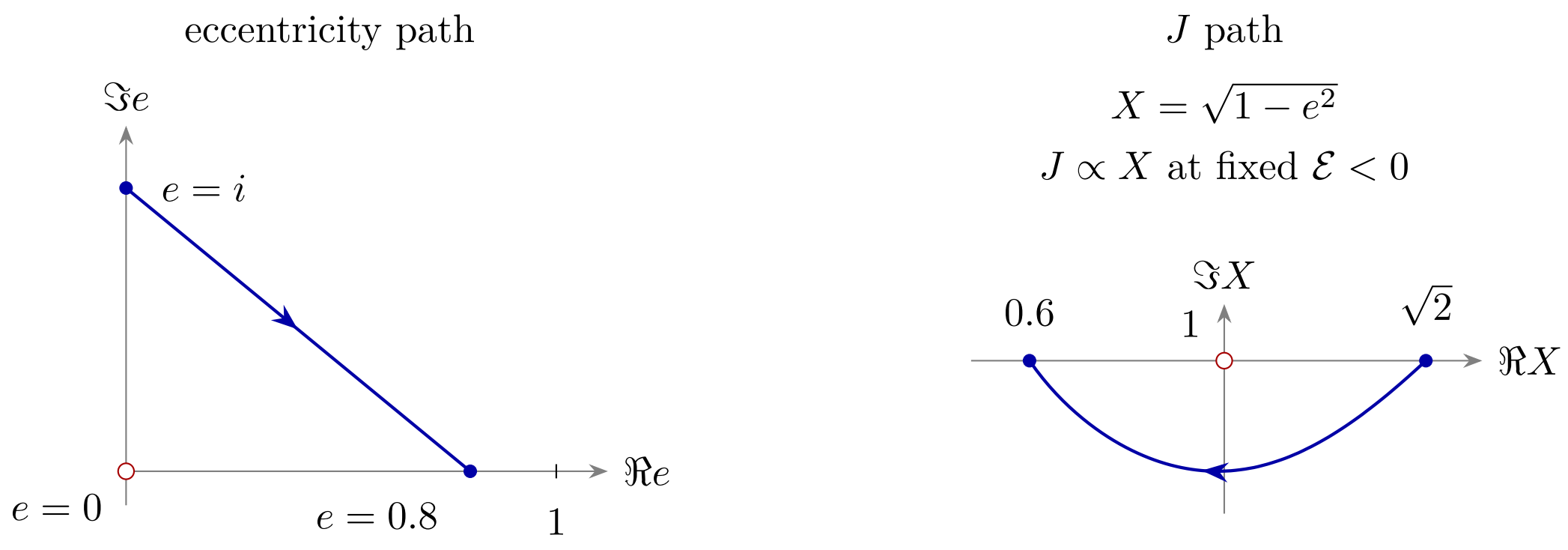

\subsection{The physical domain}
\label{app:PN-physical}
We now must carry the B2B equality from $X>1$ to physical ellipses. The
integration circle $|w|=1$ is fixed, what changes is the
orbit that it describes as we vary the eccentricity.\vskip 4pt 

We keep $\E<0$ fixed and move $e$ from imaginary
values to the desired real value $0<e<1$, along the
straight segment shown in Fig.~\ref{fig:PN-physical-path}.
This avoids $e=0$, where the two turning points coincide.
Following $X=\sqrt{1-e^2}$ with positive real part determines
the corresponding continuation of $J$. In the $X$ plane, this
path passes below $X=1$. The PN-corrected turning points are
followed continuously along the same path.\vskip 4pt

The reason this continuation is regular can be seen directly
from \eqref{eq:circle}. Throughout the parameter path, $0\leq\Re e<1$, and hence
$\Re r_N>0$ for every point of the circle. The separation
therefore never vanishes there. The chosen branch of $X$ also keeps $|\beta|<1$: the zero at
$w=\beta$ remains inside the circle and the zero at
$w=\beta^{-1}$ remains outside. Their cuts can be moved without crossing the circle. Sufficiently small PN corrections preserve these properties. At the endpoint of the map of eccentricity, the path becomes the physical orbit: $u$ increases from $0$ to $\pi$,
the radius increases from periastron to apoapsis, while from $\pi$
to $2\pi$ it decreases back to periastron. Time increases
on both passages, fully describing the radial period in \eqref{eq:bound-passage}.

\section{1PN waveforms}
\label{app:1pn}
The comparison in \S\ref{sec:1pn} starts from the scattering
spectrum \eqref{eq:1pn-sc-spectrum}. The 1PN hyperbolic waveform
is known in the time domain~\cite{DeVittori:2014psa}. Here we
obtain its instantaneous $22$ Fourier representation from the
source quadrupole~\cite{Mishra:2015bqa} and hyperbolic
motion~\cite{Cho:2018upo}, retaining the full eccentricity
dependence through 1PN. We give the source and orbital coefficients, and the reduction of the B2B waveform to the bound harmonic of~\cite{Morras:2025nbp}. 

\subsection{The Fourier prescription.}
On hyperbolic orbits, the instantaneous scattering mode
approaches constants as $t\to\pm\infty$. Its Fourier transform therefore requires a
prescription. Following Bini et al.~\cite{Bini:2024rsy},
we write the nonzero-frequency transform in terms of its
decaying time derivative:\footnote{An overall additive constant contributes only at $\omega=0$. The difference $\Delta U_{22}=U_{22}(+\infty)-U_{22}(-\infty)$ is retained in the integral of the derivative, giving the
low-frequency memory term $\ii\Delta U_{22}/\omega$.}
\begin{equation}
 S^U_{22}(\omega)=\frac{1}{-\ii\omega}
 \int_{-\infty}^{\infty}\dd t\,e^{\ii\omega t}
 \frac{\dd U^{\rm inst}_{22}(t)}{\dd t}.
 \qquad \omega\ne0.
 \label{eq:1pn-time-Fourier-check}
\end{equation}
The evaluation of this integral provides the
scattering spectrum in \eqref{eq:1pn-sc-spectrum} through 1PN, including the angular contribution represented in $\mathcal A_n$.
The frequency-domain scattering results
of~\cite{Bini:2024rsy,Bini:2024ijq} are given at finite PM order.
In \S\ref{sec:PM-bound}, we compare them with the expansion of
\eqref{eq:1pn-sc-spectrum} in powers of $G$ through $G^3$.

\subsection{Coefficients of the scattering spectrum}
\label{app:1pn-source}
The coefficients $\delta c_n$ in \eqref{eq:1pn-sc-spectrum}
receive corrections from the motion and the source quadrupole.
The orbital elements supplementing \eqref{eq:1pn-frequencies}
and \eqref{eq:et-main} follow from the hyperbolic
quasi-Keplerian motion~\cite{Cho:2018upo}. With $\E>0$ and
$a_{s,N}=GM/(2\E)$, the radius is
\begin{equation} r=a_{s,N}(1-2\E a_1^{\rm orb})(e_r\cosh v-1)\,,\quad a_1^{\rm orb}=\frac{\nu-7}{4}\,,
\end{equation}
while the angular eccentricity, $e_\phi$, modifies the Newtonian angular evolution. Writing $e_i=e-2\E\delta e_i$ at fixed $X$, their corrections
are \cite{Cho:2018upo}
\begin{equation}
 \begin{gathered}
 \delta e_r=\frac{6-\nu-5(3-\nu)X^2/4}{2e},\quad
 \delta e_\phi=\frac{6-(15-\nu)X^2/4}{2e}.
 \end{gathered}
 \label{eq:1pn-elements}
\end{equation}

The source correction enters via $Q=I_{xx}-I_{yy}-2\ii I_{xy}$. Through relative 1PN order it is given by (in harmonic coordinates), e.g.~\cite{Mishra:2015bqa},
\begin{equation}
 \begin{aligned}
 \label{eq:Q}
 Q=\mu\Bigl\{&(x-\ii y)^2
       \left[1+\frac{29-87\nu}{42}\mathbf v^2+\frac{-5+8\nu}{7}\frac{GM}{r}\right]\\
 &+\frac{-4+12\nu}{7}r\dot r\,(x-\ii y)(\dot x-\ii\dot y)
       +\frac{11-33\nu}{21}r^2(\dot x-\ii\dot y)^2\Bigr\},
 \end{aligned}
\end{equation}
where $\mathbf v=\dot{\mathbf r}$ is the relative velocity.
The explicit source corrections are already of 1PN
order and need only Newtonian positions and velocities.
The leading term $\mu(x-\ii y)^2$, however, must be evaluated
on the corrected motion. This is the origin of the two
contributions~to~$\delta c_n$.\vskip 4pt 

To obtain the scattering spectrum, we use
$U^{\rm inst}_{22}=\Cnorm\ddot Q/2$ with the quadrupole
evaluated on the orbital motion. The remaining task is to Fourier transform $Q$,
including the time measure. It is convenient to change variables to the anomaly $v$, with $w=e^v$ as before. We write
\begin{equation}
 \begin{aligned}
 \dd t&=\frac{A}{n_s}\dd v,
 &A&=e\cosh v-1,\\
 x-\ii y&=a_{s,N}B,
 &B&=e-\cosh v+X\sinh v.
 \end{aligned}
\end{equation}
Since $Q^N=\mu(x-\ii y)^2$, we find
\begin{equation}
 Q^N\dd t=\frac{\mu a_{s,N}^2}{n_s}AB^2\dd v,
 \qquad
 AB^2=\sum_{n=-3}^{3}c_nw^n.
 \label{eq:P0}
\end{equation}
These are the Newtonian coefficients $c_n$ used in the
scattering spectrum.\vskip 4pt 

The orbital shifts modify the radial scale, time measure and
position in this product. Denoting their polynomial correction
by $P_{\rm orbit}$, and the explicit source correction by
$P_{\rm source}$, we obtain
\begin{equation}
 \left.Q\dd t\right|_{\rm poly}
 =\frac{\mu a_{s,N}^2}{n_s}
 \left[AB^2-2\E(P_{\rm orbit}+P_{\rm source})\right]\dd v
\end{equation}
through 1PN, with the clock correction kept in $n_s$.\vskip 4pt

Comparing with the shifted coefficients $c_n-2\E\delta c_n$ in
\eqref{eq:1pn-sc-spectrum} yields
\begin{equation}
  \sum_{n=-3}^{3}\delta c_nw^n={} P_{\rm orbit}+P_{\rm source}\,,  \label{eq:P1-app}
  \end{equation}
  with
  \begin{equation}
 \begin{aligned}
 P_{\rm orbit}={}&2a_1^{\rm orb}AB^2
 +(\delta e_t+2\delta e_r-2\delta e_\phi)\cosh v\,B^2
 +2AB\delta e_\phi(1-e\sinh v/X),\\
 P_{\rm source}={}&-B^2\left[\frac{29-87\nu}{42}(e\cosh v+1)+\frac{-5+8\nu}{7}\right]\\
 &-B(X\cosh v-\sinh v)
 \left[\frac{-4+12\nu}{7}e\sinh v+\frac{11-33\nu}{21}(e\sinh v+X)\right].
 \end{aligned}
\end{equation}
Each factor of $w^n=e^{nv}$ multiplies the Fourier phase
and integrates to $2\mathcal B_n$, with $\mathcal B_n$
defined in \eqref{eq:sc-Bmaster}. This cancels the $1/2$
in the projection $U^{\rm inst}_{22}=\Cnorm\ddot Q/2$.

\subsection{The angular contribution and B2B trajectory}
\label{app:1pn-path}
The polynomial $P_{\rm orbit}$ accounts for the change in
angular eccentricity. The remaining precession contribution
is the logarithm in \eqref{eq:sc-Amaster}. To follow its
branch on the paths of Fig.~\ref{fig:1pn-angular-paths},
we use the quasi-Keplerian angular function underlying
\eqref{eq:1pn-R-factor}:
\begin{equation}
 R(w)=\frac{w-\beta_\phi}{1-\beta_\phi w},\qquad
 \beta_\phi=\frac{e_\phi}{1+\sqrt{1-e_\phi^2}},\qquad
 \log R(1)=0.
 \label{eq:1pn-anomaly}
\end{equation}
Here $R=e^{\ii\phi/\Korb_s}$, and the square root follows
the branch of $X$. In the logarithmic term, which already
multiplies precession, $R$ reduces to $R_0$ through 1PN.\vskip 4pt

Along the two parameter paths in \eqref{eq:1pn-zt-routes},
we carry the normalization $\log R(1)=0$ with the angular
function. In the change of variable in
\eqref{eq:1pn-angular-kernel-shift}, the integration
variable is replaced by $e^{-\ii\pi}w$, such that the the $w=1$ point moves to $w=e^{\ii\pi}$.
On the followed branch, \eqref{eq:1pn-R-factor} gives
\begin{equation}
 \log R_0(e^{\ii\pi})
 =\ii\pi+\log\frac{1+\beta}{1+\beta}
 =\ii\pi.
\end{equation}
As we discussed in the main text, reflected angular function yields $-\log R_0(w)$ plus a branch constant. Requiring it to
vanish at the transported normalization point fixes
\begin{equation}
 \left.\log R_0(e^{-\ii\pi}w)\right|_{\beta\to-\beta^{-1}}
 =\ii\pi-\log R_0(w),
\end{equation}
with the branch inherited from the continued scattering
function. For the unexpanded angular factor,
this gives
\begin{equation}
 \left.w^{\lambda}R(w)^{m\Korb}\right|_{
       \substack{\beta_\phi\to-\beta_\phi^{-1}\\
                 w\mapsto e^{-\ii\pi}w}}
 =e^{\ii\pi(m\Korb-\lambda)}w^{\lambda}R(w)^{-m\Korb}.
 \label{eq:general-1pn-clock}
\end{equation}
At a bound harmonic, $\lambda=k+m\Korb$, this phase is $(-1)^k$, while for the $U_{22}$ mode,
$k=p-2$ gives \eqref{eq:1pn-phase}. For the $\log w$ insertion, the differentiation in
\eqref{eq:1pn-angular-aligned-derivative} acts on the
relative phase as well as the two Bessel functions.
At independent $\lambda$ and fixed $z_t,\beta$, it reads
\begin{equation}
 \begin{aligned}
 2\partial_\lambda\left[
 \mathcal B_n^{\rm peri}
 -e^{\ii\pi(\lambda+n)}\mathcal B_n^{\rm apo}\right]=2\pi\ii\,\partial_\lambda J_{\lambda+n}(\ii z_t).
 \end{aligned}
 \label{eq:1pn-angular-order-connection}
\end{equation}
Setting $\lambda=p$ gives the order-derivative term in
\eqref{eq:1pn-angular-integral-connection}.\vskip 4pt

For the remaining ratio, we must follow its cuts and
branches on the Bessel contours of
Fig.~\ref{fig:1pn-angular-paths}. The exponential in
\eqref{eq:1pn-angular-w-input} determines the end directions:
\begin{equation}
 \arg w=\begin{cases}
 \pi/2+\arg(\alpha e_t),&w\to0,\\
 \pi/2-\arg(\alpha e_t),&|w|\to\infty.
 \end{cases}
 \label{eq:1pn-angular-transport-directions}
\end{equation}
Along \eqref{eq:1pn-zt-routes}, these become
$-\pi/2$ and $3\pi/2$ at $0$ and $\infty$ on the $\mathrm E$ integral; and
$3\pi/2$ and $3\pi/2$ on the $\mathrm{EJ}$ side for the reflected
substitution~\cite[\S10.9(ii)]{DLMF}.  On these parameter paths, $|X|>1$ and
$\beta^2=(1-X)/(1+X)$ has negative real part. The followed
branch therefore keeps $\beta$ above the real axis and
$\beta^{-1}$ below it. The ends can turn inside both points
near $0$ and outside both near infinity, retaining the joining
arcs without crossing a cut. The logarithmic ratio grows at
most as $\log w$ at either end and is dominated by the
Bessel exponential.\vskip 4pt

To cancel the common gray portions in
Fig.~\ref{fig:1pn-angular-paths}, the two integrals must
carry the same value of the logarithmic ratio, as well as
the same Bessel kernel. The $\log w$ term has already been
included in \eqref{eq:1pn-angular-aligned-derivative}.
For the remaining ratio, the series in
\eqref{eq:periodic-log} defines a single-valued logarithm
in $|\beta|<|w|<|\beta|^{-1}$. It vanishes at both $w=1$
and $w=-1$, so the two normalizations above select the
same branch in this region. We follow this branch from the finite parts of the paths
to their common ends. At the reference value
$\beta=\ii|\beta|$, the inner cut joins $0$ to $\beta$,
while the outer cut runs from $\beta^{-1}$ down the
negative imaginary axis. Both paths approach the
large-$|w|$ end on the left of this outer cut, as drawn
in Fig.~\ref{fig:1pn-angular-paths}. Approaching the
negative imaginary axis from that side gives
\begin{equation}
 \log\frac{1-\beta/w}{1-\beta w}
 \ \longrightarrow\
 \log\left(1+\frac{|\beta|}{|w|}\right)
 -\log\bigl(1-|\beta|\,|w|+\ii0\bigr).
 \label{eq:1pn-angular-common-log}
\end{equation}
The $+\ii0$ follows because $1-\beta w$ approaches the
real axis from above when $w$ approaches the cut from
the left. Both paths therefore reach the same value
of the logarithm, including beyond the branch point
$|w|=|\beta|^{-1}$. Near $w=0$, the denominator is
positive and the two branches likewise agree. The remaining power of $w$ also agrees: the small-end
arguments differ by $2\pi$, giving
$e^{2\pi\ii(p+n)}=1$ at integer $p$.
Thus the full integrands coincide on the common gray
portions, which cancel in the subtraction in
\eqref{eq:1pn-angular-combined-paths}.
The finite pieces form the counterclockwise loop shown
in the right panel of Fig.~\ref{fig:1pn-angular-paths}.
Both this loop and $|w|=1$ enclose the inner cut and
exclude the outer one, such that moving the arcs to the circle
crosses neither cut. The integral can therefore be
evaluated using \eqref{eq:periodic-log}, yielding
the shifted Bessels in
\eqref{eq:1pn-angular-integral-connection}. Finally, the path in \S\ref{app:PN-physical} carries
this equality to $0<e<1$. Along it, $|\beta|<1$, so
$\beta$ remains inside the circle and $\beta^{-1}$
outside. The cuts can be followed without crossing
the circle, on which the Laurent series remains
uniformly convergent.

\subsection{Bound comparison}
\label{app:published}

The B2B result contains the three contributions identified
below \eqref{eq:1pn-secular-cancellation}: the polynomial
correction, the shift of the Bessel argument and the angular
sum. Including the expansion of the common prefactor
$\omega_p^2/(\Omega_rT_r)$ gives
\begin{align}
 \left.U^{b,(1)}_{22,p}\right|_{\rm B2B}
 ={}&\left(2n_1+\frac{4k_1}{p}\right)U^{b,N}_{22,p}
 +\frac{\Cnorm\mu\epsilon p^2}{2}
 \sum_{n=-3}^3\Biggl\{\delta c_nJ_{p+n}(pe) \label{eq:1pn-map-derivatives}\\
 &+(p\delta e_t+2k_1e)c_nJ'_{p+n}(pe) +2k_1c_n\sum_{s=1}^{\infty}\frac{\beta^s}{s}
       [J_{p+n-s}(pe)-J_{p+n+s}(pe)]\Biggr\}.\nonumber
\end{align}
Here $U^{b,(1)}_{22,p}$ is the coefficient of $\epsilon$
in the expansion used in the main text.\vskip 4pt

To compare with \eqref{eq:publishedH1}--\eqref{eq:pubN1},
we express these sums in the same Bessel functions. The finite
shifts reduce directly by \eqref{eq:recurrence}. For the
angular sum, the polynomial \eqref{eq:P0} contains the factor
$e\cosh v-1$, which is proportional to the derivative of the
Fourier phase. Integrating by parts therefore gives
\begin{equation}
 \begin{aligned}
 &2\pi\ii\sum_{n=-3}^{3}c_n
       \sum_{s=1}^{\infty}\frac{\beta^s}{s}
       [J_{p+n-s}(pe)-J_{p+n+s}(pe)]\\
 &\quad=\frac1p\int_0^{2\pi\ii}\dd v\,e^{pv-pe\sinh v}
       \frac{\dd}{\dd v}
       \left\{(e-\cosh v+X\sinh v)^2
                       \log\frac{1-\beta e^v}{1-\beta e^{-v}}\right\}.
 \end{aligned}
 \label{eq:1pn-angular-ibp}
\end{equation}
where we used
$\dd(e^{pv-pe\sinh v})/\dd v=-p(e\cosh v-1)e^{pv-pe\sinh v}$. Expanding the resulting expression and using 
\eqref{eq:periodic-log} gives the angular function
$\widetilde f_p$ below, as well as additional finite ordinary Bessel terms.
Adding the latter to the polynomial and argument corrections
puts \eqref{eq:1pn-map-derivatives} in the form
\eqref{eq:1pn-continued-complete}.\vskip 4pt

The known 1PN waveform, Eq.~(86c) of~\cite{Morras:2025nbp},
uses $e_t$ and $y^2=(GM\Omega_\phi)^{2/3}/(1-e_t^2)$.
Writing its Newtonian part from \eqref{eq:HN0} as
$H_N=-U^{b,N}_{22,p}/(\Cnorm\mu\epsilon)$, it reads
\begin{equation}
 -\frac{U^b_{22,p}}{\Cnorm\mu(GM\Omega_\phi)^{2/3}}
 =H_N(e_t,p)+y^2N_1(e_t,p,\nu)+\Ord(y^4).
 \label{eq:published-bound-full}
\end{equation}
The conversion to $e,\epsilon$ is already given in
\eqref{eq:publishedH1}, and it supplies the same amplitude and
eccentricity terms as \eqref{eq:1pn-continued-complete}.
The comparison therefore reduces to the coefficients in
$N_1$, whose split into ordinary Bessels and
$\widetilde f_p$ is given in \eqref{eq:pubN1}.
The ordinary Bessel coefficients are, with
$X_t=\sqrt{1-e_t^2}$ in these expressions,
\begin{align}
 A_p^{\mathrm{1PN}}&=\frac{37}7-\frac{111}{14}e_t^2-\frac{19}{14}e_t^4
 -\nu(1-e_t^2)\left(\frac{67}{21}+\frac{17}{42}e_t^2\right)\label{eq:pubA}\\
 &\quad+pX_t\left[-\frac{113}{21}-\frac{39}7e_t^2+\frac{356}{21}e_t^4
 +\nu(1-e_t^2)\left(\frac{73}{21}+\frac{11}{21}e_t^2\right)\right]
 +\frac{2p^2}{21}(1-e_t^2)^3(1-3\nu),\nonumber\\
 B_p^{\mathrm{1PN}}&=X_t\left[\frac{37}7(1-e_t^2)
 -\nu\left(\frac{67}{21}-\frac{25}{21}e_t^2\right)\right] \label{eq:pubB}\\
 &\quad+p\left[-\frac{113}{21}+\frac{262}{21}e_t^2-\frac{23}{21}e_t^4
 +\nu(1-e_t^2)\left(\frac{73}{21}+\frac8{21}e_t^2\right)\right]\nonumber
 +\frac{2p^2}{21}(1-e_t^2)^{5/2}(1-3\nu);
\end{align}
whereas the angular function takes the functional form (with $X=\sqrt{1-e^2}, \beta =e/(1+X))$: 
\begin{align}
 \widetilde f_p(e)
 &=\frac{f_p(\beta)J_p(pe)+g_p(\beta)\beta J'_p(pe)}{4\beta^2}
  \label{eq:pubangular}
\\
 &\quad+\frac{3}{(1+\beta^2)^2}
 \sum_{s=3}^{\infty}\beta^{s-2}
 \Biggl\{\frac{24\beta^4}{(s-2)(s-1)(s+1)(s+2)}J_{p+s}(pe)
 \nonumber \\
 &\hspace{30mm}+\left[(1-\beta^2)^4+\frac2{s-2}
 -\frac{4\beta^2}{s-1}+\frac{4\beta^6}{s+1}
 -\frac{2\beta^8}{s+2}\right]J_{p-s}(pe)\Biggr\}\,,\nonumber
\end{align}
where
\begin{align}
 f_p(\beta)&=\frac{6-48\beta^2-167\beta^4+299\beta^6-136\beta^8
 +14\beta^{10}-11\beta^{12}+3\beta^{14}}{(1+\beta^2)^3} \label{eq:pubangular-coeff1}\\
 &\quad-\frac{18-54\beta^2-5\beta^4+3\beta^6-13\beta^8+3\beta^{10}}
 {(1+\beta^2)p}+\frac{12(1+\beta^4)}{p^2},\nonumber\\
 g_p(\beta)&=\frac{12-240\beta^2+130\beta^4+18\beta^6-22\beta^8+6\beta^{10}}
 {(1+\beta^2)^2} \label{eq:pubangular-coeff2}\\
 &\quad-\frac{36-96\beta^2+106\beta^4-32\beta^6+6\beta^8}
 {(1+\beta^2)p}+\frac{24(1-\beta^2)}{p^2}.
\nonumber
\end{align}
In the normalization of \eqref{eq:1pn-continued-complete},
the B2B terms with $s\geq3$ reproduce the infinite sum in
$p\widetilde f_p(e)$ given in \eqref{eq:pubangular}.
We then subtract the whole contribution $p\widetilde f_p(e)$,
including its finite $J_p$ and $J'_p$ terms, from the
bracket in \eqref{eq:1pn-continued-complete}. What remains is
\begin{equation*}
 \frac{A_p^{\mathrm{1PN}}J_p(pe)
       +B_p^{\mathrm{1PN}}eJ'_p(pe)}{e^2},
\end{equation*}
with exactly the polynomials in \eqref{eq:pubA} and
\eqref{eq:pubB}, evaluated at $e_t=e$ at this order. Together
with the common amplitude and eccentricity conversions,
this gives the full 1PN equality
\eqref{eq:1pn-analytic-match}.

\section{PM reconstruction}
\label{app:PM}
The reconstruction in \S\ref{sec:PM-1pn-bootstrap} follows
by matching the continued PM coefficients at independent
frequency. We give the coefficient relations needed for
the Newtonian and 1PN examples, then turn to the clock and
periastron contributions in \S\ref{sec:all-velocity}.

\subsection{Newtonian coefficients and weights}
\label{app:PM-newton}
To complete the Newtonian comparison in
\eqref{eq:PM-N-third-matching}, we need the coefficient
$F_N^{(3)}$ and its reflected counterpart. The exact spectrum
contains $e^{\pi\alpha/2}K_{n-\ii\alpha}(\alpha e_s)$.
Since $\alpha$ is proportional to $G$, its second-order
expansion introduces derivatives with respect to the Bessel
order in the $G^3$ waveform. The resulting coefficient
is~\cite{Bini:2024rsy,Bini:2024ijq}
\begin{align}
 F_N^{(3)}(u)={}&-(u^2+3u+2)K_0(u)
 -(u^2+5u/2+2)K_1(u)\nonumber\\
 &-\frac{u^2(2u+1)}4K_{\rm ord}(u)
 +\frac{u^2(u+1)}2K'_{\rm ord}(u),\\
 K_{\rm ord}(u)={}&2\left.\partial_q^2K_q(u)\right|_{q=0}
                  -\frac{\pi^2}{2}K_0(u).
 \label{eq:PM-N-third}
\end{align}
The reflected mode gives
\begin{align}
 F_{N,{\rm ref}}^{(3)}(u)={}&(-u^2+3u-2)K_0(u)
 +(u^2-5u/2+2)K_1(u)\nonumber\\
 &+\frac{u^2(2u-1)}4K_{\rm ord}(u)
 +\frac{u^2(u-1)}2K'_{\rm ord}(u).
 \label{eq:PM-N-third-reflected}
\end{align}

The argument $u$ reaches $xe^{-\ii\pi}$ on the energy
route and $x$ on the reflected route, as in
\eqref{eq:PMroute-table}. Applying the Bessel identity
in \eqref{eq:1pn-B-connection}, together with its derivatives
with respect to the order, gives
\begin{equation}
 \frac{\ii}{\pi}
 \bigl[F_N^{(3)}(xe^{-\ii\pi})-F_{N,{\rm ref}}^{(3)}(x)\bigr]
 =h_2(x)-\ii\pi x^2
       \bigl[(2x-1)K_0(x)+2(1-x)K_1(x)\bigr].
 \label{eq:PM-N-third-connection}
\end{equation}
The remaining $K_0,K_1$ terms cancel against the products
of the relative phase with the tree and loop in
\eqref{eq:PM-N-third-matching}, leaving $h_2(x)$.\vskip 4pt

The tree also fixes the Newtonian weights needed for the
probe reconstruction below. The PN frequency structure in
\eqref{eq:P0} supplies the factor $e\cosh v-1$ and a quadratic
polynomial in $e-\cosh v$ and $X\sinh v$. Keeping this
structure but leaving its three weights free, we write
\begin{equation}
 (e\cosh v-1)\bigl[d_1(e-\cosh v)^2
 +d_2 X(e-\cosh v)\sinh v+d_3X^2\sinh^2v\bigr].
 \label{eq:PM-newton-unknown-polynomial}
\end{equation}
Expanding in $w=e^v$ gives seven powers, $w^{-3},\ldots,w^3$.
Their Fourier integrals are $2\mathcal B_n$, with
$\mathcal B_n$ defined in \eqref{eq:sc-Bmaster}.
Their continued sum, normalized as in
\eqref{eq:PM-newton-main} and expanded at fixed $x$, must
reproduce the leading coefficient supplied by the tree in
\eqref{eq:PM-N-tree-matching}. Using the identities in
\eqref{eq:recurrence}, this condition becomes
\begin{equation}
 (d_3-d_2x)I_0(x)+[d_2-(d_1+d_3)x]I_1(x)
 =(1-2x)I_0(x)+2(1-x)I_1(x).
 \label{eq:allv-newton-weight-match}
\end{equation}
The right-hand side is $h_0(x)$ in \eqref{eq:hs}.
Matching the constant and linear terms fixes
$d_1=d_3=1$, $d_2=2$, recovering
$(e\cosh v-1)(e-\cosh v+X\sinh v)^2$.
Its seven coefficients are the $c_n$ in \eqref{eq:Ncoeffs},
which also multiply the angular function at 1PN.

\subsection{Matching the 1PN weights}

\label{app:direct-PM}\label{app:PM-nu}
\label{app:PM-class}\label{app:PM-dictionary}
\paragraph{Beyond the probe limit.} To determine the weights of $A^{(1)},B^{(1)}$ in
\eqref{eq:nu-class}, we first express the continued tree
in the variables of the ansatz. At large $X$ and fixed $x$, the continued tree contribution
in \eqref{eq:PM-tree-discontinuity}, with the normalization
of \eqref{eq:PM-reduced-normalization}, gives the first
bracket below. The energy, argument and period conversions in
\eqref{eq:PM-shifted-argument}--\eqref{eq:PM-loop-phase}
add $h_0/4-xh'_0/4$ from the Newtonian waveform. At fixed $\lambda$, the supplied amplitude and eccentricity
terms in \eqref{eq:nu-regular-completion} give
$h_0/12+7xh'_0/8$ at leading order. Expressing the Newtonian term at fixed $x$ using
\eqref{eq:PM-loop-phase} adds $-xh'_0/8$. Subtracting their sum, $h_0/12+3xh'_0/4$, leaves
\begin{align}
 &-\frac27\bigl[(6x^2-3x+2)I_0(x)+(6x^2-6x+4)I_1(x)\bigr]
       +\frac{h_0(x)}6-xh'_0(x)\nonumber\\
 &\hspace{1cm}=\frac{12x^2+22x-17}{42}I_0(x)
              +\frac{6x^2+8x+25}{21}I_1(x).
 \label{eq:PM-tree-comparison}
\end{align}

These are the two polynomials matched in
\eqref{eq:PM-tree-weights}. Expanding their Bessel factors
as in \eqref{eq:PM-loop-from-order} gives the predicted
coefficient of $1/X$:
\begin{equation}
 -\frac{x(12x^2+22x-17)}{42}
       \left[K_0(x)-\frac{\ii\pi}{2}I_0(x)\right]
 +\frac{x(6x^2+8x+25)}{21}
       \left[K_1(x)+\frac{\ii\pi}{2}I_1(x)\right].
 \label{eq:PM-regular-loop}
\end{equation}
The continued $G^2$ waveform, including the conversions in
\eqref{eq:PM-shifted-argument}--\eqref{eq:PM-loop-phase},
gives this expression without further weights.
For the next order, the scattering coefficient needed in
\eqref{eq:nu-scattering} is~\cite{Bini:2024ijq}
\begin{align}
 P_3^\varsigma(u)={}&(24\varsigma u^3+100u^2+180\varsigma u+88)K_0(u)
 +(24u^3+88\varsigma u^2+158u+88\varsigma)K_1(u)\nonumber\\
 &+(12u^4+20\varsigma u^3+11u^2)K_{\rm ord}(u)
 +(-12\varsigma u^4-26u^3-22\varsigma u^2)K'_{\rm ord}(u).
 \label{eq:nu-polynomials}
\end{align}

Its coefficient contains two powers of the PM expansion
parameter, whose 1PN conversion is
\begin{equation}
 \left.\frac{GM}{bp_\infty^2}\right|_{\rm peri,apo}
 =\mp\frac{\ii}{X}
 \left[1+\frac{1+5\nu}{8}\epsilon+\Ord(\epsilon^2)\right].
 \label{eq:PMdictionary}
\end{equation}

At $1/X^2$, the six tree weights also contribute through
the second-order expansion of their Bessels. Subtracting
these terms and the supplied orbital contributions from
the continued $G^3$ coefficient leaves
$(6x^2-73x+67)[I_0(x)+I_1(x)]/21$.
This is the matching condition \eqref{eq:nu-last}: its
constant and linear terms determine the four remaining
weights, while its quadratic term tests the tree values.

\pagebreak[3]
\paragraph{The probe contribution at 1PN.}\label{app:tb-bootstrap}
For the probe contribution in \S\ref{sec:PM-reconstruction},
the frequency ansatz contains both the ordinary Bessels and
the angular function. We start from
\eqref{eq:PM-HN-definition}, include its conservative
orbital and clock corrections, and allow a 1PN polynomial
correction and an angular term with unknown weights.
Their frequency dependence is supplied by the PN structure
in \eqref{eq:1pn-sc-spectrum}.\vskip 4pt

The required orbital and clock corrections follow from
\eqref{eq:1pn-endpoint-dictionary}, \eqref{eq:et-main} and \eqref{eq:1pn-elements}. Their probe values are
\cite{Cho:2018upo,Memmesheimer:2004cv}
\begin{equation}
 \begin{gathered}
 n_1=-\frac{15}{8},\qquad k_1=\frac3{X^2},\qquad
 a_1^{\rm orb}=-\frac74,\\
 \delta e_t=\frac{17X^2/4-2}{2e},\qquad
 \delta e_r=\delta e_\phi=\frac{6-15X^2/4}{2e}.
 \end{gathered}
 \label{eq:tb-conservative-input}
\end{equation}
They fix the change in the Newtonian polynomial through
$P_{\rm orbit}$ in \eqref{eq:P1-app}, the Bessel argument
through $e_t$, and the normalization through $\Omega_r$.
For the angular term, we use the continuation of
$\sum_n c_n\mathcal A_n$ under \eqref{eq:PM-continuous-choice},
with the branches established in App.~\ref{app:1pn} and
the Newtonian weights fixed by
\eqref{eq:allv-newton-weight-match}. We multiply its
precession contribution by an unknown constant $\zeta$,
so that $\zeta=1$ gives the conservative precession
coefficient in \eqref{eq:1pn-sc-spectrum}.
The relative clock phase in \eqref{eq:PM-loop-phase} is
supplied by conservative dynamics and multiplies the
Newtonian waveform. We match at independent
$\lambda=\omega/\Omega_r$, with $\epsilon>0$ and
$X=j\sqrt\epsilon$ on the continued side.\vskip 4pt

The polynomial correction is treated as in the previous case.
The finite shifts of the Bessel index reduce by
\eqref{eq:recurrence} to $J_\lambda$ and $J'_\lambda$,
giving, in the normalization of
\eqref{eq:PM-reduced-normalization},
\begin{equation}
 \epsilon\,
 \frac{A^{(0)}(e,\lambda)J_\lambda(\lambda e)
       +B^{(0)}(e,\lambda)eJ'_\lambda(\lambda e)}{e^2}.
 \label{eq:tb-frequency-ansatz}
\end{equation}
The PN polynomial sector supplies the powers of $X$ and
$\lambda$ multiplying these functions. Keeping their weights
independent, we write
\begin{equation}
 \begin{aligned}
 A^{(0)}(e,\lambda)
 &=a_0^{(0)}\lambda^2X^4+a_1^{(0)}\lambda X^3
   +a_2^{(0)}X^2+a_3^{(0)}\lambda X+a_4^{(0)},\\
 B^{(0)}(e,\lambda)
 &=b_0^{(0)}\lambda^2X^3+b_1^{(0)}\lambda X^2
   +b_2^{(0)}X+b_3^{(0)}\lambda.
 \end{aligned}
 \label{eq:tb-polynomial-class}
\end{equation}
There are nine polynomial constants, in addition to
the angular weight $\zeta$. At fixed $x$, $\lambda=x/X$ in this
explicit 1PN correction and $e=\ii\sqrt{X^2-1}$ on the
reference sheet. The first three terms of each polynomial
then contribute at leading order, while those containing
$a_3^{(0)},a_4^{(0)},b_3^{(0)}$ first contribute at $1/X^2$.
\vskip 4pt

The leading information comes from the probe tree spectrum
\cite{Bini:2024rsy,Bini:2024ijq},
\begin{align}
 \left.S^U_{2,2\varsigma}\right|_{1\rm PN,\nu,\text{tree}}
 =\frac{\nu\Cnorm GM^2p_\infty}{21}\Bigl\{
 &(38u^2-2\varsigma u+78)K_0(u)\nonumber\\
 &+[38\varsigma u^2+17u-138\varsigma]K_1(u)\Bigr\}.
 \label{eq:PM-tree-full}
\end{align}
Applying \eqref{eq:PM-continuous-choice} to these two modes,
with the conversions in
\eqref{eq:PM-shifted-argument}--\eqref{eq:PM-loop-phase},
puts their continued combination in the variables and
normalization of the ansatz, including the Newtonian
conversion term $-7h_0/4+xh'_0/4$ at leading order in $1/X$.\vskip 4pt 

The angular contribution starts at $1/X$, so the leading
coefficient contains only the polynomial weights and the
supplied orbital and clock corrections. To isolate the
weights, we subtract the latter contribution,
\begin{equation}
 \begin{aligned}
 &-\frac{15}{8}\bigl[2h_0(x)-xh'_0(x)\bigr]
 +\left(\frac{17x^2}{4}-5x+\frac92\right)I_0(x)\\
 &\hspace{24mm}
 +\left(\frac{17x^2}{4}-\frac{57x}{8}+\frac{21}{4}\right)I_1(x).
 \end{aligned}
 \label{eq:tb-conservative-tree}
\end{equation}
The first bracket comes from normalization and frequency
conversion; the other terms come from $P_{\rm orbit}$ and
the Bessel argument. The continued tree then leaves
\begin{equation}
 \left(-\frac{2x^2}{21}+\frac{22x}{21}-\frac9{14}\right)I_0(x)
 +\left(-\frac{2x^2}{21}+\frac{23x}{21}-\frac97\right)I_1(x).
 \label{eq:tb-tree-target}
\end{equation}
On the ansatz side, the leading coefficient of
\eqref{eq:tb-frequency-ansatz} is
$-(a_0^{(0)}x^2+a_1^{(0)}x+a_2^{(0)})I_0(x)
 -(b_0^{(0)}x^2+b_1^{(0)}x+b_2^{(0)})I_1(x)$.
Matching the constant, linear and quadratic terms for
independent $x$ gives
\begin{equation}
 \begin{aligned}
 (a_0^{(0)},a_1^{(0)},a_2^{(0)})
 &=\left(\frac2{21},-\frac{22}{21},\frac9{14}\right),\\
 (b_0^{(0)},b_1^{(0)},b_2^{(0)})
 &=\left(\frac2{21},-\frac{23}{21},\frac97\right).
 \end{aligned}
 \label{eq:tb-tree-matching}
\end{equation}
At the next order, $1/X$, these six weights generate the
Bessel terms illustrated in \eqref{eq:PM-loop-from-order}.
The angular function also enters, while the three remaining
polynomial weights still do not contribute. We therefore
subtract the tree-generated Bessel terms and the supplied
orbital and clock corrections from the continued loop to
determine $\zeta$. At this order the angular integrals
reduce to exponentials: the two routes in
\eqref{eq:PMroute-table} give $e^x$ and $e^{-x}$.
Equating their coefficients gives
\begin{equation}
 (\zeta-1)\left[
 \left(-6x+6-\frac3x\right)e^x-\frac3x e^{-x}
 \right]=0,
 \qquad \zeta=1.
 \label{eq:tb-loop-angular-weight}
\end{equation}
Both exponential coefficients give the precession
normalization $\zeta=1$. The ordinary Bessel terms agree
with those fixed by the tree and require no further weight.\vskip 4pt

The $G^3$ coefficient fixes
$a_3^{(0)},a_4^{(0)},b_3^{(0)}$, which enter at $1/X^2$.
From the continued result of~\cite{Bini:2024ijq}, we
subtract the second-order Bessel expansion of the tree
weights, the angular contribution with $\zeta=1$, and
the supplied orbital and clock terms. This includes the
ordinary Bessel terms generated by the angular function.
The remaining equation is
\begin{equation}
 \begin{aligned}
 &-(a_3^{(0)}x+a_4^{(0)})I_0(x)-b_3^{(0)}xI_1(x)\\
 &\qquad=\left(\frac{x}{3}-\frac9{14}\right)I_0(x)
              +\frac{2x}{7}I_1(x).
 \end{aligned}
 \label{eq:tb-third-matching}
\end{equation}
The constant and linear terms fix
$a_3^{(0)}=-1/3$, $a_4^{(0)}=9/14$ and $b_3^{(0)}=-2/7$.
Combining them with the tree values gives
\begin{equation}
 \begin{aligned}
 A^{(0)}&=\frac{2\lambda^2X^4}{21}
       -\frac{22\lambda X^3}{21}-\frac{\lambda X}{3}
       +\frac9{14}(X^2+1),\\
 B^{(0)}&=\frac{2\lambda^2X^3}{21}
       -\frac{23\lambda X^2}{21}-\frac{2\lambda}{7}
       +\frac{9X}{7}.
 \end{aligned}
 \label{eq:tb-fitted-weights}
\end{equation}
We combine these polynomials with the Newtonian expression,
the supplied orbital corrections and the angular term
with $\zeta=1$, then evaluate the resulting frequency
expression at
$\lambda=p+2\epsilon k_1$, as in \eqref{eq:PMfreqshift}.
The frequency prefactor, Bessel order and argument are
shifted together. The cancellations in
\eqref{eq:1pn-order-shift}--\eqref{eq:1pn-secular-cancellation}
then put the result in the three sums of
\eqref{eq:1pn-map-derivatives}: the polynomial correction,
the argument shift and the periodic angular term.\vskip 4pt

To compare with \eqref{eq:published-bound-full}, we use
\eqref{eq:1pn-angular-ibp} to express the angular sum as
$\widetilde f_p(e)$ and finite ordinary Bessel terms.
Adding the latter to the fitted polynomial and orbital
corrections gives the same coefficients of $J_p,J'_p$
as the bound harmonic after the variable conversion
in \eqref{eq:publishedH1}. Its angular function has the
normalization fixed by the loop. The scattering
weights, within the supplied PN frequency class, therefore give
the full probe contribution $H_1^{(0)}(e,p)$ at finite
eccentricity.

\subsection{The relativistic clock and angle}
\label{app:PM-allv}
To obtain the clock coefficients used in
\eqref{eq:allv-reference-integral}, we start from the 1PM
radial momentum~\cite{Kalin:2019rwq},
\begin{equation}
 p_r^2=P_\infty^2+\frac{2C}{r}-\frac{J^2}{r^2},
 \qquad P_\infty=\frac{\mu p_\infty}{\Gamma},\qquad
 C=\frac{GM\mu^2(2\gamma^2-1)}{\Gamma}.
 \label{eq:allv-radial-action}
\end{equation}
Here $P_\infty$ is the physical relative momentum and
$E=M\Gamma$ the total centre-of-mass energy. The relation
$\dd t/\dd r=\partial_Ep_r$, with $J$ and the masses held
fixed, gives the scattering time. At large outgoing separation,
$p_r=P_\infty+C/(P_\infty r)+\cdots$. The first term gives
the time proportional to $r$, while the $1/r$ term gives
a logarithm with coefficient
\begin{equation}
 \partial_E(C/P_\infty)
 =\partial_E\!\left[\frac{GM\mu(2\gamma^2-1)}{p_\infty}\right]
 =-\frac{GE\gamma(3-2\gamma^2)}{p_\infty^3}.
 \label{eq:allv-log-time-derivative}
\end{equation}
Since $v$ grows as $\log r$, this is the coefficient of $v$
in the clock. Hence $\alpha_{\rm 1PM}=-\omega\partial_E(C/P_\infty)$,
which gives \eqref{eq:allv-delay-coefficient} and identifies
the time-delay contribution to the loop.\vskip 4pt 

The coefficient of $\sinh v$ follows from the same time
integral before the large-separation expansion. Setting
$t=0$ at periastron, using $b=J/P_\infty$ and
$\partial_E\gamma=\Gamma/\mu$, gives
\begin{equation}
 t(v)=\sqrt{b^2+\frac{C^2}{P_\infty^4}}\,
       \partial_E P_\infty\sinh v
       +\partial_E(C/P_\infty)v,
 \qquad
 \partial_E P_\infty=\frac{\gamma}{p_\infty}
                   -\frac{\nu p_\infty}{\Gamma^2}.
 \label{eq:allv-reference-clock}
\end{equation}
Multiplying the coefficient of $\sinh v$ by $\omega$ gives
$z_{t,\rm 1PM}$. At fixed $b$ its leading term is independent
of $G$, while $C^2$ first contributes at $G^2$, yielding
\eqref{eq:allv-flight-time-main}. In the Newtonian limit,
the two clock coefficients reduce to $\alpha=\omega/n_s$
and $z_t=\alpha e_s$.\vskip 4pt

Together with the conservative precession coefficient
\eqref{eq:allv-periastron-full}, these clock coefficients
supply the higher-PM contributions considered next.

\subsection{Higher-PM terms from the clock}
The clock insertions in \eqref{eq:allv-clock-insertions}
act on the lower-PM waveform inside its Fourier integral.
We write the instantaneous probe coefficients as
\begin{equation}
 \left.S^U_{22}\right|_{G^k, {\rm probe}}
 =\frac{\Cnorm\mu GM}{p_\infty}
 \left(\frac{GM}{bp_\infty^2}\right)^{k-1}
 \int_{-\infty}^{\infty}\dd\tau\,e^{\ii u\tau}
 f_k(\tau;p_\infty).
 \label{eq:allv-PM-kernel-normalization}
\end{equation}
Here $f_k$ includes the anomaly measure, and $f_1,f_2$
are the tree and loop coefficients. Applying
\eqref{eq:allv-clock-insertions} with one insertion and
$v=\mathrm{arcsinh}(\tau/\gamma)$ gives
$\gamma(3-2\gamma^2)\,\dd(vf_1)/\dd\tau$ in $f_2$.
This includes both the prefactor and the Bessel-order shift;
the integration by parts uses the Fourier prescription
\eqref{eq:1pn-time-Fourier-check}. Subtracting it leaves
\begin{equation}
 f_2^{\rm rem}=f_2-\gamma(3-2\gamma^2)
                   \frac{\dd}{\dd\tau}(vf_1).
 \label{eq:allv-loop-clock-subtraction}
\end{equation}
At $G^5$, the four insertions on the tree and three on
this loop remainder give
\begin{equation}
 \left.f_5\right|_{\text{tree+loop clock}}
 =\frac{[\gamma(3-2\gamma^2)]^4}{4!}
       \frac{\dd^4}{\dd\tau^4}(v^4f_1)
 +\frac{[\gamma(3-2\gamma^2)]^3}{3!}
       \frac{\dd^3}{\dd\tau^3}(v^3f_2^{\rm rem}).
 \label{eq:allv-G5-clock-operation}
\end{equation}
When all four derivatives act on $f_1$, the first term
retains the $v^4$ scaling. The coefficient of $v^3$, however, also receives contributions
from both terms. Keeping their $\gamma$ dependence until
the PN expansion, we obtain the two highest logarithmic
powers compared in the main text. Through relative 2PN,
their coefficients agree with those in the Fourier integrals
of~\cite{Geralico:2026kbm,Geralico:2026efi}: the highest
power at $G^2$--$G^5$, and the next at $G^3$--$G^5$.

The lower powers at $G^5$ also receive two insertions on
the $G^3$ remainder and one on the $G^4$ remainder, defined
by subtracting their lower-order clock contributions as in
\eqref{eq:allv-loop-clock-subtraction}. Including these
inputs gives the coefficients of every positive power of
$\mathrm{arcsinh}\tau$ through 2PN in the same scattering
result. The term without this logarithm requires independent
$G^5$ waveform information.

\subsection{The square of the precession factor}
The square in \eqref{eq:allv-periastron-expansion} first
contributes at 2PN, so it needs only leading precession
and the Newtonian angular function:
\begin{equation}
 2(\Korb_s-1)^2(\log R)^2
 =\frac{18}{j^4}(\log R_0)^2+\cdots.
 \label{eq:allv-periastron-square}
\end{equation}
This factor multiplies the Newtonian source inside the
Fourier integral underlying \eqref{eq:mass-scatter-full}. In the probe limit,
$j=bp_\infty/(GM)$, so
$j^{-4}=p_\infty^4[GM/(bp_\infty^2)]^4$.
The precession square already contributes at $G^4$ through
the straight-line source. To extract its $G^5$ contribution,
we expand the Newtonian encounter one order further. At leading PM order,
$\log R_0=\ii\arctan\tau$, giving a term proportional to $(\arctan\tau)^2$. We put the frequency factors in the same integral form as
\eqref{eq:allv-PM-kernel-normalization}, using the integration
by parts in \eqref{eq:allv-clock-insertions}. Derivatives of
the angular factor lower its power. The coefficient that
still multiplies $(\arctan\tau)^2$ is therefore the
Newtonian tree kernel, yielding 
\begin{equation}
 \left.f_5\right|_{\mathrm{2PN},\,\arctan^2}
 =18p_\infty^4\,
 \frac{\tau^2+3-2\ii\tau(\tau^2+2)}{(1+\tau^2)^{3/2}}
 (\arctan\tau)^2.
 \label{eq:allv-periastron-square-scattering}
\end{equation}
It agrees with the corresponding coefficient in the probe
scattering result of~\cite{Geralico:2026kbm,Geralico:2026efi}.

\subsection{The bound clock and angular terms}
\label{app:PM-bound-clock-angle}
To compare the angular terms of \S\ref{sec:allv-bound}
with the known bound harmonic, we expand the factor in
\eqref{eq:allv-bound-joint-factor} together with the known
1PN polynomial, at fixed $e_t$.
The conversion $X^2=X_t^2+2\epsilon e_t\delta e_t+\cdots$
changes the $j^{-2}$ periastron contribution to
\begin{equation}
 \left.(\Korb-1)\right|_{j^{-2}}
 =\epsilon k_1+\epsilon^2 k_2 +\Ord(\epsilon^3),
 \qquad k_1=\frac3{X_t^2},\qquad
  k_2=\frac{3(2\nu-5)}{4X_t^2}
                    -\frac{6e_t\delta e_t}{X_t^4}.
 \label{eq:allv-bound-periastron-energy}
\end{equation}
Here $X_t^2=1-e_t^2$, and $\delta e_t$ is given in
\eqref{eq:et-main}. The first term in $k_2$
is the energy correction, $k_{2,\gamma}$, while the second comes
from expressing leading precession in terms of $e_t$.\vskip 4pt 

Besides the mixed and squared terms identified in
\S\ref{sec:allv-bound}, the frequency prefactor in
\eqref{eq:allv-bound-joint-factor} supplies
$8\epsilon^2k_1^2\mathcal L/p$. Leading precession also
multiplies the 1PN polynomial, whose coefficient at fixed
$e_t$ is $\delta c_n-\delta e_t\,\dd c_n/\dd e_t$.
Combining these contributions gives
 \begin{align}
  \label{eq:allv-bound-joint-input}
 \left.\delta U^b_{22,p}\right|_{\mathrm{B2B},\,\mathcal L,\mathcal L^2}
 ={}&\frac{\Cnorm\mu\epsilon^3p^2}{2}
 \oint_{|w|=1}\frac{\dd w}{2\pi\ii w}\,
 w^p e^{-pe_t(w-w^{-1})/2}\times\\
 &\sum_{n=-3}^{3}w^n\Biggl\{
 2k_1\left(\delta c_n-\delta e_t\frac{\dd c_n}{\dd e_t}\right)\mathcal L
 +c_n\left[
 \left(4n_1k_1+2k_2+\frac{8k_1^2}{p}\right)\mathcal L
 +2k_1^2\mathcal L^2\right]\Biggr\}\,,\nonumber
 \end{align}
where only terms containing $\mathcal L$ or $\mathcal L^2$
are displayed.\vskip 4pt 

In the bound harmonic of~\cite{Tessmer:2010ii}, the same
angular factors multiply the waveform amplitude. We denote the Newtonian
and 1PN parts as $Q_0+\epsilon Q_1$.\footnote{These are the terms
inside braces in their Eq.~(207), with the common precession
factor removed and coefficients given through Eq.~(218).
Here $\epsilon=2|E|$ and $\eta=\nu$, so $Q_1$ is half
their coefficient of $|E|$.} In the common Fourier variable $w=e^v$, we have
$1-e_t\cos u=-A$ and $\ii\sin u=\sinh v$,
with $A=e_t\cosh v-1$.
The Newtonian amplitude becomes\footnote{We keep only one factor of $|E|$ in the
1PN term denoted as $\alpha_{[42]}$ in their expression Eq. (211), which we believe has a typo.}
\begin{equation}
 Q_0(w)=2-e_t^2
 +\frac{e_t^2-2+2X_t e_t \sinh v}{-A}
 -\frac{2X_t^2}{A^2}
 +\frac{2X_t^4-2X_t^3e_t \sinh v}{(-A)^3}.
 \label{eq:allv-bound-published-square}
\end{equation}
The bound harmonic in \cite{Tessmer:2010ii} also contains the 2PN clock
correction beyond the 1PM clock used here, and the
$j^{-4}$ periastron contribution. We separate these terms
to compare the contributions generated by the same
conservative input in \eqref{eq:allv-bound-joint-input}.
The additional clock correction shifts the anomaly
in $Q_0$ at fixed Fourier phase, while the periastron term changes
the angular exponential. Their respective contributions
to the Fourier integrand of
$-\delta U^b_{22,p}/(\Cnorm\mu\epsilon^3)$ are
\begin{equation}
 \frac{3(5-2\nu)}{2X_t e_t^2}\,
       w\frac{\dd Q_0}{\dd w}\mathcal L,
 \qquad
 -\frac{15(7-2\nu)}{2X_t^4e_t^2}\,A Q_0\mathcal L.
 \label{eq:allv-bound-separate-conservative-terms}
\end{equation}
After subtracting them, the bound expression contains
the same conservative input as
\eqref{eq:allv-bound-joint-input} and reads
\begin{equation}
 \begin{aligned}
 \left.\delta U^b_{22,p}\right|_{\text{Ref.\cite{Tessmer:2010ii}},\,\mathcal L,\mathcal L^2}
 ={}&\Cnorm\mu\epsilon^3
 \oint_{|w|=1}\frac{\dd w}{2\pi\ii w}\,
 w^p e^{-pe_t(w-w^{-1})/2}\frac{A}{e_t^2}\\
 &\times\left\{\left[2k_1Q_1(w)+2k_2 Q_0(w)\right]
                  \mathcal L+2k_1^2Q_0(w)\mathcal L^2\right\}.
 \end{aligned}
 \label{eq:allv-bound-joint-target}
\end{equation}
To compare the two integrals, we transfer the derivatives
in the waveform to the Fourier factor, as in
\eqref{eq:1pn-angular-ibp}. For the $\mathcal L^2$ term
only the Newtonian amplitude enters. Its relation to the
polynomial $\sum_n c_nw^n=AB^2$ in \eqref{eq:P0}, with
$B=e_t-\cosh v+X_t\sinh v$, is
\begin{equation}
 \frac{A}{e_t^2}Q_0(w)
 =\frac{w}{2}\frac{\dd}{\dd w}
 \left[\frac{w}{A}\frac{\dd B^2}{\dd w}\right].
 \label{eq:allv-bound-log-reduction}
\end{equation}
Inserting this identity in \eqref{eq:allv-bound-joint-target}
allows two integrations by parts. The Fourier factor and
$\mathcal L$ are periodic for integer $p$, so there is no
boundary term. When both derivatives act on the Fourier
factor, they give $p^2AB^2/2$, reproducing the coefficient
of $2k_1^2\mathcal L^2$ in
\eqref{eq:allv-bound-joint-input}. Further integrations
by parts, using the Fourier derivative in
\eqref{eq:1pn-angular-ibp}, reduce this Laurent polynomial
to a constant and a term proportional to $w$ multiplying
$\mathcal L^2$. These are the two terms displayed in
\eqref{eq:allv-bound-precession-square-waveform}.\vskip 4pt

 To finish the angular comparison, we collect the terms
linear in $\mathcal L$. These come from the explicit
contribution $[2k_1Q_1+2k_2Q_0]\mathcal L$ in
\eqref{eq:allv-bound-joint-target}, but also from the
precession square: whenever integration by parts
differentiates $\mathcal L^2$, it produces
$2\mathcal L\,\dd\mathcal L/\dd w$. Both contributions
must therefore be included when comparing with the
linear part of \eqref{eq:allv-bound-joint-input}.

The clock--precession term is identified after accounting
for leading precession acting on the known 1PN polynomial
and for the frequency correction $8k_1^2c_n/p$.
The remaining coefficient of the Newtonian polynomial is
$2k_2+4n_1k_1$. Since $k_2$ includes the conversion from
$e$ to $e_t$, \eqref{eq:allv-bound-periastron-energy} gives
\begin{equation}
 2k_2+4n_1k_1
 =-\frac{12e_t\delta e_t}{X_t^4}
   +2k_{2,\gamma}+4n_1k_1.
\end{equation}
The first term is this eccentricity conversion. The
other two are the periastron energy correction and the
mixed clock-precession term. They multiply the same
Newtonian polynomial, and we land in the same
frequency integral as in 
\eqref{eq:allv-bound-clock-precession-waveform}, with
coefficients $2k_{2,\gamma}$ and $4n_1k_1$, respectively.
Together with the precession square in \eqref{eq:allv-bound-precession-square-waveform}, these
are the three angular contributions yielding the
bound harmonic for $p\geq1$, $0<e_t<1$ and arbitrary $\nu$.
\vskip 4pt

The terms without $\mathcal L$ do not separate in the
same way. For example, the angular-eccentricity correction
in \eqref{eq:allv-bound-angle-expansion} contains
\begin{equation}
 \partial_{e_t}\log R_0(v;e_t)
 =-\frac{\sinh v}{X_tA},
 \qquad
 AB^2\,\partial_{e_t}\log R_0
 =-\frac{B^2\sinh v}{X_t}.
 \label{eq:allv-bound-angular-conversion}
\end{equation}
The denominator cancels against the Newtonian polynomial,
leaving a finite sum of powers of $w$, whose integrals
are ordinary Bessels. Terms generated when the integrations
by parts remove $\mathcal L$ also contribute to these
Bessel coefficients. The bound waveform combines them
with the pure clock correction and further orbital and
radiative contributions at 2PN. Comparing this part
therefore requires the remaining radiative frequency
dependence; the angular comparison alone does not isolate
the pure clock correction. Preliminary work completing
this dependence in the probe limit reproduces the full
instantaneous 2PN bound harmonic at finite eccentricity.
We will present this reconstruction elsewhere.
\subsection{The extended frequency function}
\label{app:allv-energy-basis}

The additional function $\mathfrak F$ in
\eqref{eq:allv-extra-master} illustrates the frequency
dependence needed beyond these clock and precession terms.
Its two denominator factors can be separated without
expanding in velocity: acting with $-\partial_u^2+\gamma^2$
on the Fourier exponential supplies $\tau^2+\gamma^2$,
cancelling the second denominator. The remaining transform
is $2K_0(u)$ for $u>0$. Therefore, the continuation can be obtained
from the familiar Bessel identity through the equation
\begin{equation}
 (-\partial_u^2+\gamma^2)\mathfrak F=2K_0(u),\qquad
 \mathfrak F(0,\gamma)=
 \frac{2\,\mathrm{arcsinh}(p_\infty)}{\gamma p_\infty},\qquad
 \partial_u\mathfrak F(0,\gamma)=0.
 \label{eq:allv-master-equation}
\end{equation}
The initial values follow from the defining integral and fix
the homogeneous solution. Under $p_\infty\to\ii s$, the
first becomes $2\arcsin s/(\gamma s)$ on the followed branch.
Its contribution is proportional to $\cosh(\gamma u)$,
which has the same value on the two routes
$u=xe^{-\ii\pi}$ and $u=x$ and therefore cancels in their
difference. The source $2K_0$, however, differs by
$2\pi\ii I_0(x)$ according to \eqref{eq:PM-K01}.
The difference consequently obeys the same differential
operator with this source and zero initial value and derivative.
Solving it gives \eqref{eq:conclusion-master-continuation}.
\bibliographystyle{JHEP}
\bibliography{b2b_waveforms_IV}
\end{document}